\documentclass[longauth]{aa}

\usepackage{graphicx}
\usepackage{adjustbox,rotating}
\usepackage{lmodern}

\usepackage{threeparttable}

\usepackage{txfonts}

\providecommand{\bjdtdb}{\ensuremath{\rm {BJD_{TDB}}}}
\providecommand{\feh}{\ensuremath{\left[{\rm Fe}/{\rm H}\right]}}
\providecommand{\teff}{\ensuremath{T_{\rm eff}}}

\providecommand{\mj}{\ensuremath{\,M_{\rm J}}}
\providecommand{\rj}{\ensuremath{\,R_{\rm J}}}

\providecommand{\fave}{\langle F \rangle}
\providecommand{\fluxcgs}{10$^9$ erg s$^{-1}$ cm$^{-2}$}

\newcommand{\logg}{\mbox{$\log g_*$}\xspace}
\newcommand{\vsini}{\mbox{$v \sin i_{*}$}\xspace}
\newcommand{\kms}{\mbox{km\,s$^{-1}$}\xspace}
\newcommand{\ms}{\mbox{m\,s$^{-1}$}\xspace}

\newcommand{\mjup}{\mbox{$\mathrm{M_{\rm Jup}}$}\xspace}
\newcommand{\rjup}{\mbox{$\mathrm{R_{\rm Jup}}$}\xspace}

\newcommand{\rstar}{\mbox{$R_{*}$}\xspace}

\newcommand{\msol}{\mbox{$\mathrm{M_\odot}$}\xspace}
\newcommand{\rsol}{\mbox{$\mathrm{R_\odot}$}\xspace}

\newcommand{\lsol}{\mbox{$\mathrm{L_\odot}$}\xspace}

\usepackage{hyperref}
\begin{document} 


\title{Populating the brown dwarf and stellar boundary: Five stars with transiting companions near the hydrogen-burning mass limit}

\titlerunning{Five transiting objects near the H-burn limit}
\authorrunning{N. Grieves, et al.}

   \author{Nolan Grieves \inst{\ref{inst-geneva}}\fnmsep\thanks{\email{nolangrieves@gmail.com}}
    \and Fran\c{c}ois Bouchy \inst{\ref{inst-geneva}}
    \and Monika Lendl \inst{\ref{inst-geneva}}
    \and Theron Carmichael \inst{\ref{inst-harvard},\ref{inst-cfa}}
    \and Ismael Mireles \inst{\ref{inst-NM}}
    \and Avi Shporer \inst{\ref{inst-mit}}
    \and Kim K. McLeod \inst{\ref{inst-wellesley}}
    \and Karen A. Collins \inst{\ref{inst-cfa}}
    \and Rafael Brahm \inst{\ref{inst-adolfo},\ref{inst-mifa}}
    \and Keivan G. Stassun \inst{\ref{inst-vandy}}
    \and Sam Gill \inst{\ref{inst-war},\ref{inst-warexo}}
    \and Luke G. Bouma \inst{\ref{inst-princ}}
    \and Tristan Guillot \inst{\ref{inst-cotedazur}}
    \and Marion Cointepas \inst{\ref{inst-geneva},\ref{inst-grenob}}
    \and Leonardo A. Dos Santos \inst{\ref{inst-geneva}}
    \and Sarah L. Casewell \inst{\ref{inst:leic}}
    \and Jon M. Jenkins \inst{\ref{inst-ames}}
    \and Thomas Henning \inst{\ref{inst-maxpl}}
    \and Louise D. Nielsen \inst{\ref{inst-geneva}}
    \and Angelica Psaridi \inst{\ref{inst-geneva}}
    \and St\'ephane Udry \inst{\ref{inst-geneva}}
    \and Damien S\'egransan \inst{\ref{inst-geneva}}
    \and Jason D. Eastman \inst{\ref{inst-cfa}}
    \and George Zhou \inst{\ref{inst-cfa}}
    \and Lyu Abe \inst{\ref{inst-cotedazur}}
    \and Abelkrim Agabi \inst{\ref{inst-cotedazur}}
    \and Gaspar Bakos \inst{\ref{inst-princ},\ref{inst-prince2}}
    \and David Charbonneau \inst{\ref{inst-cfa}}
    \and Kevin I. Collins \inst{\ref{inst-georgemason}}
    \and Knicole~D.~Colon \inst{\ref{inst-goddard}}
    \and Nicolas Crouzet \inst{\ref{inst-noodwijk}}
    \and Georgina Dransfield \inst{\ref{inst-birmingham}}
    \and Phil Evans \inst{\ref{inst-elsauce}}
    \and Robert~F.~Goeke \inst{\ref{inst-mit}}
    \and Rhodes Hart \inst{\ref{inst-squeensland}}
    \and Jonathan M. Irwin \inst{\ref{inst-cfa}}
    \and Eric L.N. Jensen \inst{\ref{inst-swarthmore}}
    \and Andr\'es Jord\'an \inst{\ref{inst-adolfo},\ref{inst-mifa}}
    \and John F. Kielkopf \inst{\ref{inst-louisville}}
    \and David W. Latham \inst{\ref{inst-cfa}}
    \and Wenceslas Marie-Sainte \inst{\ref{inst-conco}}
    \and Djamel M\'ekarnia \inst{\ref{inst-cotedazur}}
    \and Peter Nelson \inst{\ref{inst-ellenbank}}
    \and Samuel N. Quinn \inst{\ref{inst-cfa}}
    \and Don J. Radford \inst{\ref{inst-brierfield}}
    \and David~R.~Rodriguez \inst{\ref{inst-stsci}}
    \and Pamela Rowden \inst{\ref{inst-RAS}}
    \and Fran\c{c}ois--Xavier Schmider \inst{\ref{inst-cotedazur}}
    \and Richard P. Schwarz  \inst{\ref{inst-patashnik}}
    \and Jeffrey C. Smith \inst{\ref{inst-seti},\ref{inst-ames}}
    \and Chris Stockdale \inst{\ref{inst-hazelwood}}
    \and Olga Suarez \inst{\ref{inst-cotedazur}}
    \and Thiam-Guan Tan \inst{\ref{inst-PEST}}
    \and Amaury H.M.J. Triaud \inst{\ref{inst-birmingham}}
    \and William Waalkes \inst{\ref{inst-colorado}}
    \and Geof Wingham \inst{\ref{inst-mtstuart}}
    }
     \institute{
   Observatoire de Gen{\`e}ve, Universit{\'e} de Gen{\`e}ve, 51 Ch. des Maillettes, 1290 Sauverny, Switzerland \label{inst-geneva}
    \and 
    Harvard University, Cambridge, MA 02138 \label{inst-harvard}
    \and
    Center for Astrophysics, Harvard \& Smithsonian, 60 Garden Street, Cambridge, MA 02138, USA \label{inst-cfa}
    \and
    Department of Physics and Astronomy, University of New Mexico, 210 Yale Blvd NE, Albuquerque, NM 87106, USA \label{inst-NM}
    \and
    Department of Physics and Kavli Institute for Astrophysics and Space Research, Massachusetts Institute of Technology, Cambridge, MA 02139, USA \label{inst-mit}
    \and 
    Department of Astronomy, Wellesley College, Wellesley, MA 02481, USA \label{inst-wellesley}
    \and
    Facultad de Ingenier{\'i}a y Ciencias, Universidad Adolfo Ib{\'a}\~{n}ez, Av. Diagonal las Torres 2640, Pe\~{n}alol{\'e}n, Santiago, Chile \label{inst-adolfo}
    \and
    Millennium Institute for Astrophysics, Chile \label{inst-mifa}
    \and
    Vanderbilt University, Department of Physics \& Astronomy, 6301 Stevenson Center Lane, Nashville, TN 37235, USA \label{inst-vandy}
    \and
    Department of Physics, University of Warwick, Gibbet Hill Road, Coventry, CV4 7AL, UK \label{inst-war}
    \and 
    Centre for Exoplanets and Habitability, University of Warwick, Gibbet Hill Road, Coventry, CV4 7AL, UK \label{inst-warexo}
    \and
   Department of Astrophysical Sciences, Princeton University, 4 Ivy Lane, Princeton, NJ 08540, USA \label{inst-princ}
     \and
    Universit\'e C\^ote d'Azur, Observatoire de la C\^ote d'Azur, CNRS, Laboratoire Lagrange, Bd de l'Observatoire, CS 34229, 06304 Nice cedex 4, France \label{inst-cotedazur}
    \and
   University of Grenoble Alpes, CNRS, IPAG, F-38000 Grenoble, France \label{inst-grenob}
   \and
    School of Physics and Astronomy, University of Leicester, LE1 7RH, UK \label{inst:leic}
    \and NASA Ames Research Center, Moffett Field, CA 94035, USA \label{inst-ames}
    \and Max-Planck-Institut f\"{u}r Astronomie, K\"{o}nigstuhl  17, 69117 Heidelberg, Germany \label{inst-maxpl}
    \and Institute for Advanced Study, Princeton, NJ 08540 \label{inst-prince2}
    George Mason University, 4400 University Drive, Fairfax, VA, 22030 USA \label{inst-georgemason}
    \and 
    NASA Goddard Space Flight Center, Exoplanets and Stellar Astrophysics Laboratory (Code 667), Greenbelt, MD 20771, USA \label{inst-goddard}
    \and
    European Space Agency (ESA), European Space Research and Technology Centre (ESTEC), Keplerlaan 1, 2201 AZ Noordwijk, The Netherlands\label{inst-noodwijk}
    \and
    School of Physics \& Astronomy, University of Birmingham, Edgbaston, Birmingham B15 2TT, United Kingdom \label{inst-birmingham}
    \and
    El Sauce Observatory, Coquimbo Province, Chile \label{inst-elsauce}
    \and
    Centre for Astrophysics, University of Southern Queensland, Toowoomba, QLD, 4350, Australia \label{inst-squeensland}
    \and
    Dept.\ of Physics \& Astronomy, Swarthmore College, Swarthmore PA 19081, USA \label{inst-swarthmore}
    \and
    Department of Physics and Astronomy, University of Louisville, Louisville, KY 40292, USA \label{inst-louisville}
    \and Concordia Station, IPEV/PNRA, Antarctica \label{inst-conco}
    \and
    Ellinbank Observatory, Australia \label{inst-ellenbank}
    \and Brierfield Observatory, New South Wales, Australia \label{inst-brierfield}
    \and Space Telescope Science Institute, 3700 San Martin Drive, Baltimore, MD, 21218, USA \label{inst-stsci}
    \and Royal Astronomical Society, Burlington House, Piccadilly, London W1J 0BQ, UK \label{inst-RAS}
    \and
    Patashnick Voorheesville Observatory, Voorheesville, NY 12186, USA \label{inst-patashnik}
    \and SETI Institute, Mountain View, CA 94043, USA \label{inst-seti}
    \and Hazelwood Observatory, Australia \label{inst-hazelwood}
    \and Perth Exoplanet Survey Telescope, Perth, Western Australia \label{inst-PEST}
    \and Department of Astrophysical and Planetary Sciences, University of Colorado, Boulder, CO 80309, USA \label{inst-colorado}
    \and Mt. Stuart Observatory, New Zealand\label{inst-mtstuart}
    }
   \date{Received XX; accepted XX}


 \cleardoublepage
    
    \abstract{We report the discovery of five transiting companions near the hydrogen-burning mass limit in close orbits around main sequence stars originally identified by the Transiting Exoplanet Survey Satellite (TESS) as TESS Objects of Interest (TOIs): TOI-148, TOI-587, TOI-681, TOI-746, and TOI-1213. Using TESS and ground-based photometry as well as radial velocities from the CORALIE, CHIRON, TRES, and FEROS spectrographs, we found the companions have orbital periods between 4.8 and 27.2\,days, masses between 77 and 98\,$\mjup$, and radii between 0.81 and 1.66\,$\rjup$. These targets have masses near the uncertain lower limit of hydrogen core fusion ($\sim$73-96\,$\mjup$), which separates brown dwarfs and low-mass stars. We constrained young ages for TOI-587 (0.2\,$\pm$\,0.1 Gyr) and TOI-681 (0.17\,$\pm$\,0.03 Gyr) and found them to have relatively larger radii compared to other transiting companions of a similar mass. Conversely we estimated older ages for TOI-148 and TOI-746 and found them to have relatively smaller companion radii. With an effective temperature of 9800\,$\pm$\,200\,K, TOI-587 is the hottest known main-sequence star to host a transiting brown dwarf or very low-mass star. We found evidence of spin-orbit synchronization for TOI-148 and TOI-746 as well as tidal circularization for TOI-148. These companions add to the population of brown dwarfs and very low-mass stars with well measured parameters ideal to test formation models of these rare objects, the origin of the brown dwarf desert, and the distinction between brown dwarfs and hydrogen-burning main sequence stars.}

   \keywords{Stars: low-mass, brown dwarfs; binaries: eclipsing}
\maketitle

%

 

   

%

\section{Introduction}

Brown dwarfs are objects with masses in between giant planets and low-mass stars. They are often defined with a lower limit of $\sim$13\,$\mjup$, the approximate mass at which an object can begin to ignite deuterium fusion in its core, and with an upper limit of $\sim$80\,$\mjup$, the approximate mass at which an object becomes sufficiently massive to fuse hydrogen nuclei into helium nuclei within its core: the principal characteristic of a main-sequence star. However, these boundaries are not clear-cut as the exact masses where deuterium and hydrogen fusion occur depend on the chemical composition of the object \citep[e.g.,][]{Baraffe2002,Spiegel2011,Dieterich2014}. A defining characteristic of brown dwarfs is their relative low occurrence rate ($\lessapprox$1\%) in close orbits ($\lessapprox$5 AU) around main-sequence stars compared to giant planets and other stars, or the 'brown dwarf desert' \citep[e.g.,][]{MarcyButler2000,GretherLineweaver2006,Sahlmann2011,Santerne2016,Grieves2017}, with recent studies finding a dry desert for periods\,$<$\,100\,days \citep[e.g.,][]{Kiefer2019,Kiefer2021}.


The relative lack of brown dwarf companions may be related to a transition of the formation mechanisms required to form giant planets and low-mass stars. In this case, lower mass brown dwarfs may form similar to giant planets via core accretion \citep{Pollack1996} or disk instability \citep{Cameron1978,Boss1997} and higher mass brown dwarfs may form similar to stars from gravitational collapse and turbulent fragmentation of molecular clouds \citep{PadoanNordlund2004,HennebelleChabrier2008}. The boundary of these formation mechanisms is unclear and certainly depends on an object's initial environment. 


Using a statistical study of 62 brown dwarfs \citet{MaGe2014} found the ``driest'' part of the desert in the mass range 35\,$\mjup$\,$<$\,$M_b$\,sin\,$i$ $<$ 55\,$\mjup$ with periods less than 100 days. \citet{MaGe2014} also suggest 42.5\,$\mjup$ may represent a transition between brown dwarfs that formed more similar to giant planets and those that formed more similar to main-sequence stars, as they found that brown dwarfs with masses above 42.5\,$\mjup$ have an eccentricity distribution more consistent with binaries. However, \citet{MaGe2014} were limited by a small sample size and more brown dwarfs have been found even in the driest part of the desert \citep[e.g.,][]{Persson2019,Carmichael2019}. Other studies have also suggested two separate populations for lower and higher mass brown dwarfs based on metallicity and eccentricity distributions \citep[e.g.,][]{Maldonado2017,Kiefer2021}. Conversely, some studies have suggested a more continuous formation between giant planets and low-mass stars where giant planets range from 0.3 - 60 or 73\,$\mjup$ based on a continuum of their mass-density relation \citep{HatzesRauer2015,Persson2019}. While \citet{Whitworth2018} argues brown dwarfs should not be distinguished from hydrogen-burning stars as they have more similarities to stars than planets. 

\begin{figure}
  \centering
  \includegraphics[width=0.45\textwidth]{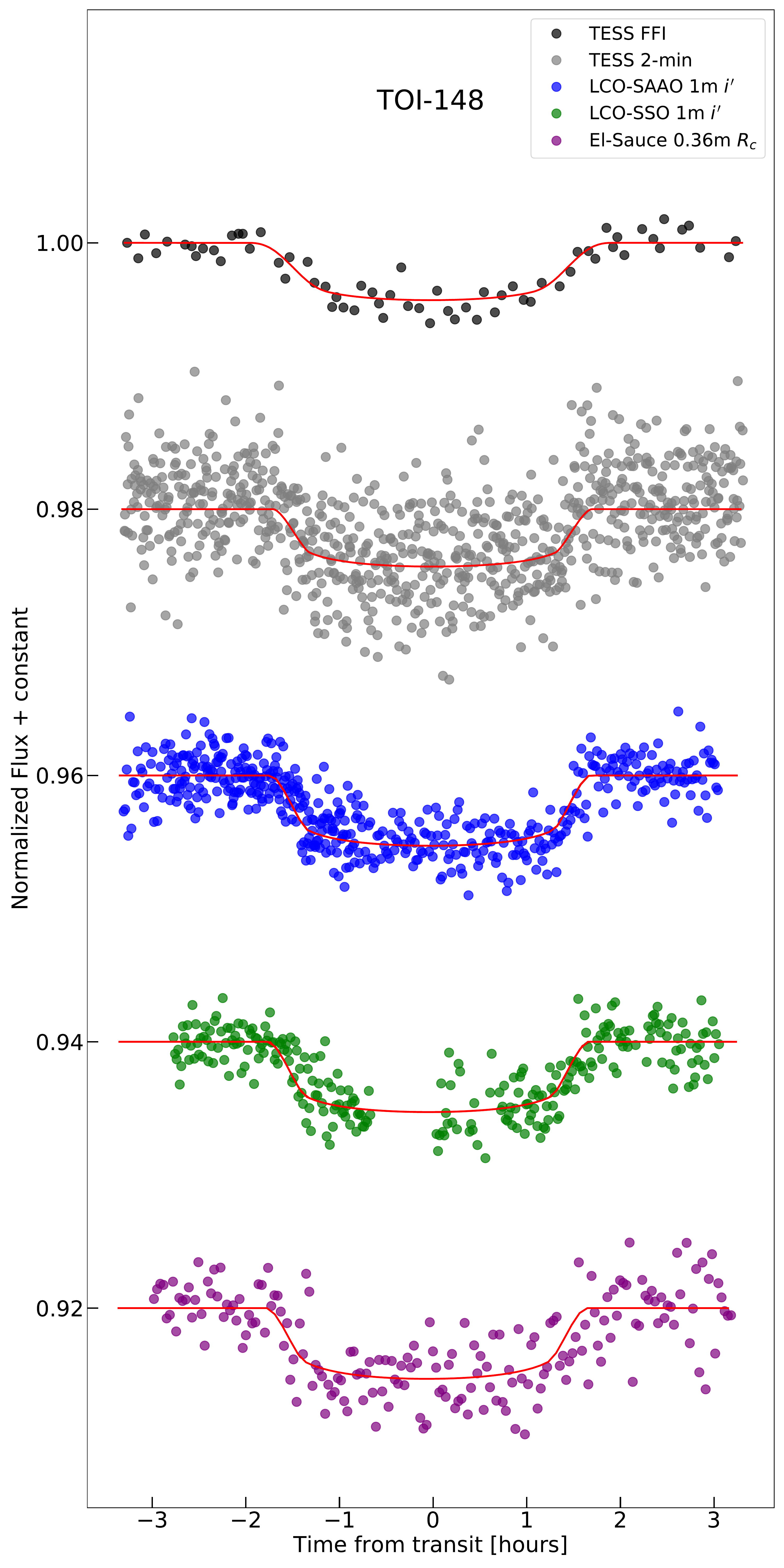}
    \includegraphics[width=0.5\textwidth]{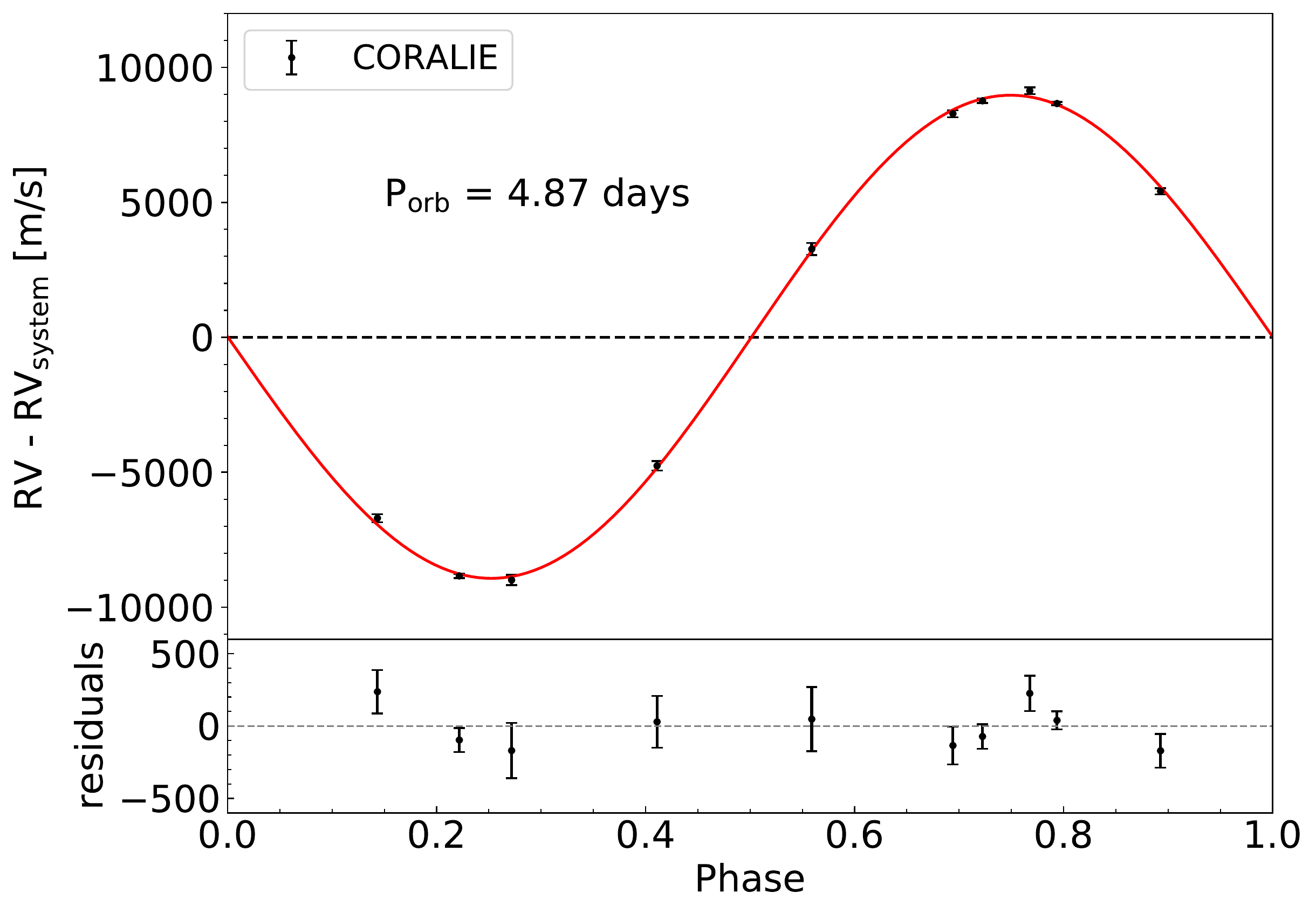}
  \caption{TOI-148 light curves (\textit{top}) and RVs (\textit{bottom}) phased to the companion's orbital period. The red lines in the upper plot show the best-fit transit model to each photometry data set from our EXOFASTv2 analysis described in Section \ref{sec:exofast}. In the lower plot the red line shows the best-fit Keplerian model to the RVs from our EXOFASTv2 analysis.}
  \label{fig:148phase}
\end{figure}

\begin{figure}
  \centering
  \includegraphics[width=0.47\textwidth]{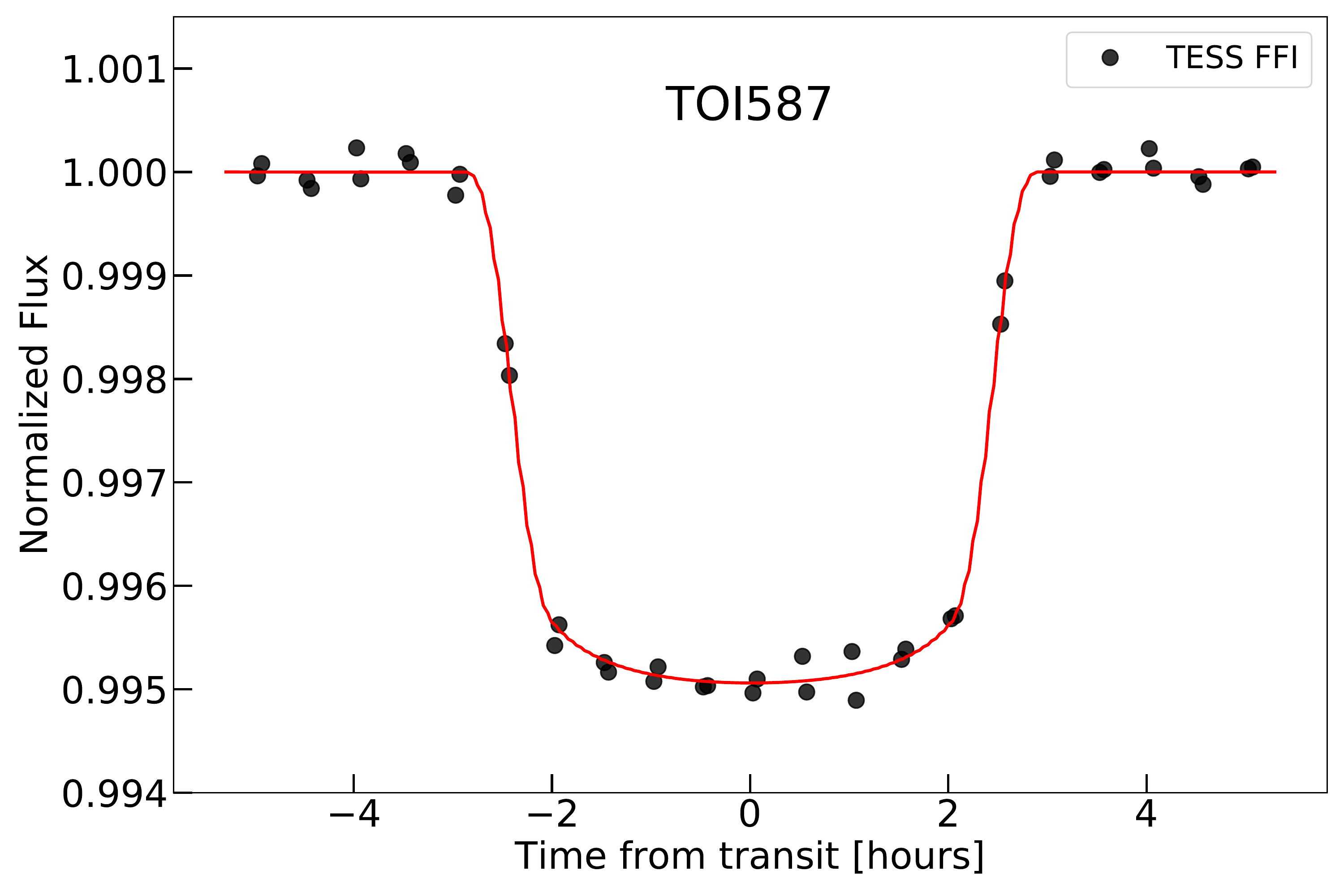}
    \includegraphics[width=0.47\textwidth]{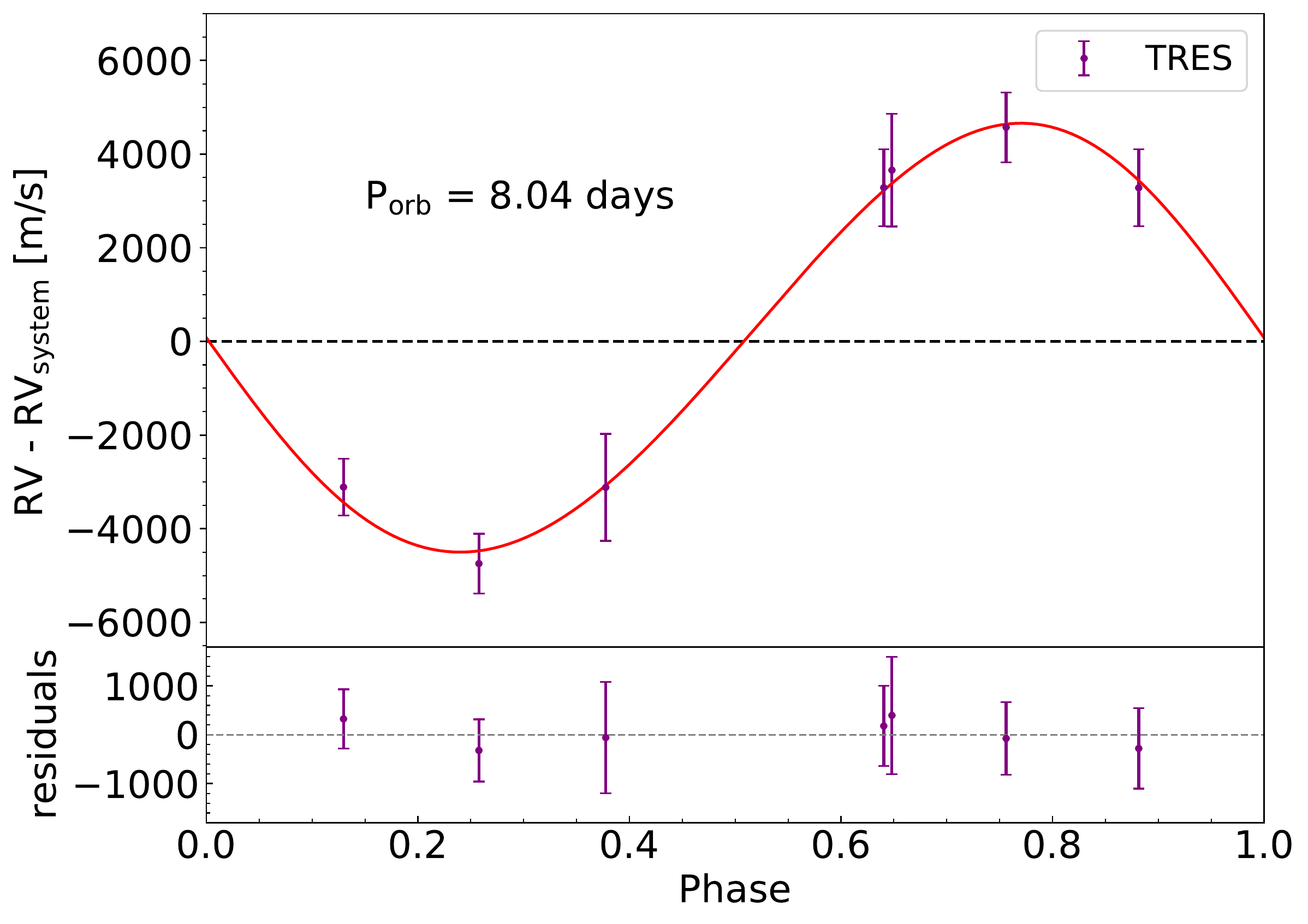}
  \caption{TOI-587 light curves (\textit{top}) and RVs (\textit{bottom}) phased to the companion's orbital period. The red lines display our best-fit models from EXOFASTv2 as described in Figure \ref{fig:148phase}.}
  \label{fig:587phase}
\end{figure}

Individual well-characterized brown dwarfs and very low-mass stars provide crucial insight into these possible formation scenarios. Particularly transiting brown dwarfs whose radii can be precisely determined allowing for a better interpretation of models, unlike brown dwarfs detected via radial velocities only, whose upper mass limits are not clearly defined and may actually be low-mass stars \citep[e.g.,][]{Kiefer2021}. Space-based photometric missions are ideal to find these rare objects given their robust photometric precision and long uninterrupted observations, and recent missions such as $Kepler$ \citep{Borucki2010} and $K2$ \citep{Howell2014} have led to successful follow-up mass measurements of transiting brown dwarfs \citep[e.g.,][]{Bayliss2017,Canas2018,Carmichael2019,Persson2019}. The currently operating space-based all-sky Transiting Exoplanet Survey Satellite \citep[TESS;][]{Ricker2015} has continued to populate the sparse desert with well-characterized brown dwarfs and very low-mass stars \citep[e.g.,][]{Subjak2020, Carmichael2020,Carmichael2021,Mireles2020}.

We report the discovery of five transiting companions with masses close to the upper boundary of the brown dwarf regime that were each first identified as TESS Objects of Interests (TOIs): TOI-148, TOI-587, TOI-681, TOI-746, and TOI-1213. Each object has robust photometric and spectroscopic measurements allowing precisely determined characteristics, and given their close proximity to the hydrogen-burning limit they are ideal to test current formation models and comparisons between brown dwarfs and low-mass stars. In Section \ref{sec:obs} we describe our photometry, spectroscopy, and imaging observations. In Section \ref{sec:analysis} we detail our analysis of the systems including the host stars and their companions. In Section \ref{sec:disc} we discuss our results and in Section \ref{sec:conc} we give our conclusions.



\section{Observations} \label{sec:obs}

\begin{figure}
  \centering
  \includegraphics[width=0.47\textwidth]{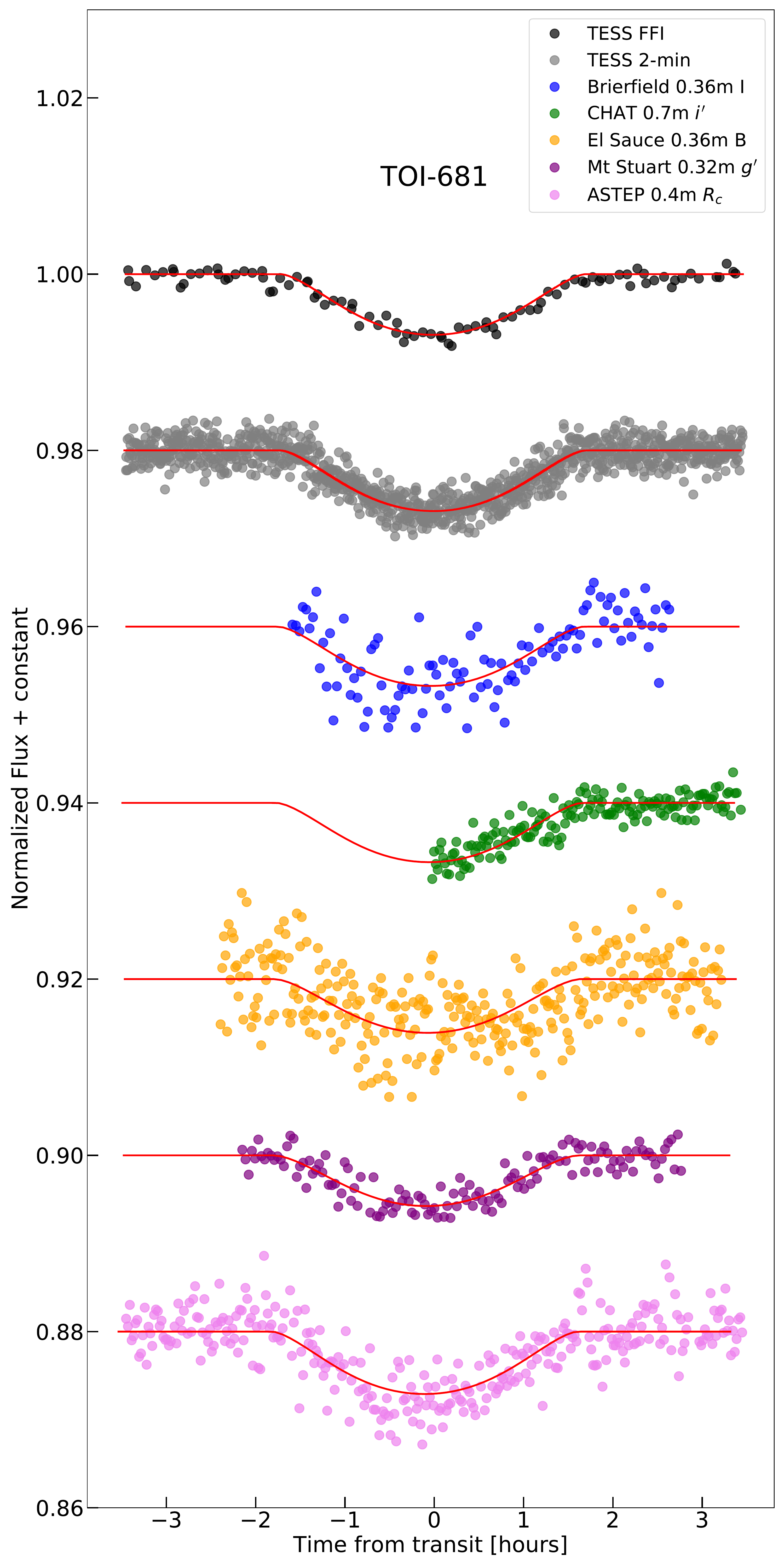}
    \includegraphics[width=0.47\textwidth]{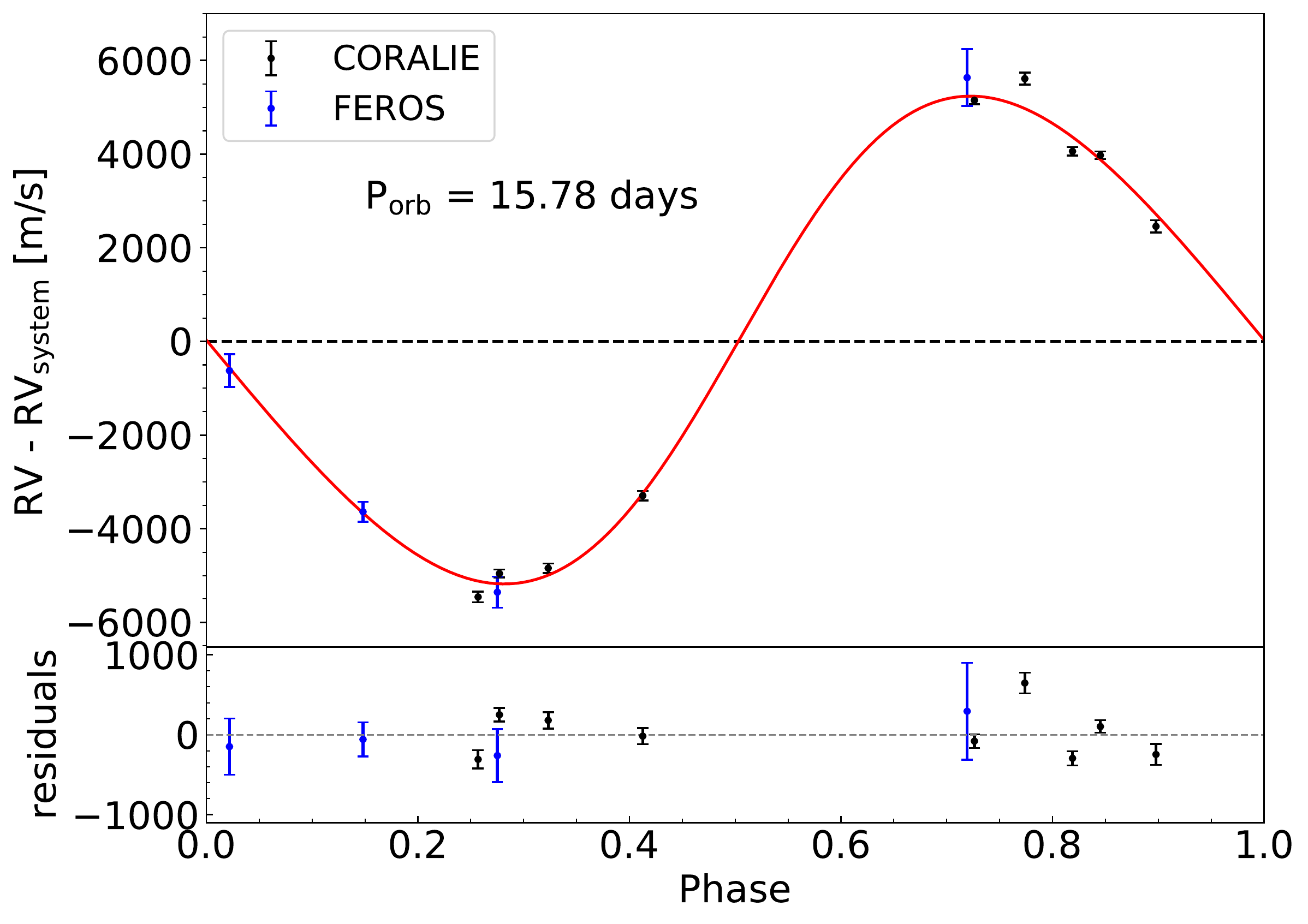}
  \caption{TOI-681 light curves (\textit{top}) and RVs (\textit{bottom}) phased to the companion's orbital period. The red lines display our best-fit models from EXOFASTv2 as described in Figure \ref{fig:148phase}.}
  \label{fig:681phase}
\end{figure}

\begin{figure}
  \centering
  \includegraphics[width=0.47\textwidth]{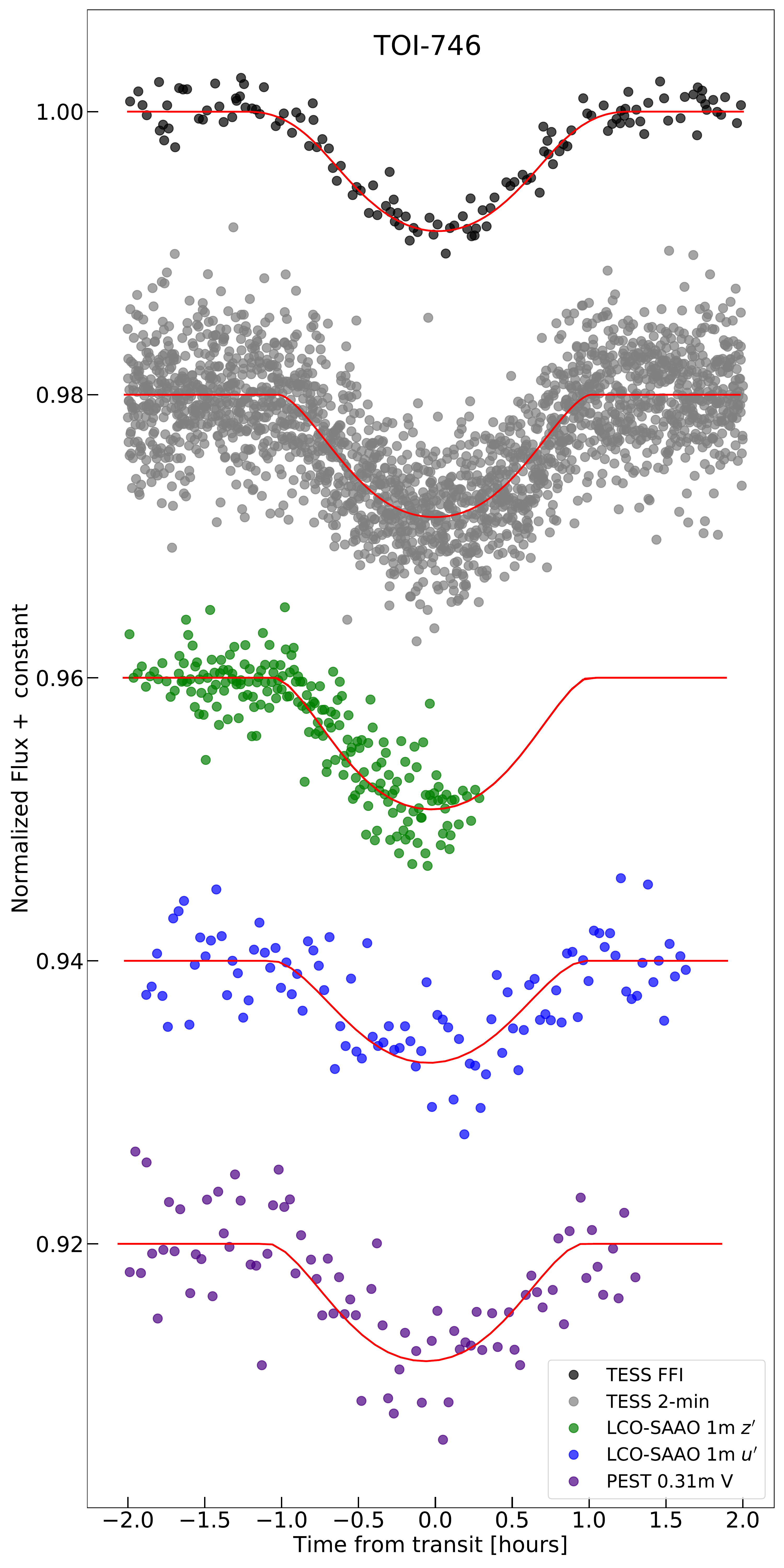}
    \includegraphics[width=0.47\textwidth]{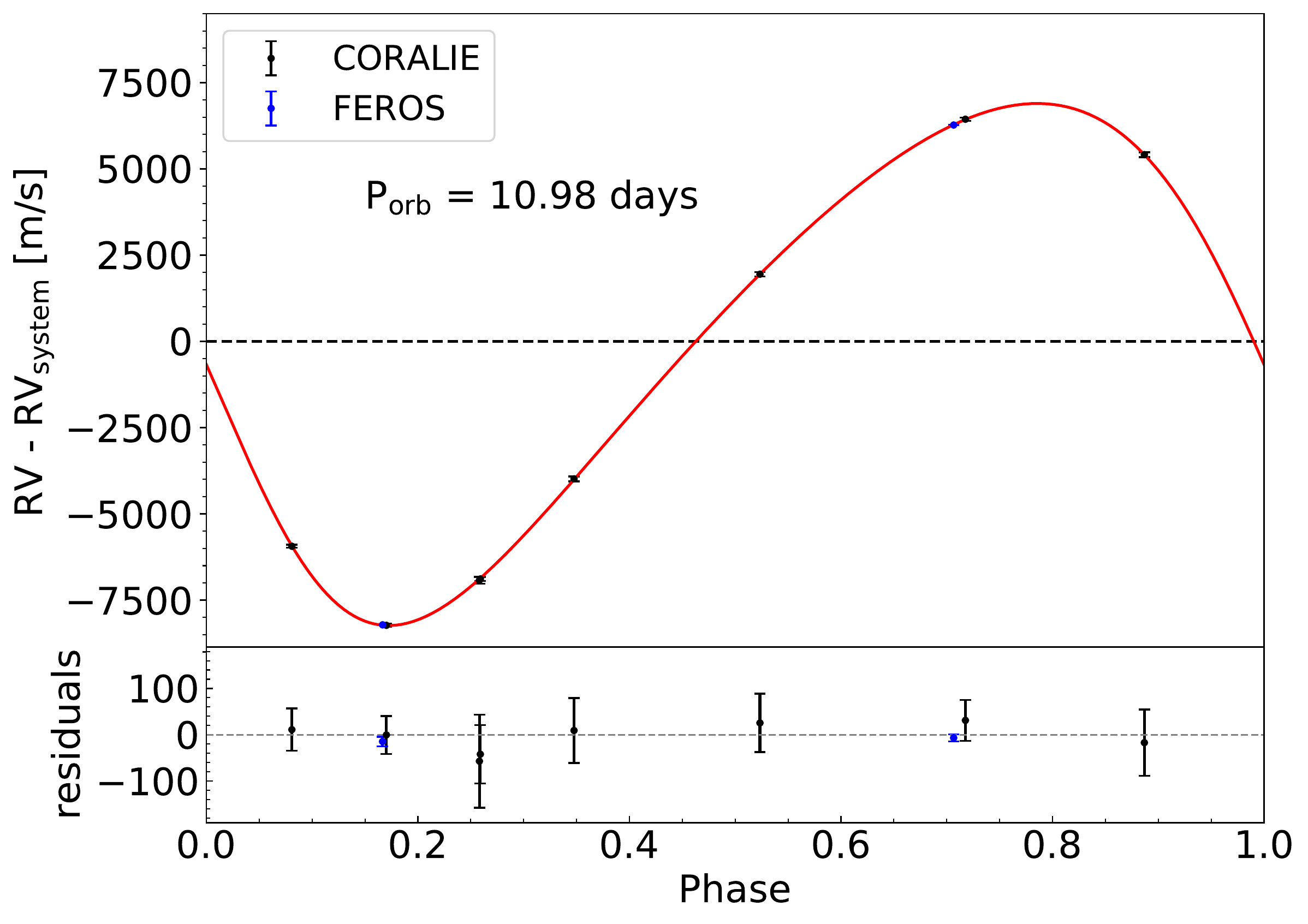}
  \caption{TOI-746 light curves (\textit{top}) and RVs (\textit{bottom}) phased to the companion's orbital period. The red lines display our best-fit models from EXOFASTv2 as described in Figure \ref{fig:148phase}.}
  \label{fig:746phase}
\end{figure}

\begin{figure}
  \centering
  \includegraphics[width=0.47\textwidth]{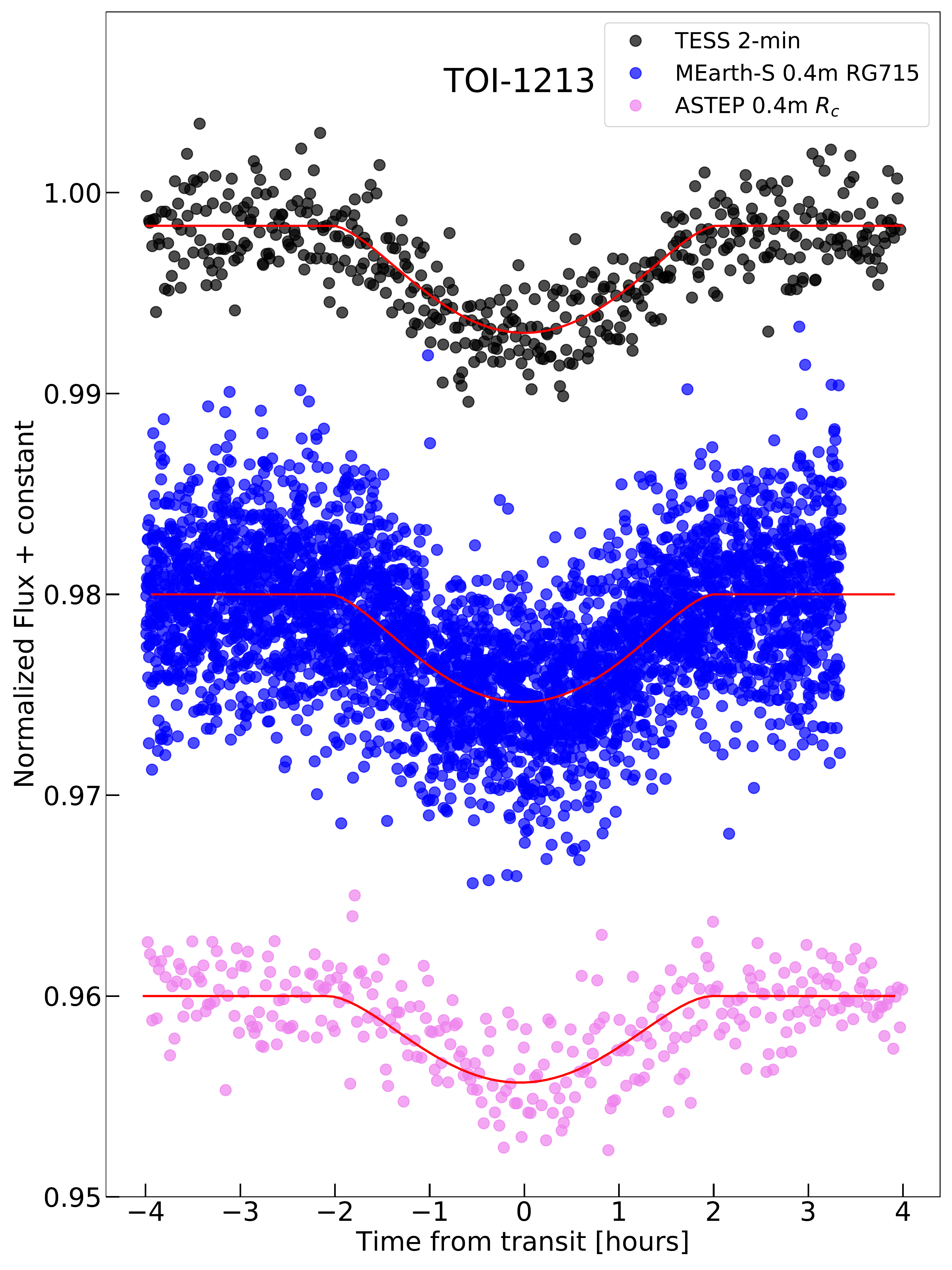}
  \includegraphics[width=0.47\textwidth]{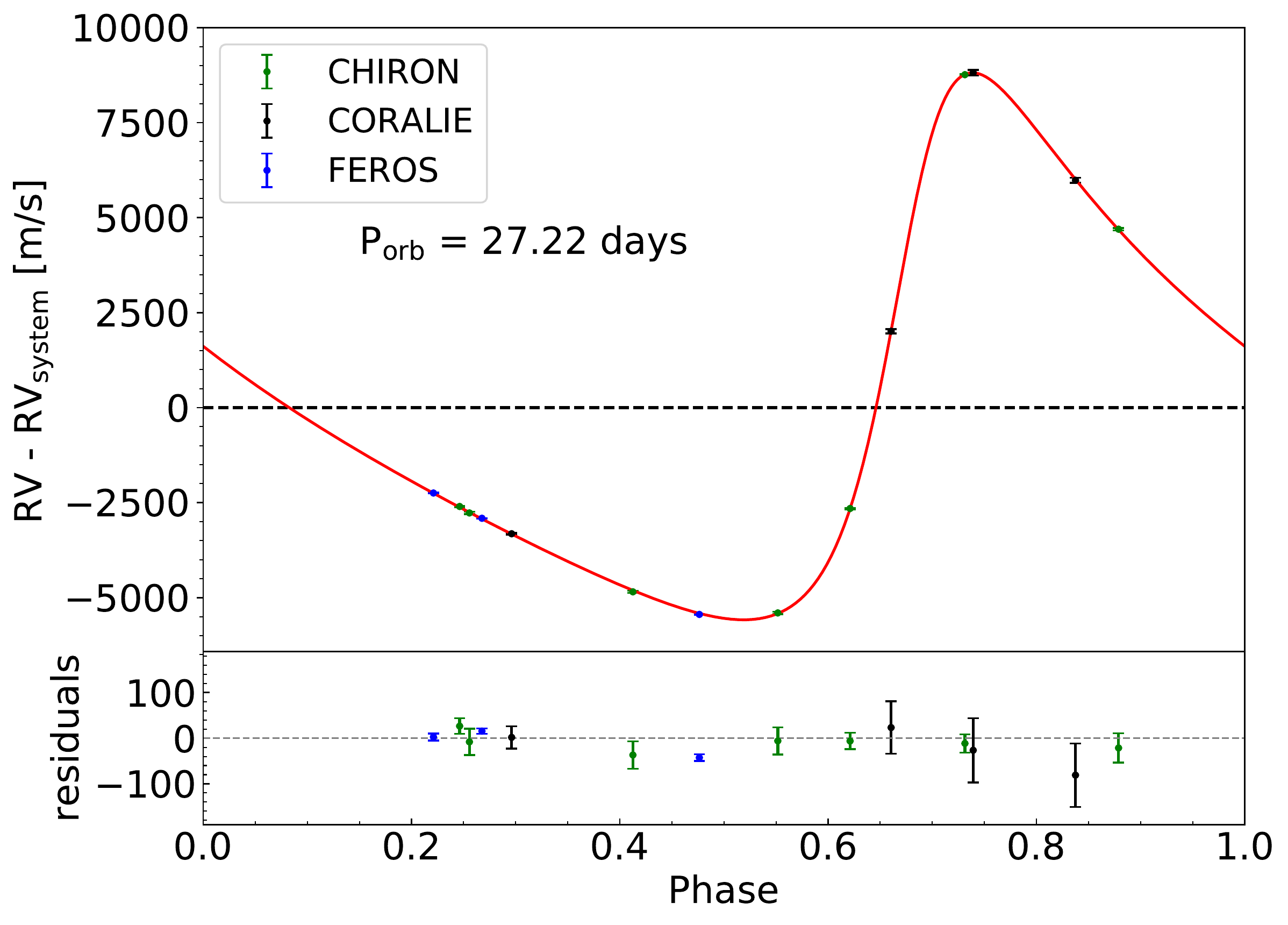}
  \caption{TOI-1213 light curves (\textit{top}) and RVs (\textit{bottom}) phased to the companion's orbital period. The red lines display our best-fit models from EXOFASTv2 as described in Figure \ref{fig:148phase}.}
  \label{fig:1213phase}
\end{figure}

\begin{figure*}
  \centering
  \includegraphics[width=0.4\textwidth]{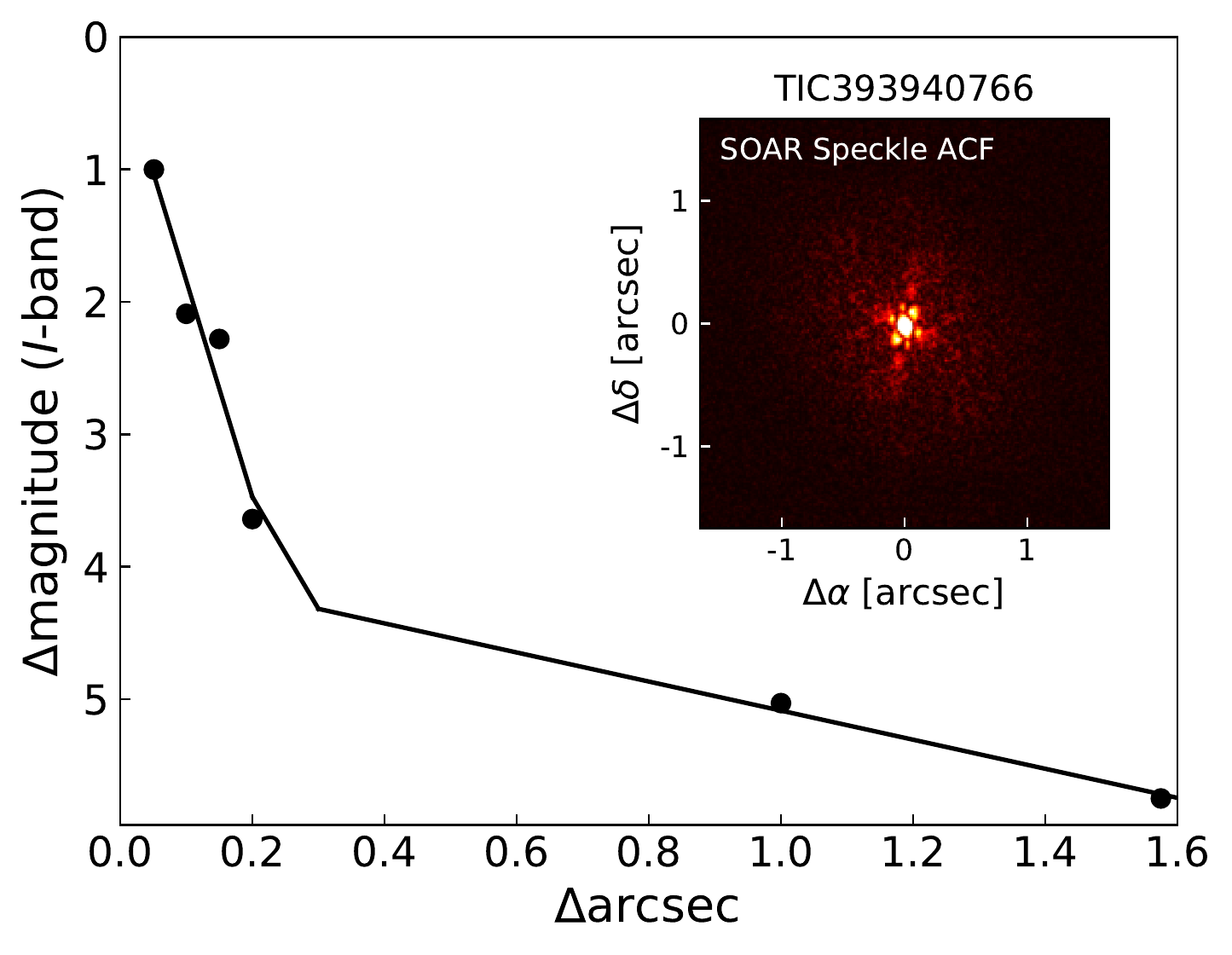}
  \includegraphics[width=0.4\textwidth]{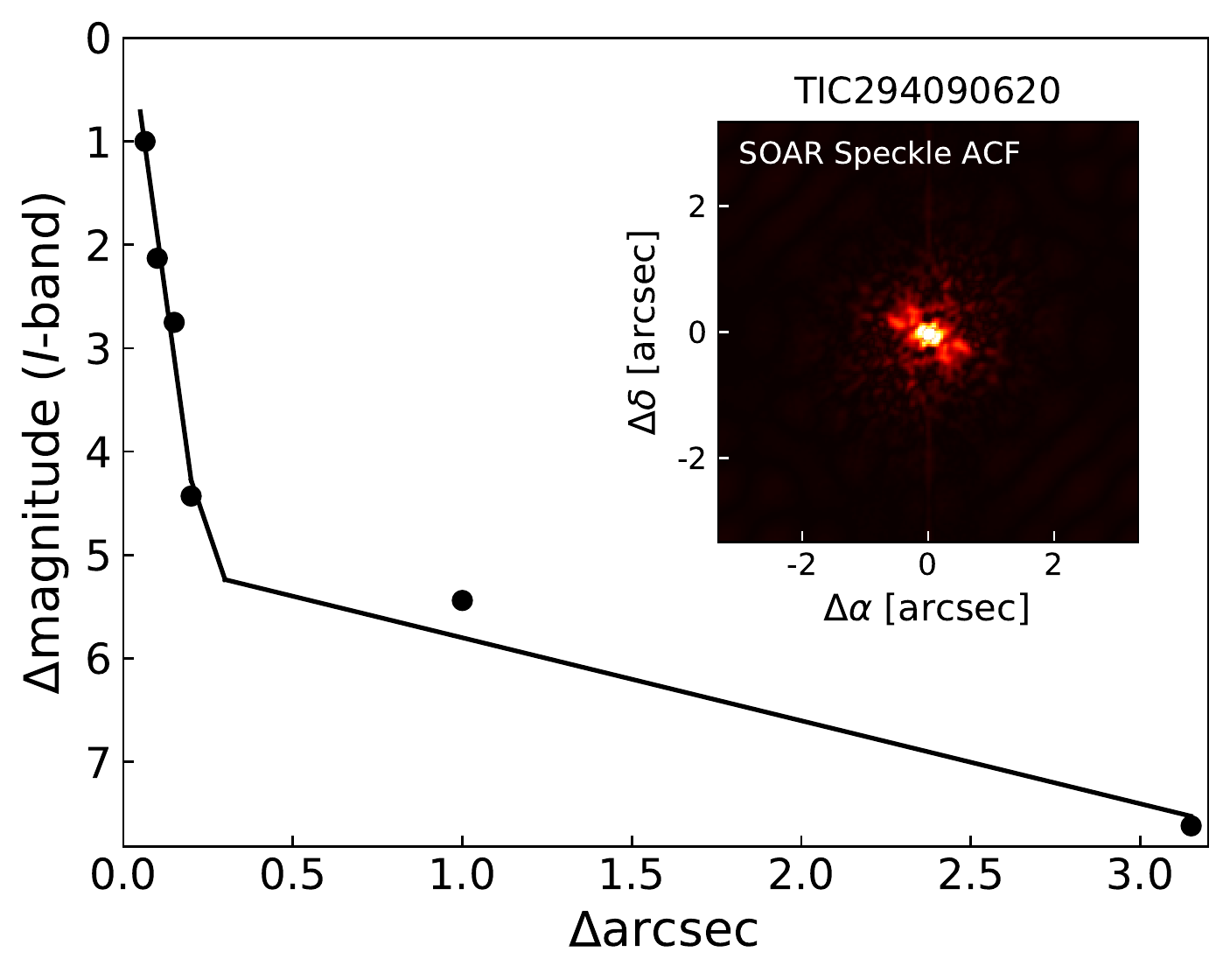}
  \includegraphics[width=0.4\textwidth]{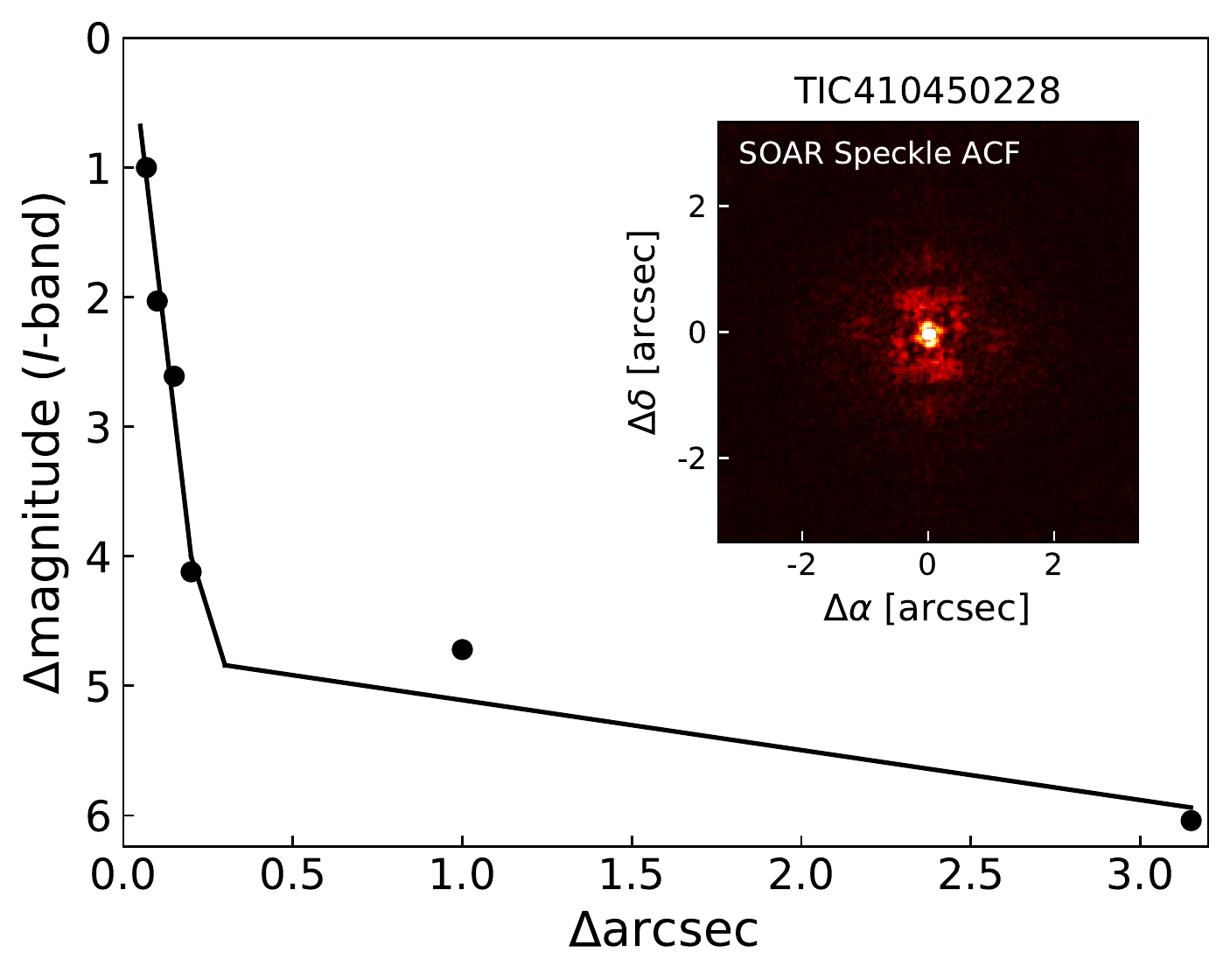}
  \includegraphics[width=0.4\textwidth]{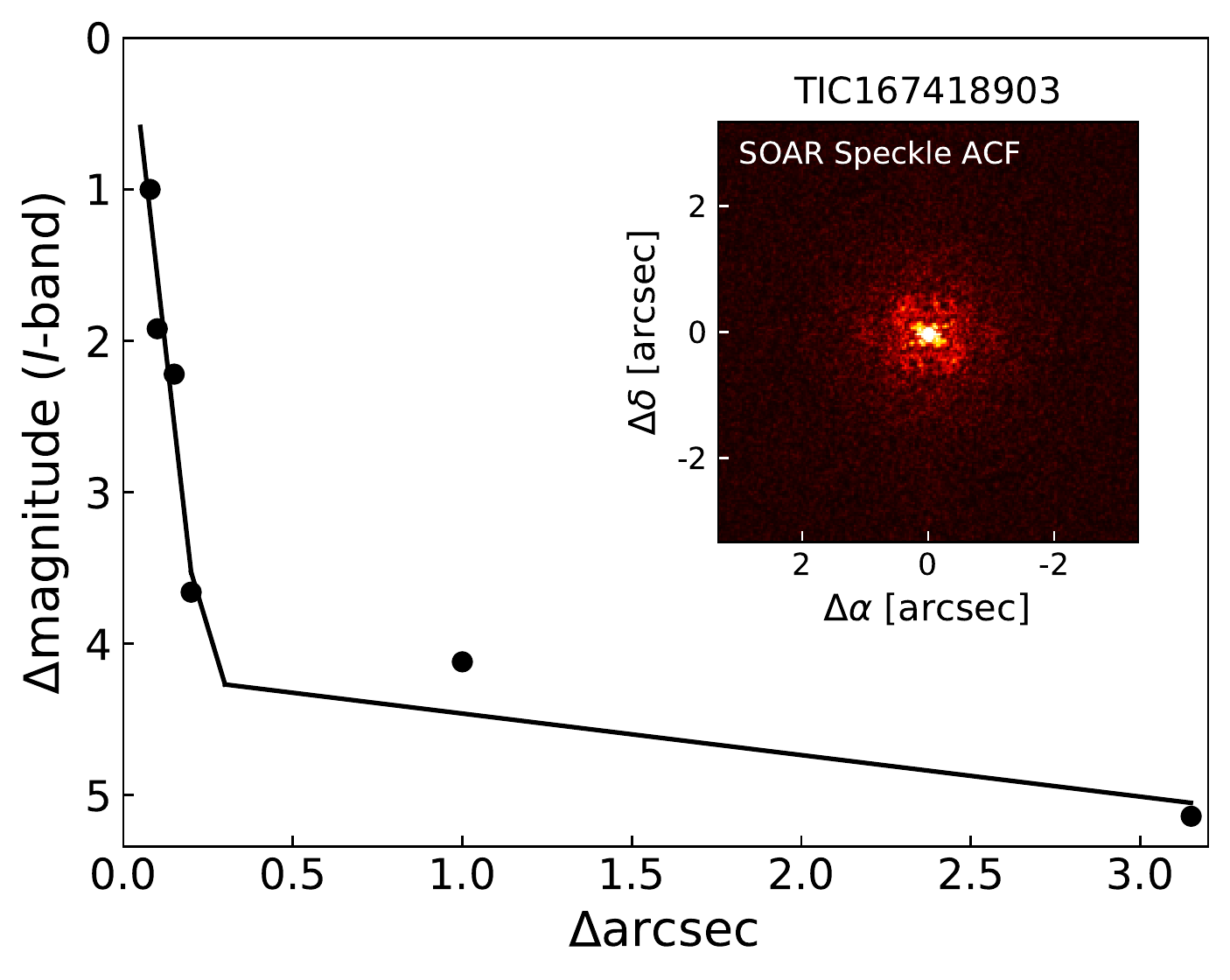}
  \includegraphics[width=0.4\textwidth]{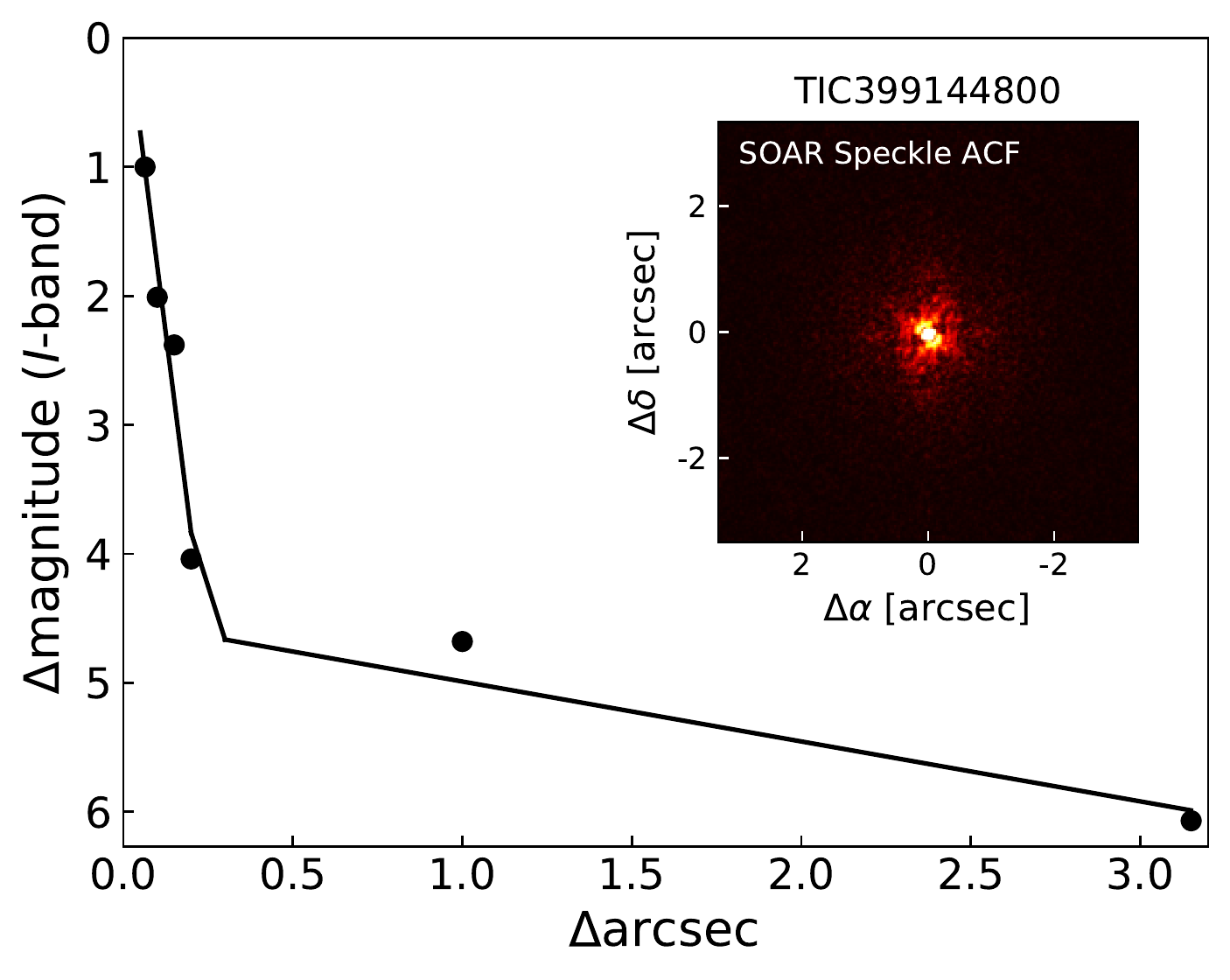}
  \caption{SOAR/HRCam speckle interferometry imaging with I-band autocorrelation functions \citep{Ziegler2020} for TOI-148 (TIC 393940766; \textit{top left}), TOI-587 (TIC 294090620; \textit{top right}), TOI-681 (TIC 410450228; \textit{middle left}), TOI 746 (TIC 167418903; \textit{middle right}), and TOI 1213 (TIC 399144800; \textit{bottom}). The 5$\sigma$ contrast curves with a linear fit are shown with black solid lines. The auto-correlation functions obtained in I-band are shown within the contrast curve plots.}
  \label{fig:soar}
\end{figure*}

\subsection{TESS photometry}

In this section we describe the TESS observations for each star. We refer the reader to tables \ref{tab:TOI-148} - \ref{tab:TOI-1213} for the stellar parameters and properties of each star. We display phased TESS photometry for all five of our stars with transiting companions as well as transit observations from all available ground-based photometry, described in Section \ref{sec:sg1}, in Figures \ref{fig:148phase} - \ref{fig:1213phase}. We obtained TESS photometry from exomast\footnote{https://exo.mast.stsci.edu}, which is part of the  Barbara A. Mikulski Archive for Space Telescopes (MAST) astronomical data archive hosted by the Space Telescope Science Institute. 

TOI-148 (TIC 393940766) was observed by TESS in Sector 1 (UT 2018 July 25 to UT 2018 August 22) and the light curve was derived from Full Frame Image (FFI) observations with a cadence of 30 minutes and processed by the Quick Look Pipeline \citep[QLP;][]{Huang2020a,Huang2020b}. QLP transits were identified with a high signal to noise ratio every 4.87 days for a total of six transits detected in the QLP data. TOI-148 was observed again by TESS in Sector 28 (UT 2020 July 31 to UT 2020 August 25) with 2-minute cadence exposures. The 2-minute data were processed by the Science Processing Operations Center \citep[SPOC;][]{Jenkins2016} pipeline which produces two light curves per sector called Simple Aperture Photometry (SAP) and Presearch Data Conditioning Simple Aperture Photometry \citep[PDCSAP;][]{Smith2012,Stumpe2012,Stumpe2014}. We use the PDCSAP light curves for our analysis. The SPOC pipeline detected the transit signatures in the transit search pipeline \citep{Jenkins2002,Jenkins2010} and these transit signatures were fit to a limb-darkened transit model \citep{Li2019}. The transit fits passed all of the data validation module’s diagnostic tests \citep{Twicken2018}, including the difference image centroiding test, which localized the source of the transit signal to within 1.23\,$\pm$\,2.6 arcsec. This difference image centroiding test complements the speckle interferometry results in Section \ref{sec:highresimaging}. A total of four transits were detected in the 2-minute cadence data; we note a $\sim$5 day gap in the middle of the TESS Sector 28 data. No secondary transit or eclipse was detected above 3.3$\sigma$ (426 ppm), and no additional transit signals were detected in the multiplanet transit search.


TOI-587 (TIC 294090620) was observed in TESS Sector 8 (UT 2019 February 2 to UT 2019 February 28) with FFI observations. Two transits with a depth of $\sim$0.5\% were identified in the QLP processed light curve with a separation 8.0 days. 

TOI-681 (TIC 410450228) was observed in TESS Sectors 8 (UT 2019 February 2 - 27), 9 (UT 2019 February 28 to UT 2019 March 26), 27 (UT 2020 July 5 - 30), and 31 (UT 2020 October 22 to UT 2020 November 18) with 2-minute cadence exposures. The 2-minute data were processed by SPOC which identified two 1\% deep transits 15.8 days apart. TOI-681 was also observed with FFI observations in Sectors 1 (UT 2018 July 25 to UT 2018 August 22), 4 (UT 2018 October 19 to UT 2018 November 15), 5 (UT 2018 November 15 to UT 2018 December 11), 7 (UT 2019 January 7 to UT 2019 February 1), 10 (UT 2019 March 26 to UT 2019 April 21), and 11 (UT 2019 April 22 to UT 2019 May 20). The transit signature passed all the data validation diagnostic tests and was localized to within 0.726\,$\pm$\,2.5 arcsec in the multisector 8 and 9 search. No statistically weak secondary was detected above the 3.3 sigma (426 ppm) level, and no further transiting planet signatures were detected in the multiplanet search. This star was also observed in sector 27 but only one transit was observed. It was re-observed in sector 31, passed all the data validation diagnostic tests, including the difference image centroiding that localized the source to within 0.958\,$\pm$\,2.5 arcsec, and no statistically significant weak secondaries were detected above the 2.8$\sigma$ (629 ppm) level.

TOI-746 (TIC 167418903) was observed by TESS in Sectors 1-13 (UT 2018 July 25 to UT 2019 July 17) with FFI observations as well as 2-minute cadence exposures in Sectors 11 - 13 (UT 2019 May 21 to UT 2019 July 17) and 28 - 33 (UT 2020 July 31 to UT 2021 January 13). The FFI observations of all sectors were processed by the QLP and the 2-minute data were processed with the SPOC pipeline, both showing a transit-like event occurring every 10.98 days. We detected 17 transits in the FFI data and 11 transits in the 2-minute data.

\begin{table*}
\centering
    \begin{tabular}{llllll}
    \hline\hline
    Telescope & Location & Date & Filter & Aperture radius & Coverage\\
           &           & [UTC]&        &   [arcsec]      & \\
    \hline


{\it  TOI 148.01 (TIC 393940766)}\\
\hline
LCO-SAAO 1m    & South Africa   & 2020-06-07   &  $i^\prime$   & 4.7 & ingress \\
LCO-SSO 1m     & Australia      & 2020-06-27   &  $i^\prime$   & 5.1 & ingress \\
LCO-SSO 1m     & Australia      & 2020-07-02   &  $i^\prime$   & 4.7 & egress  \\
El-Sauce 0.36m & Chile          & 2020-07-12   &  $R_c$        & 5.9 & full \\
LCO-SAAO 1m    & South Africa   & 2020-07-21   &  $i^\prime$   & 4.3 & full \\[2mm] 

\hline

{\it  TOI 681.01 (TIC 410450228)}\\
\hline
Brierfield 0.36m & Australia      & 2019-10-26  & I            & 10.3  & full \\
CHAT 0.7m        & Chile          & 2019-11-26  & $i^\prime$   & 8.4   & egress \\
El Sauce 0.36m   & Chile          & 2019-11-27  & B            & 7.4   & full \\
Mt. Stuart 0.32m & New Zealand    & 2020-04-01  & $g^\prime$   & 4.4   & full \\
ASTEP 0.4m       & Antarctica     & 2020-06-19  & $R_c$        & 10.2  & full \\[2mm] 

\hline
{\it  TOI 746.01 (TIC 167418903)}\\
\hline
LCO-SAAO 1m & South Africa        &  2019-02-18  & $z^\prime$   & 6.2   & ingress \\
LCO-SAAO 1m & South Africa        &  2019-03-01  & $u^\prime$   & 5.9   & full \\
LCO-SAAO 1m & South Africa        &  2019-04-25  & $z^\prime$   & 5.9   & ingress \\
PEST 0.31m  & Australia           &  2020-10-25  & V            & 7.4   & full \\[2mm] 

\hline
{\it TOI 1213.01 (TIC 399144800)}\\
\hline
MEarth-S 0.4m & Chile               &  2020-02-23  &  RG715       & 5.0   & full\\
ASTEP 0.4m       & Antarctica     & 2020-08-04  & $R_c$        & 10.2  & full \\[2mm]

\hline
\end{tabular}
\begin{tablenotes}
\item Observatory acronyms: ASTEP--Antarctic Search for Transiting ExoPlanets; CHAT--Chilean-Hungarian Automated Telescope; LCO--Las Cumbres Observatory;  PEST--Perth Exoplanet Survey Telescope
\end{tablenotes}
\caption{Summary of Ground-based Photometric Follow-up Observations.}
\label{table:SG1-obs}
\end{table*}

TOI-1213 (TIC 399144800) was observed by TESS in Sectors 10 and 11 (UT 2019 March 26 to UT 2019 May 21) with 2-minute cadence exposures. The 2-minute data was processed by the SPOC pipeline, which showed one transit-like event in each sector, spaced 27.214 days apart. The multisector data validation report for sectors 10 and 11 indicated that the transit signature passed all the DV diagnostic tests, localized the source to within 3.8\,$\pm$\,2.1 arcsec, and detected no weak secondaries above 3.1$\sigma$ (435 ppm), and no additional transiting planet signatures were found.

\subsection{Ground-based photometric follow-up} \label{sec:sg1}

%
%
%

We acquired ground-based time-series photometry of TOI-148, TOI-681, TOI-746, and TOI-1213 as part of the \textit{TESS} Follow-up Observing Program (TFOP)\footnote{https://tess.mit.edu/followup}. The TFOP's Sub Group 1 (SG1; seeing-limited photometry) group includes observers at more than a hundred telescopes distributed around the world. TFOP SG1 partners choose targets to follow up using the {\tt TESS Transit Finder}, which is a customized version of the {\tt Tapir} software package \citep{Jensen:2013}.  Photometric data are extracted by each observer most often using the {\tt AstroImageJ} ({\tt AIJ}) software package \citep{Collins:2017}. Data and analysis notes are uploaded to the {\it ExoFOP-TESS}\footnote{https://exofop.ipac.caltech.edu/tess/} repository and submitted to the SG1 team for validation.

In the case of the objects described here, we used the observations to rule out nearby eclipsing binaries as sources of the TESS signal, confirm the events on target, determine the \textit{TESS} photometric deblending factors for each field, place constraints on transit depth differences across optical filter bands, and refine the \textit{TESS} ephemerides by extending the time baselines. We summarize the observations that resulted in data useful for fitting in Table \ref{table:SG1-obs} and we plot the resulting light curves in Figures \ref{fig:148phase}, \ref{fig:681phase}, \ref{fig:746phase}, and \ref{fig:1213phase}.

The optimal photometric aperture radii used to extract light curves for these targets depends on the combination of pixel size, seeing disk size, focus (some observations are intentionally defocused to avoid saturation in the case of very bright targets), and the presence of on-sky neighbors visible in the images.  We checked the apertures listed in Table \ref{table:SG1-obs} against the {\it Gaia\/} DR2 catalog to determine the extent to which they might be contaminated by other stars.  In the case of TOI-681, there is a Gaia neighbor 7” away, and so inside most of the apertures used; however, at 6.4 magnitudes fainter than the target, the contribution is negligible within the photometric precision.  Similarly, in the case of TOI-746, there is a possible Gaia blend at a separation of 6 arcsec, just at the edge of the apertures used, but at 8.8 magnitudes fainter it is insignificant.  We describe checks for smaller-separation neighbors in Section \ref{sec:highresimaging}.

\subsection{Spectroscopic follow-up and radial velocities}

\begin{table*} 
    \centering
    \begin{tabular}{cccccc}
        \hline\hline
        Star & $\teff$ & $\feh$ & $\logg$  & $\vsini$ & Source  \\
         &  [K] & dex & [cgs] & [km$^{-1}$] &   \\
        \hline
        TOI-148 & 5836 $\pm$ 286 & -0.28 $\pm$ 0.28 & 4.11 $\pm$ 0.37 & 10.1 $\pm$ 0.8 & CORALIE \\     
        TOI-587 & 10400 $\pm$ 300 & 0.07 $\pm$ 0.12 & 4.20 $\pm$ 0.30 & 34.0 $\pm$ 2.0 & TRES \\
        TOI-681 & 7297 $\pm$ 45 & -0.12 $\pm$ 0.05 & 4.32 $\pm$ 0.14 & 30.8 $\pm$ 0.8 & GALAH DR2 \\
        TOI-746 & 5593 $\pm$ 215 & 0.01 $\pm$ 0.29 &  4.30 $\pm$ 0.29 & 6.1 $\pm$ 1.2 & CORALIE \\ 
        TOI-1213 & 5430 $\pm$ 215 & 0.28 $\pm$ 0.16 & 4.47 $\pm$ 0.26 & 4.0 $\pm$ 1.2 & CORALIE \\ 
        \hline
        
    \end{tabular}
    \begin{tablenotes}
    \item TOI-148, TOI-746, and TOI-1213 parameters were obtained from CORALIE spectra by combining the analysis results of both SpecMatch-Emp \citep{Yee2017} and with an iSpec \citep{Blanco-Cuaresma2014} analysis. The high temperature of TOI-587 precluded using SpecMatch-Emp and iSpec analysis and we use a separate analysis described in Section \ref{sec:specanal} on the TRES spectra. For TOI-681 we use GALAH DR2 results \citep{Buder2018}. 
    \end{tablenotes}
    \caption{Table of stellar parameters derived from spectra. \textbf{We refer the reader to the final adopted parameters for each star presented in tables \ref{tab:TOI-148} - \ref{tab:TOI-1213}}.}

    \label{tab:spectra_param}
\end{table*}

We obtained spectra and radial velocities of the five stars using observations from the 1.5 m SMARTS/CHIRON, Euler 1.2 m/CORALIE, MPG/ESO 2.2 m/FEROS, and 1.5 m SAO Tillinghast/TRES facilities. We summarize the radial velocity (RV) measurements of each star in Table \ref{tab:rv} and each star's phased radial velocities with their best Keplerian fits (see Section \ref{sec:analysis}) are displayed in Figures \ref{fig:148phase} - \ref{fig:1213phase}.

For TOI-148, TOI-681, TOI-746, and TOI-1213, we obtained RVs with the high resolution CORALIE spectrograph on the Swiss 1.2 m Euler telescope at La Silla Observatory, Chile \citep{Queloz2001}. CORALIE has a resolution of $R$ $\sim$ 60,000 and is fed by two fibers: a 2 arcsec on-sky science fiber encompassing the star and another fiber that can either connect to a Fabry-P\'erot etalon for simultaneous wavelength calibration or on-sky for background subtraction of sky flux. We observed all four stars in the simultaneous Fabry-P\'erot wavelength calibration mode. The spectra were reduced with the CORALIE standard reduction pipeline and RVs were computed for each epoch by cross-correlating with a binary G2 mask \citep{Pepe2002}.

We obtained 11 CORALIE observations for TOI-148 from UT 2018 September 23 to UT 2019 July 15. Nine CORALIE observations of TOI-681 were obtained from UT 2019 May 19 to UT 2020 March 15. We obtained eight CORALIE observations for TOI-746 from UT 2019 October 21 to UT 2020 March 12. TOI-1213 was observed four times with CORALIE from UT 2020 February 4 to UT 2020 March 17.

We used the TRES instrument on Mt. Hopkins, Arizona to obtain reconnaissance spectra for TOI-587. TRES has a resolving power of $R\sim 44\,000$ and covers a wavelength range of 390 to 910 nm. We use multiple echelle orders for each spectrum to measure a relative RV at each phase in the orbit of the transiting companion. We visually review each individual order to omit those with low signal-to-noise per resolution element (S/N) and we remove obvious cosmic rays. Each order is cross-correlated with the highest S/N spectrum of the star and then the average RV of all the orders per spectrum is taken as the RV of the star at a given orbital phase. The spectra of TOI-587 were taken between UT 2019 April 16 and UT 2019 April 30 with exposure times of 150\,s and 360\,s, giving a signal-to-noise per resolution element between 66 and 117.

\begin{figure*}
  \centering
  \includegraphics[width=0.85\textwidth]{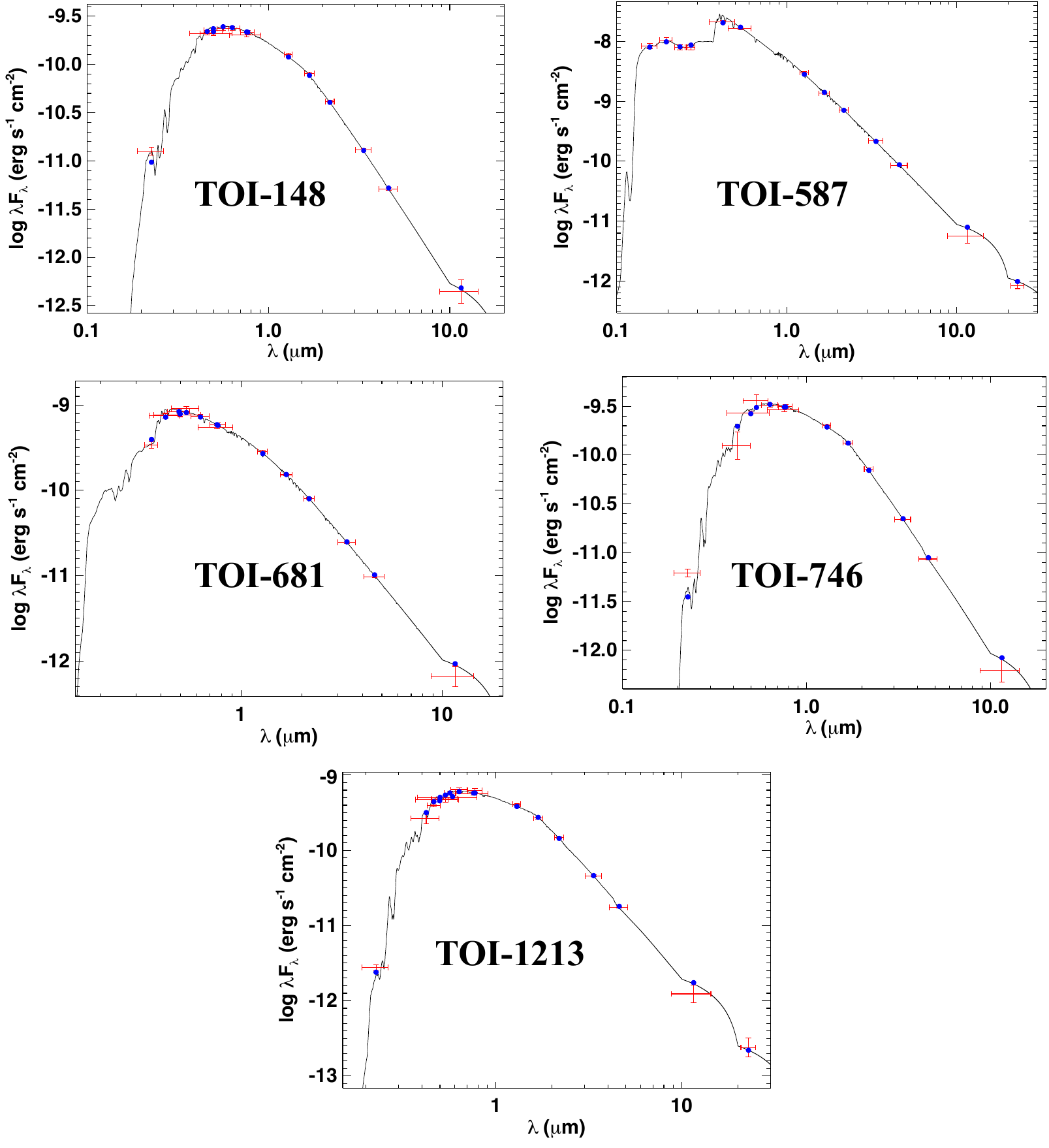}
  \caption{Spectral energy distributions for TOI-148 (top left), TOI-587 (top right), TOI-681 (middle left), TOI-746 (middle right), TOI-1213 (bottom). Red symbols represent the observed photometric measurements, where the horizontal bars represent the effective width of the bandpass. Blue symbols are the model fluxes from the best-fit Kurucz atmosphere model (black).}
  \label{fig:sed}
\end{figure*}


TOI-618, TOI-746, and TOI-1213 were monitored with the FEROS spectrograph \citep{Kaufer:99} mounted on the MPG 2.2m telescope in the ESO La Silla Observatory. These observations were performed between June of 2019 and March of 2020 in the context of the Warm gIaNts with tEss (WINE) collaboration which focuses in the discovery and orbital characterization of transiting planets with period longer than $\approx$ 10 days \citep[e.g.,][]{jordan:2020,brahm:2020,schlecker:2020}.

We obtained between two and four spectra per target on different epochs in order to determine if they presented radial velocity variations consistent to those produced by an orbiting planetary companion. All observations were executed with the simultaneous calibration technique, where a second fiber is illuminated with a Thorium-Argon lamp to trace instrumental velocity drifts during the exposure. The exposure times adopted were 400\,s, 1200\,s, and 900s\,, for TOI-618, TOI-746, and TOI-1213, respectively. The FEROS data were processed from raw images to precision radial velocities with the \texttt{ceres} pipeline \citep{ceres}. We used a G2-type binary mask as a template to compute the radial velocities through the cross-correlation technique. We also obtained line bisector span measurements from the cross-correlation function. These observations showed that these three systems were presenting radial velocity variations with amplitudes larger than those produced by planet mass companions, and therefore were dropped from the queue of the WINE project.

TOI-1213 was observed with the CTIO High Resolution spectrometer \citep[CHIRON;][]{Tokovinin2013,Paredes2021}, mounted on the CTIO 1.5-meter Small and Moderate Aperture Research Telescope System (SMARTS) telescope. CHIRON is a fiber-fed high resolution echelle spectrograph, with a resolving power of R = 80,000 and a wavelength range of 410\,nm to 870\,nm. The target was observed by CHIRON 7 times, three times between UT 2020 February 7 and UT 2020 March 9, and four times between UT 2020 December 8 and UT 2020 December 25. Each observation was composed of 3 x 20 minute exposures. The spectral extraction was performed by the default CHIRON pipeline \citep{Piskunov2002}, and the RVs were derived from the CHIRON spectra by a least-squares deconvolution technique \citep{Zhou2020}, and are listed in Table \ref{tab:rv}.




\subsection{High-resolution imaging}\label{sec:highresimaging}

We used speckle imaging from the high resolution camera \citep[HRCam;][]{Tokovinin2018} mounted on the southern astrophysical research (SOAR) 4.1 m telescope in Cerro Pach\'{o}n, Chile to verify there are no stars close to our targets that would significantly contaminate the transit or RV signals we observe. The relatively large $21''$ pixels of TESS can result in photometric contamination causing astrophysical false positives such as a background or nearby eclipsing binary stars if they fall within the same TESS image profile as the target. Close contaminants can also lead to inaccurate estimates of the transit depth from a diluted transit in a blended light curve. 

All five targets were observed with HRCam, which uses the speckle interferometry techniques in a visible bandpass similar to that of TESS. \citet{Ziegler2020} provides a description of HRCam observations of TESS targets with data reduction described in \citet{Tokovinin2018}. SOAR speckle imaging was obtained on UT 2018 September 24 for TOI-148, UT 2019 May 18 for both TOI-587 and TOI-681, UT 2020 January 7 for TOI-746, and UT 2019 December 12 for TOI-1213. We detect no nearby starts within $3''$ for all five stars. We show the 5-$\sigma$ detection sensitivity and the speckle auto-correlation function from the SOAR observations in Figure \ref{fig:soar}.

\section{Analysis} \label{sec:analysis}

\subsection{Spectral analysis} \label{sec:specanal}
We derived stellar parameters including effective temperature $\teff$, surface gravity $\logg$, and metalicity $\feh$ using SpecMatch-Emp \citep{Yee2017} on CORALIE spectra for TOI-148, TOI-746, and TOI-1213. Spectra were coadded onto a common wavelength axis to increase signal-to-noise prior to spectral analysis. SpecMatch-Emp uses a large library of stars with well-determined parameters to match the input spectra and derive spectral parameters. We use a spectral region that includes the Mg I b triplet (5100 - 5400 $\AA$) to match our spectra. SpecMatch-Emp uses $\chi^{2}$ minimisation and a weighted linear combination of the five best matching spectra in the SpecMatch-Emp library to determine $\teff$, $\logg$, and $\feh$.


We also determine $\teff$, $\logg$, and $\feh$ using coadded CORALIE spectra for TOI-148, TOI-746, and TOI-1213 with the analysis package iSpec \citep{Blanco-Cuaresma2014}. We used the synthesis method to fit individual spectral lines of the coadded spectra. For TOI-148, TOI-746, and TOI-1213 we used the radiative transfer code SPECTRUM \citep{Gray1994} to generate model spectra with MARCS model atmospheres \citep{Gustafsson2008}, version 5 of the GES (GAIA ESO survey) atomic line list provided within iSpec and solar abundances from \citet{Asplund2009}. Using the same method as \citet{Grieves2021}, we combine the iSpec analysis results and SpecMatch-Emp results for TOI-148, TOI-746, and TOI-1213. In order to create wide uncertainties we include the entire uncertainty range of both results that includes the lowest and highest uncertainty values of both methods. We present these spectroscopically derived parameters in Table \ref{tab:spectra_param}.



TOI-587 is a very hot star which is outside the range of temperatures in the spectral library of SpecMatch-Emp. TOI-587 also falls outside the valid temperature range for the MARCS model atmospheres and we instead use the ATLAS9 model atmospheres \citet{Castelli2004}.  Macroturbulence was estimated using equation (5.10) from \citet{Doyle2014} and microturbulence was accounted for at the synthesis stage using equation (3.1) from the same source. The H$\alpha$, NaID and Mg I b lines were used to infer the effective temperature $\teff$ and gravity $\logg$ while FeI and FeII lines were used to determine the metallicity $\feh$ and the projected rotational velocity \vsini. Trial synthetic model spectra were fit until an acceptable match to the data was found. Uncertainties were estimated by varying individual parameters until the model spectrum was no longer well-matched to the data.  


For TOI-681 we use stellar parameters from GALAH DR2 \citep{Buder2018} spectroscopy as we find these parameters more robust given their higher signal-to-noise ratio (S/N) compared to our CORALIE spectra. We do not find GALAH DR2 data for the other stars. Our wavelet analysis \citep{Gill2018,Gill2019} is also able to determine surface rotational velocity $\vsini$ from spectra which we apply to TOI-148, TOI-587, TOI-746, and TOI-1213 and present in Table \ref{tab:spectra_param}. For TOI-681 we present the $\vsini$ from GALAH DR2. We also examined the spectra and their cross-correlation functions (CCFs) with a binary template \citep[e.g.,][]{Pepe2002} when available and do not find any evidence that the targets are double-lined spectroscopic binaries.


\subsection{Spectral Energy Distribution analysis} \label{sec:sed}

\begin{table*} 
\centering
    \begin{tabular}{cccccccc}
        \hline\hline
        Star & $\chi_\nu^2$ & $\teff$ & $\feh$ & $\logg$  & $A_V$  &  $F_{\rm bol}$ & $R_\star$  \\
         &  &  [K] & dex & [cgs] & mag & $10^{-10}$ [erg~s$^{-1}$~cm$^{-2}$] & R$_\odot$ \\
        \hline
        TOI-148 & 1.4 & 5975 $\pm$ 150 & -0.50 $\pm$ 0.25 & 4.25 $\pm$ 0.1 & 0.03 $\pm$ 0.03 & 3.405 $\pm$ 0.079 & 1.192 $\pm$ 0.068 \\     
        TOI-587 & 0.9 & 9800 $\pm$ 200 & -0.10 $\pm$ 0.15 & 4.15 $\pm$ 0.15 & 0.02 $\pm$ 0.02 & 252.9 $\pm$ 5.9 & 2.031 $\pm$ 0.092 \\  
        TOI-681 & 1.0 & 7390 $\pm$ 150 & -0.25 $\pm$ 0.25 & 4.20 $\pm$ 0.15 & 0.22 $\pm$ 0.02 & 12.41 $\pm$ 0.14 & 1.586 $\pm$ 0.067 \\  
        TOI-746 & 1.6 & 5700 $\pm$ 150 & -0.25 $\pm$ 0.25 & 4.40 $\pm$ 0.10 & 0.18 $\pm$ 0.02 & 5.126 $\pm$ 0.059 & 0.957 $\pm$ 0.051 \\  
        TOI-1213 & 1.2 & 5675 $\pm$ 175 & 0.00 $\pm$ 0.25 & 4.45 $\pm$ 0.10 & 0.34 $\pm$ 0.03 & 10.47 $\pm$ 0.12 & 0.951 $\pm$ 0.059 \\  
        \end{tabular}
        \begin{tabular}{ccccccc}
        \hline\hline
        Star & $M_\star^a$ &  $M_\star^b$  &  $P_{\rm rot}/\sin i$  &  $\tau_\star^c$  &  $R'_{\rm HK}$  & $\tau_\star^d$  \\
        & M$_\odot$  & M$_\odot$ & [d]  & [Gyr]  & dex  & [Gyr] \\
        \hline
        TOI-148 & 0.94 $\pm$ 0.26 & 1.03 $\pm$ 0.06 & 5.97 $\pm$ 0.47 & 0.55 $\pm$ 0.07 & -- & -- \\     
        TOI-587 & 2.12 $\pm$ 0.29 & 2.32 $\pm$ 0.14 & 3.02 $\pm$ 0.18 & -- & -- & -- \\  
        TOI-681 & 1.45 $\pm$ 0.25 & 1.51 $\pm$ 0.09 & 2.61 $\pm$ 0.13 & -- & -- & -- \\  
        TOI-746 & 0.84 $\pm$ 0.12 & 0.98 $\pm$ 0.06 & 7.9 $\pm$ 1.6 & 0.52 $\pm$ 0.06 & $-4.50 \pm 0.05$ & 0.60 $\pm$ 0.23 \\  
        TOI-1213 & 0.93 $\pm$ 0.15 & 1.01 $\pm$ 0.06 & 12.0 $\pm$ 3.6 & 1.1 $\pm$ 0.2 & -- & -- \\  
        \hline
    \end{tabular}
\begin{tablenotes}
\item $^a$Stellar mass inferred from $R_\star$ and spectroscopic \logg. $^b$Stellar mass estimated via empirical relations of \citet{Torres2010}. $^c$System age estimated from inferred rotation period via empirical relations of \citet{Mamajek2008}. $^d$System age estimated from inferred $R'_{\rm HK}$ via empirical relations of \citet{Mamajek2008}. The ages derived from stellar rotation and activity are likely underestimated due to the stellar rotations likely being affected by the companions as discussed in Section \ref{sec:circsync}.
\end{tablenotes}
    \caption{Table of stellar parameters derived from SED fitting and empirical relations. \textbf{We refer the reader to the final adopted parameters for each star presented in tables \ref{tab:TOI-148} - \ref{tab:TOI-1213}}.}
\label{tab:sed}
\end{table*} 

As an independent check on the derived stellar parameters, we performed an analysis of the broadband spectral energy distributions (SEDs) of each star, together with the {\it Gaia\/} DR2 parallax in order to constrain the basic stellar parameters and to determine an empirical measurement of the stellar radius, following the procedures described in \citet{Stassun2016,Stassun2017,Stassun2018}. We pulled the {\it GALEX} NUV and FUV fluxes, the $B_T V_T$ magnitudes from {\it Tycho-2}, the $BVgri$ magnitudes from APASS, the $JHK_S$ magnitudes from {\it 2MASS}, the W1--W4 magnitudes from {\it WISE}, and the $G G_{\rm BP} G_{\rm RP}$ magnitudes from {\it Gaia}. For the hottest source we also pulled the TD1/wide UV fluxes from the TD1 satellite \citep{Thompson1978}. Together, the available photometry spans the full stellar SED over the wavelength range 0.15--22~$\mu$m (see Figure~\ref{fig:sed}). We performed a fit using Kurucz stellar atmosphere models, with the priors on effective temperature ($T_{\rm eff}$), surface gravity ($\log g$), and metallicity ([Fe/H]) from the spectroscopic analysis. The remaining parameter is the extinction ($A_V$), which we limited to the maximum line-of-sight extinction from the Galactic dust maps of \citet{Schlegel1998}. 

The resulting fits (Figure~\ref{fig:sed}) have small reduced $\chi^2$ values reported in Table~\ref{tab:sed}, which also summarizes the other resulting stellar parameters: Integrating the (unreddened) model SED gives the bolometric flux at Earth, $F_{\rm bol}$; taking the $F_{\rm bol}$ and $T_{\rm eff}$ together with the {\it Gaia\/} DR2 parallax, adjusted by $+0.08$~mas to account for the systematic offset reported by \citet{StassunTorres2018}, gives the stellar radius, $R_\star$. We can then infer the stellar mass, $M_\star$, empirically from $R_\star$ and $\log g$, as well as estimate it via the empirical relations of \citet{Torres2010}. 

We can also estimate the stellar rotation periods from the spectroscopic $v\sin i$ together with the above $R_\star$, giving an upper limit of $P_{\rm rot} / \sin i$. From the excess GALEX UV emission, when present, we can also estimate the chromospheric activity indicator, $R'_{\rm HK}$ via the empirical relations of \citet{Findeisen2011}. In turn, these provide another estimate of the stellar rotation period via the empirical rotation-activity relations of \citet{Mamajek2008}. Finally, these activity measures provide an estimate of the stellar age, $\tau_\star$ again via the empirical activity-age relations of \citet{Mamajek2008}. However, we note the ages derived from stellar rotation and activity are likely underestimated due to the stellar rotations likely being affected by the companions (e.g., tidal locking) as discussed in Section \ref{sec:circsync}.


\subsection{TOI-681 kinematic analysis and cluster membership}
\label{sec-toi681age}
TOI-681 has been previously reported to be a member of the open star cluster NGC 2516 \citep{GaiaCollaboration2018,Cantat-Gaudin2018,KounkelCovey2019,Meingast2021}. We reassessed its position and kinematics relative to NGC 2516 as follows. First, we collected the NGC 2516 members reported by \citet{Cantat-Gaudin2018}. To define a set of reference ``neighborhood'' stars, we then queried Gaia DR2 for stars within 4 standard deviations of the mean values of NGC 2516's right ascension, declination, and parallax. The corresponding positions, proper motions, and Gaia DR2 radial velocities are shown in Figure~\ref{fig:toi681kin}.

TOI-681 is well within the cloud of cluster members (dark black points in Figure~\ref{fig:toi681kin}) in each dimension, rather than being within the outlying neighborhood (gray points). While TOI 681 did not have a Gaia DR2 RV, we measured a barycenter-corrected velocity with CORALIE of 22.120$^{+0.096}_{-0.087}$\,\kms and with FEROS of 23.75$\pm$0.14\,\kms. This is an independent line of support for the cluster membership, as the mean (Gaia-derived) cluster RV is 24.1\,\kms.
Independent age indicators for TOI-681 such as the photospheric lithium abundance or the rotation period are not applicable due to the stellar type
($T_{\rm eff}\,\approx\,7400\,{\rm K}$).  Nonetheless, given the six-dimensional
position and kinematic overlap with the other cluster members, we proceed under
the assumption that TOI-681 is a member of NGC 2516.

Reported ages of NGC 2516 vary between 100\,Myr and 300\,Myr
\citep{Jeffries1997,Jeffries1998,Randich2018,GaiaCollaboration2018,KounkelCovey2019}.
While determining absolute ages for clusters is a challenging problem
\citep{soderblom_ages_2014}, NGC 2516 appears to be slightly
older than the Pleiades based on the main sequence turn-off and
gyrochronology (\citealt{cummings_2018};
\citealt{fritzewski_rotation_2020}; L.~Bouma et al., in preparation).
The current consensus Pleiades age based on the main sequence turn-off
and the lithium depletion boundary is $125\pm20$\,Myr (see
\citealt{soderblom_ages_2014}).  We therefore expect the
age of NGC\,2516 to be within the interval of 140 to 200\,Myr. We adopt $170\pm25$\,Myr and use this value as the age of TOI-681. This is older than the absolute model-averaged age of $\approx90$\,Myr
determined by
\citet{Randich2018} using isochrones, which we suspect might be explained by the presence
of blue-stragglers on the main sequence turn-off
\citep{cummings_2018}.



\subsection{Global modeling}

\subsubsection{Modeling with EXOFASTv2} \label{sec:exofast}

\begin{figure}
  \centering
  \includegraphics[width=0.45\textwidth]{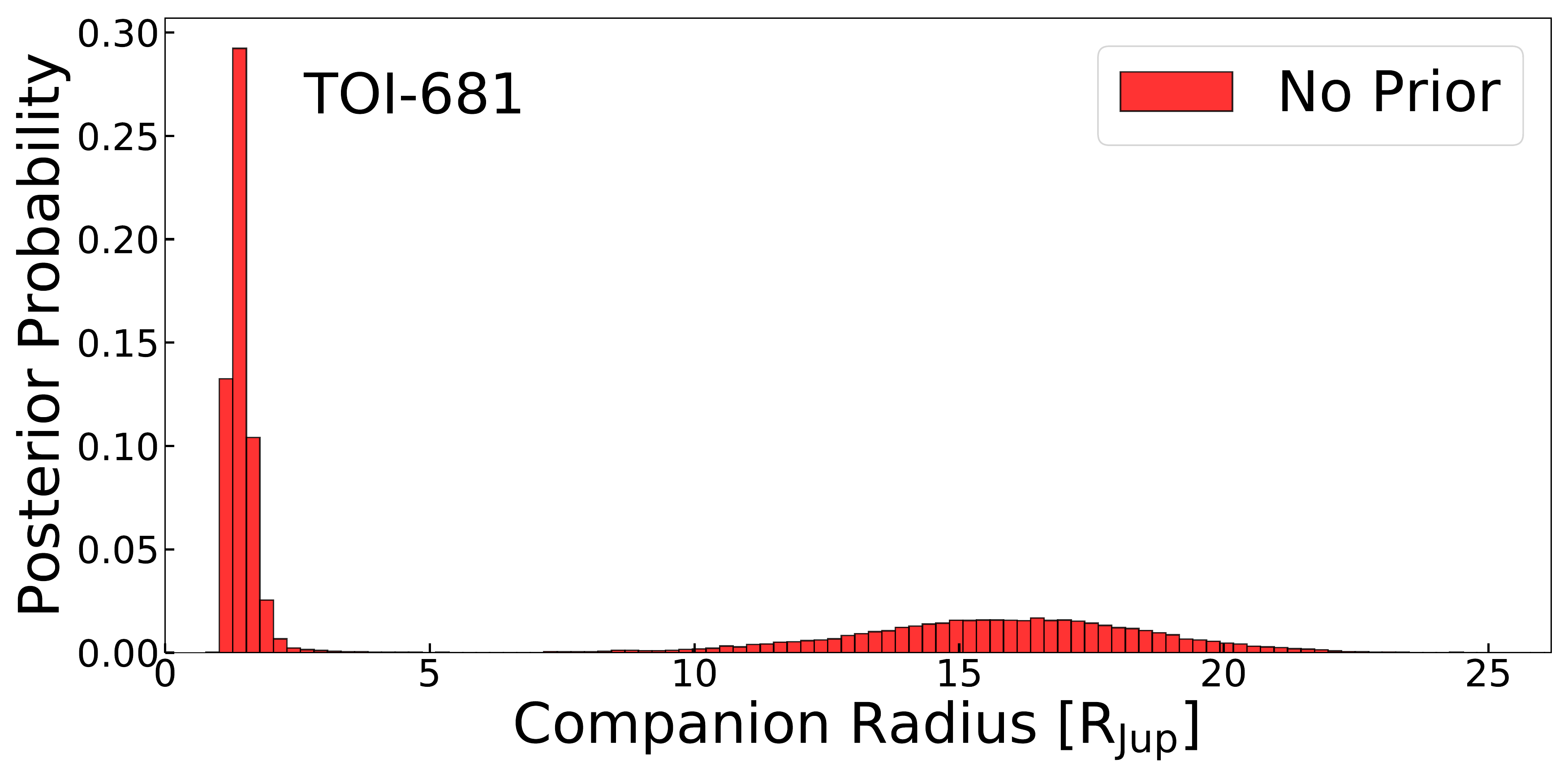}
  \includegraphics[width=0.45\textwidth]{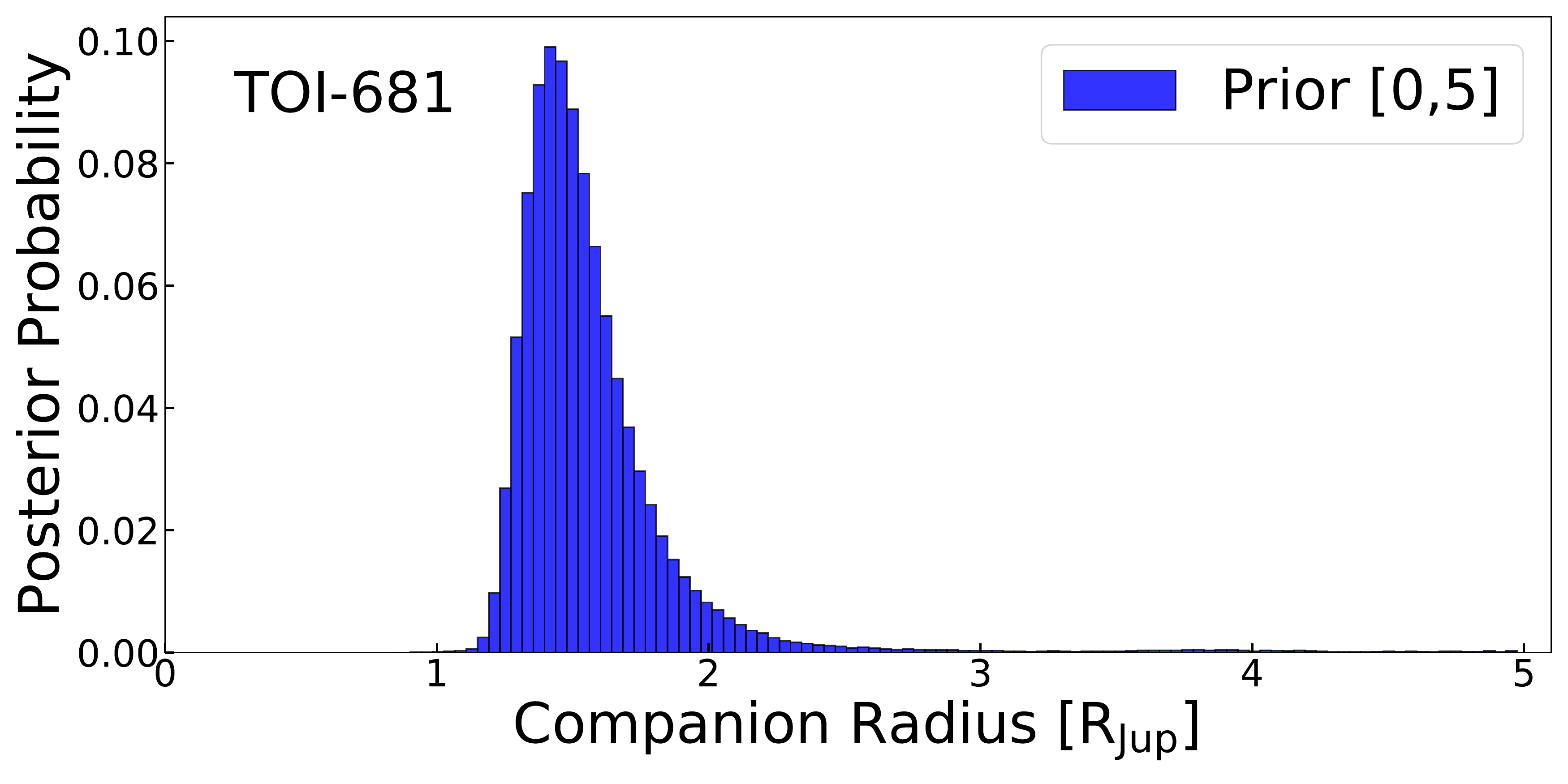}
  \includegraphics[width=0.45\textwidth]{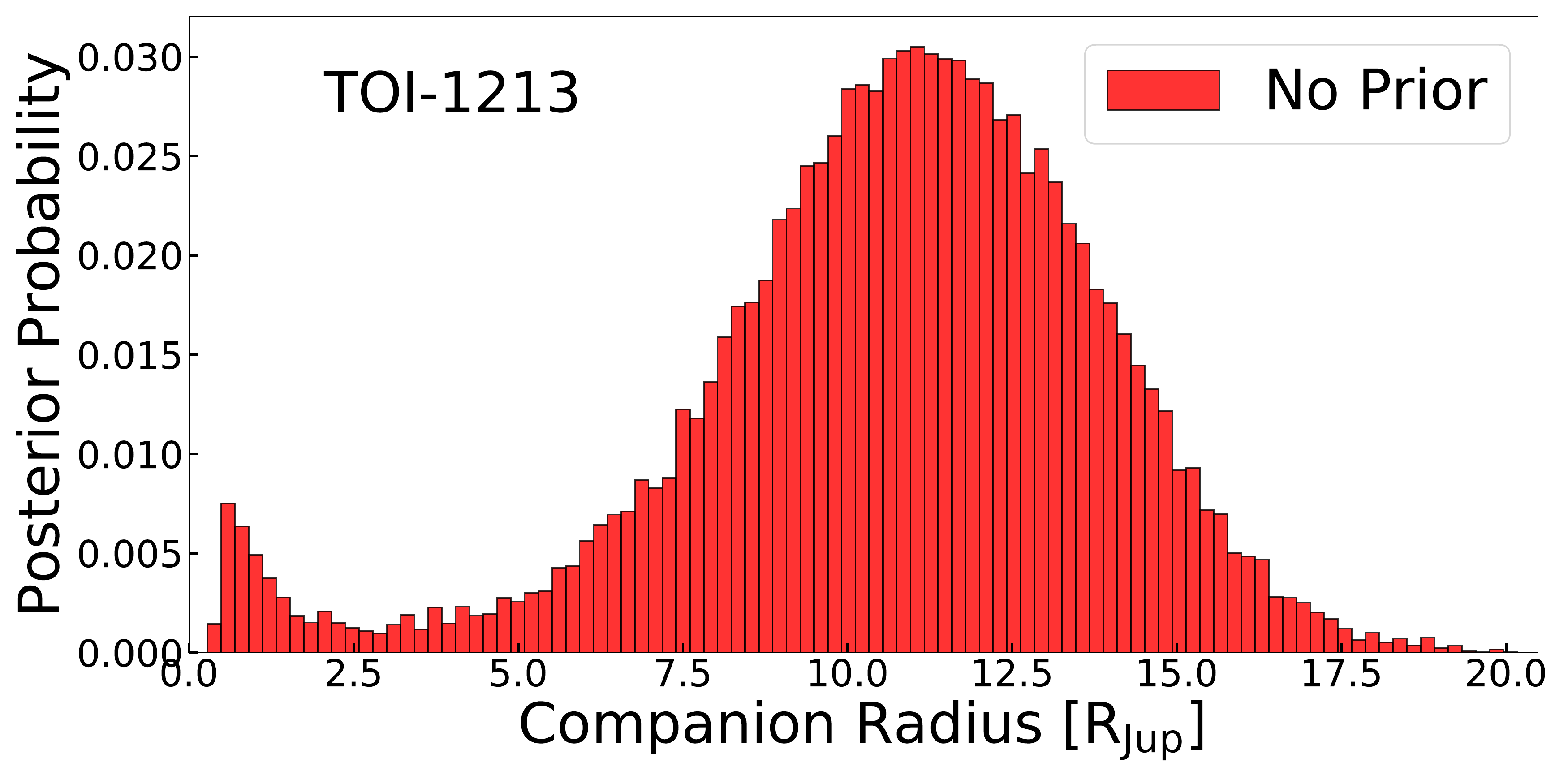}
  \includegraphics[width=0.45\textwidth]{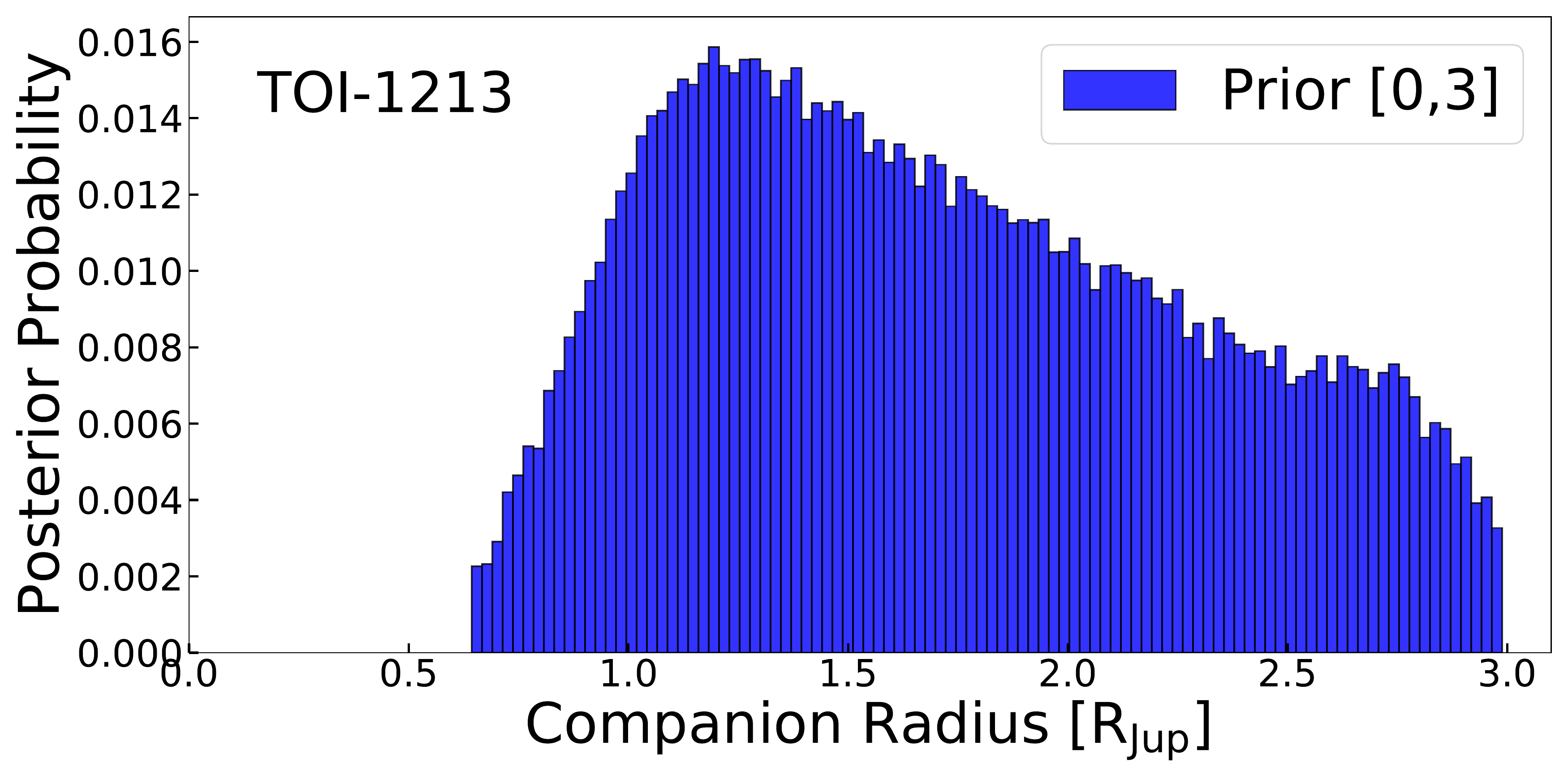}
  \caption{EXOFASTv2 posterior distributions of the companion radii without (red) and with (blue) upper limits on the companion radius for TOI-681 and TOI-1213 as described in Section \ref{sec:grazing}.}
  \label{fig:rppost}
\end{figure}

We derive companion parameters and final stellar parameters using EXOFASTv2 \citep{Eastman2013,Eastman2017,Eastman2019}. A full description of EXOFASTv2 is given in \citet{Eastman2019}, which can fit any number of transit and RV sources while exploring the vast parameter space through a differential evolution Markov Chain coupled with a Metropolis-Hastings Monte Carlo sampler. A built-in Gelman-Rubin statistic \citep{GelmanRubin1992,Gelman2003,Ford2006} is used to check the convergence of the chains. For each fit we use 2 $\times$ $n_{\text{\rm{parameters}}}$ walkers, or chains, and run until the fit passes the default convergence criteria for EXOFASTv2 that is described in \citet{Eastman2019}. For each star we simultaneously fit the RVs and photometry and determine stellar parameters using the Modules for Experiments in Stellar Astrophysics (MESA) Isochrones \& Stellar Tracks \citep[MIST;][]{Paxton2015,Choi2016,Dotter2016} isochrones. 

We place Gaussian priors on $\rstar$ and $\teff$ from the SED analysis described in Section \ref{sec:sed} and do not fit the SED again within the EXOFASTv2 global model. We place Gaussian priors on $\feh$ from the spectroscopic analysis described in Section \ref{sec:specanal}. For TOI-681 we use the NGC 2516 cluster age discussed in Section \ref{sec-toi681age} as a prior. We place upper boundaries on the companion radius for TOI-681 and TOI-1213 due to the grazing nature of their transits, which we discuss further in Section \ref{sec:grazing}.

EXOFASTv2 inherently applies priors on the quadratic limb darkening by interpolating the \citet{ClaretBloemen2011} limb darkening models at each step in $\logg$, $\teff$, and $\feh$. We use the inherent EXOFASTv2 limb darkening fitting without applying priors on any filters for TOI-148, TOI-681, TOI-746, and TOI-1213. However, given TOI-587's high $\teff$ we disable the limb darkening table interpolation using the {\tt noclaret} option in EXOFASTv2 and apply Gaussian priors of $u_{1}$ = 0.15 $\pm$ 0.1 for the linear limb-darkening coefficient and $u_{2}$ = 0.25 $\pm$ 0.1 for the quadratic limb-darkening coefficient for the TOI-587 TESS FFI lightcurve. To account for smearing in the 30-minute TESS full frame exposures, we specify an exposure time of 30 minutes and average over 10 data points to integrate a model over the exposure time equivalent to a midpoint Riemann sum \citep{Eastman2019}.

Specific priors used for each EXOFASTv2 fit are displayed in Tables \ref{tab:TOI-148} - \ref{tab:TOI-1213} as well as the final results for both the host star and companion parameters. All other fitted and derived parameters from our EXOFASTv2 model have conservative physical boundaries that are detailed in Table 3 of \citet{Eastman2019}, which also gives a thorough explanation of each parameter. The final EXOFASTv2 transit and RV Keplerian fits are displayed as the red lines in Figures \ref{fig:148phase} - \ref{fig:1213phase}.


\subsubsection{TOI-681 and TOI-1213 grazing transits} \label{sec:grazing}

TOI-681 and TOI-1213 both have v-shaped light curves which suggests that their companions have grazing transits. In such cases where transiting companions have a grazing geometry there is a degeneracy between the companion radius and impact parameter of the transit and the upper limit of the companion radius is unconstrained by the light curve. For TOI-681 we find that the posterior distribution of both the companion radius and impact parameter are bimodal with the lower companion radius and impact parameter being favored. We show the unconstrained posterior distribution of the companion radius in Figure \ref{fig:rppost}. Given this clear bimodality and a more probable lower-radius solution both physically and statistically, we remodeled TOI-681 with very conservative boundaries of 0 and 5 $\rjup$ for the companion radius. We show the new companion radius posterior distribution in Figure \ref{fig:rppost} that finds a clear peak at $\sim$1.5 $\rjup$. 

\begin{figure}
  \centering
  \includegraphics[width=0.48\textwidth]{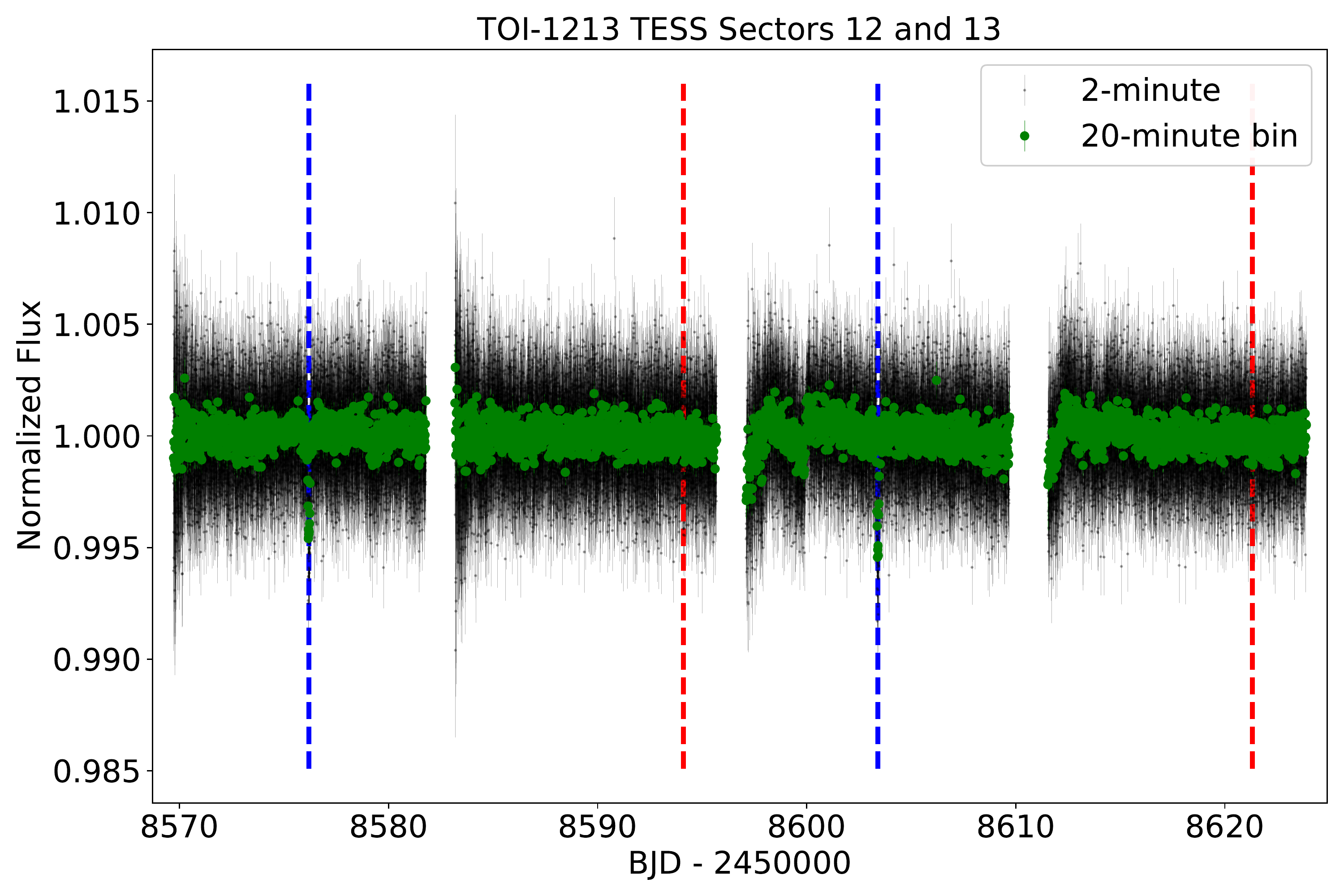}
  \includegraphics[width=0.48\textwidth]{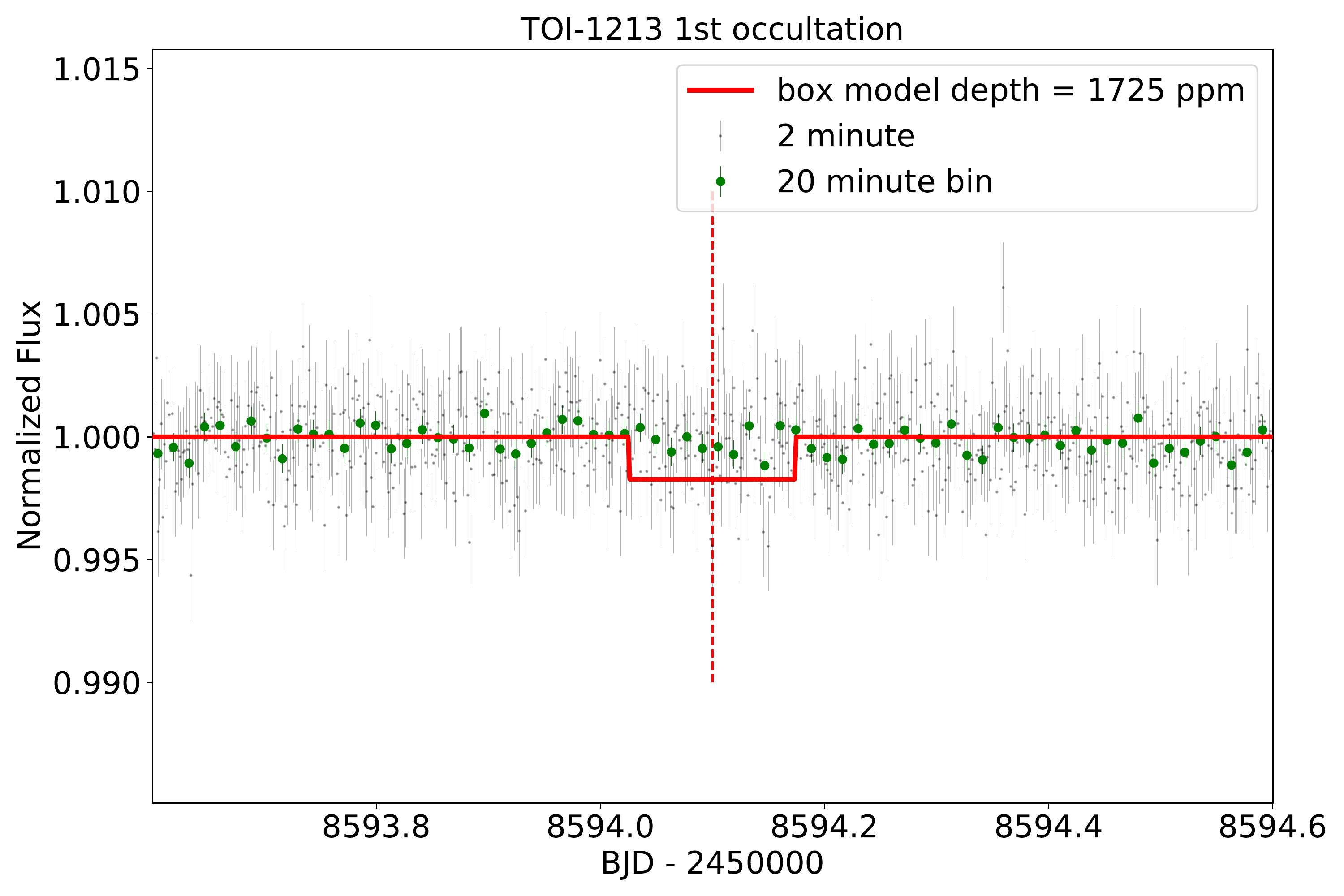}
  \caption{\textit{Top}: Full light curve of TOI-1213 for TESS sectors 12 and 13. The blue dashed vertical lines display the times of transit and the red dashed vertical lines show the expected occultation times which we determined to be at a phase of 0.6581. \textit{Bottom}: Zoom in of the time where the first occultation of TOI-1213 should occur with a box model with a depth equal to the standard deviation of the TESS photometry during the total eclipse duration (0.149 days) calculated from our EXOFASTv2 model.}
  \label{fig:toi1213tess}
\end{figure}

The posterior distribution for the companion radius of TOI-1213 does not clearly favor the lower-radius solution, and the most probable solution is at a much larger radius than physically expected. We therefore put a tighter upper boundary of 3.0 $\rjup$ for TOI-1213 when remodeling the system based on the physical limit of nondetections of the occultations in the TESS light curve. From our EXOFASTv2 model we find an eclipse impact parameter of 0.403$^{+0.041}_{-0.027}$ showing the system should be aligned to see an occultation in the lightcurve if it is within the TESS detection limits. As displayed in Figure \ref{fig:toi1213tess} we do not find any occultations of TOI-1213b within the precision of the TESS light curve, which allows us to put an upper limit on the occultation depth and thus an upper limit on the companion radius. The lower panel of Figure \ref{fig:toi1213tess} shows the first occultation and displays a box model of an occultation with a depth equal to the standard deviation (1725 ppm) of the TESS data during the total eclipse duration (0.149$^{+0.015}_{-0.014}$ days) calculated from our EXOFASTv2 model. We set this standard deviation of 1725 ppm as our detection limit and the lower limit of the occultation depth. To turn this occultation depth limit into an upper limit on the companion radius, we obtained simulated PHOENIX spectra \citep{Husser2013} of a G6V star for TOI-1213 and M7.5V for TOI-1213b (from our calculated mass of 97.5 $\mjup$) and integrated the total flux of each spectrum over the TESS bandpass of 600 - 1000 nm. We then multiply this total flux of the modeled G6V star by the area of TOI-1213 $\pi\rstar^{2}$. The companion radius can then be computed by setting the occultation depth to the TESS precision:
\begin{equation}
\mathrm{Depth}_{\mathrm{occ}} = \frac{\mathrm{Flux}_{b} \pi R_{b}^{2}}{\mathrm{Flux}_{A} \pi R_{A}^{2}}.
\end{equation}
With a lower limit on occultation of 1725 ppm, we find an upper limit on the companion radius of 3.00 $\rjup$. We set 3.0 $\rjup$ as the upper limit on TOI-1213b and display the companion radius with and without boundaries in Figure \ref{fig:rppost}. 




\section{Discussion} \label{sec:disc}

The five transiting companions analyzed here have masses around the hydrogen-burning mass limit, the upper boundary of brown dwarfs and lower boundary of main sequence stars, which is generally adopted as 80\,$\mjup$ \citep[e.g.,][]{MarcyButler2000,GretherLineweaver2006}. However, this border depends on initial formation conditions including the initial radius of the object, the efficiency of convection in the outermost layers, opacity, metallicity, and the initial abundance of deuterium \citep[e.g.,][]{Chabrier1997,Baraffe2002}. \citet{Dieterich2018} summarizes previous model predictions for the stellar-substellar boundary which range from 73.3\,-\,96.4\,$\mjup$ \citep[e.g.,][]{Burrows2001}, whose predictions differ from observed populations \citep[e.g.,][]{Dieterich2014} and differ from the $\sim$70\,$\mjup$ boundary estimate by \citet{Dupuy2017} using astrometric masses of ultracool binaries. 

This brown dwarf and stellar boundary is important as small changes in mass can cause vastly different lives for these objects, where low-mass M-dwarfs may burn hydrogen for up to trillions of years \citep[e.g.,][]{Adams1997} compared to brown dwarfs that will only have a short-lived deuterium burning stage of less than a billion years \citep[e.g.,][]{Spiegel2011} before cooling and shrinking. The exact population these five objects belong to is uncertain, but we can put them into context with other transiting brown dwarfs and very low-mass stars as discussed in Section \ref{sec:13-150comp}. We first explore the possible tidal effects on these individual systems in Section \ref{sec:circsync}.

\subsection{Tidal circularization and spin-orbit synchronization} \label{sec:circsync} 





The gravitational and tidal interactions between a star and a close-in orbiting companion (another star, brown dwarf, or planet) have varying effects on the rotation, orbit, and momentum axes of the system. In terms of timescales, the first effect that may be induced is spin-orbit synchronization, or tidal locking, where the stellar rotation period and orbital period become equal from tidal torques \citep[e.g.,][]{Zahn1977,Zahn1989,Witte2002,Mazeh2008}. A system with an eccentric orbit can reach a tidally locked state when the stellar rotation period becomes equal to the equilibrium rotation period, which can be predicted by tidal models \citep{Fleming2019}. A star's rotation period is generally expected to reduce over time from magnetic braking \citep{Skumanich1972}; however, this affect may be slowed down due to tidal torques \citep[e.g.,][]{Verbunt1981,Fleming2018, Fleming2019}. On longer timescales and particularly for short orbital periods (e.g., P$_{\rm{orb}}$\,$\lesssim$\,10\,days) the orbital eccentricity may be damped and the orbit may become circular \citep[e.g., tidal circularization,][]{Goldreich1966,Hut1981,Adams2006}. Evolution of the system's obliquity, or the angle between the stellar spin axis and the orbital axis occurs even more slowly than eccentricity dampening \citep[e.g.,][]{Hut1981,Winn2005,BarkerOgilvie2009}. Here we examine possible tidal effects including eccentricity dampening or circularization, stellar rotation spin up, and spin-orbit synchronization of the brown dwarf and low-mass star systems presented in this work.

From our spectroscopic analysis we find a \vsini = 10.1\,$\pm$\,0.1\,\kms for TOI-148, which is on the upper end of \vsini distributions for stars of similar mass \citep[e.g.,][]{Robles2008}, indicating TOI-148's stellar rotation period may be affected by tidal torques with its companion. With this \vsini we put an upper limit on the stellar rotation period ($P_{rot}$ $\leq$ $2\pi\rstar/\vsini$) of 5.97\,$\pm$\,0.47\,days for TOI-148, which is smaller than expected given its older age of 7.7\,$\pm$\,3.7\,Gyr in comparison to other rotational periods of older stars with similar masses \citep[e.g.,][]{Lorenzo-Oliveira2019}. Additionally the upper limit of the stellar rotation period is less than one day larger than the orbital period of 4.87 days for TOI-148b, suggesting that TOI-148 may be close to a tidally locked state. We find the eccentricity of TOI-148 $e$ = 0.005$^{+0.006}_{-0.004}$ to be insignificant (using the significance test from \citet{Lucy1971} we find P($e$ $>$ 0) = 0.63, which fails the 5\% significance level) also suggesting TOI-148 has undergone tidal circularization.


TOI-681 is relatively hot ($\teff$\,=\,7440$^{+150}_{-140}$\,K) and fast rotating ($P_{\rm rot}/\sin i$\,=\,2.61\,$\pm$\,0.13\,days) star, which can create enhanced magnetic fields from spin-orbit tidal synchronization for short period systems and has been thought to cause inflated radii of similar low-mass companions \citep[e.g.,][]{Chabrier2007,Mazeh2008}. However, given its relatively long period of 15.78 days and young 0.170\,$\pm$\,0.025\,Gyr age TOI-681b is unlikely to have reached spin-orbit synchronization. TOI-587 is also a young 0.2\,$\pm$\,0.1\,Gyr star that is unlikely to have undergone spin-orbit synchronization with $P_{\rm rot}/\sin i$ = 3.02\,$\pm$\,0.18\,days and $P_{orb}$ = 8.04 days.

TOI-746 is an older 6.5$^{+4.3}_{-3.9}$\,Gyr star with a companion on a moderately eccentric $e$ = 0.199\,$\pm$\,0.003 orbit. As discussed, tidal torques can drive a star's rotation rate toward a tidally locked state where the tidal torques fix the rotation period $P_{rot}$ to the equilibrium rotation period $P_{eq}$. Tidal models can predict $P_{eq}$ including the “constant phase lag” \citep[CPL;][]{Ferraz-Mello2008,Heller2011} equilibrium tidal model. Following \citet{Barnes2017} and \citet{Fleming2019} the CPL model permits a 1:1 and 3:2 spin-orbit state by:
\begin{equation}
\centering
    P_{eq}^{\text{CPL}} = 
    \begin{cases}
      P_{eq} \text{ if } e < \sqrt{1/19} \\
      \frac{2}{3}P_{eq} \text{ if } e \geq \sqrt{1/19}. \\
    \end{cases}  
\end{equation}
Using the second case we find $P_{eq}$ = 7.32 days for TOI-746 and with an upper limit on the rotation period of $P_{\rm rot}/\sin i$ = 7.9\,$\pm$\,1.6\,days TOI-746 may be in a supersychronous 3:2 spin–orbit state. We also consider the system may be in a pseudo-synchronous rotation state \citep{Hut1981} that approximates spin-orbit synchronization around the time of periastron. \citet{Zimmerman2017} explored possible pseudosynchronization for heartbeat binary stars with $Kepler$ lightcurves but found generally that their sample clustered around $\frac{3}{2}$ times the pseudosynchronization period indicating that they have plateaued prematurely in their synchronization. We roughly estimate this using Kepler's third law to determine what the period would be if the companion was in a circular orbit at the periastron distance:
\begin{equation}
    P_{\mathrm{peri}} = P_{\mathrm{orb}} (1 - e)^{3/2}.
\end{equation}
Assuming this circular orbit at periastron we find $P_{\mathrm{peri}}$ = 7.88\,$\pm$\,0.05 days which is very similar to the upper limit of the rotation period suggesting TOI-746 may be in a pseudo-synchronous rotation at periastron.

TOI-1213 is also relatively older 5.3$^{+4.2}_{-3.4}$\,Gyr star with a companion on an eccentric $e$ = 0.498$^{+0.003}_{-0.002}$ orbit. We find $P_{eq}$ = 18.14 days for TOI-1213, and with $P_{\rm rot}/\sin i$ = 12.0\,$\pm$\,3.6\,days TOI-1213 does not appear to have spin-orbit synchronization. However, assuming a circular orbit at periastron we find $P_{\mathrm{peri}}$ = 9.64\,$\pm$\,0.09  days, which is within uncertainties of the upper limit of the stellar rotation period suggesting TOI-1213 may be in a pseudo-synchronous rotation at periastron.

\subsection{Discoveries in context with transiting brown dwarfs and very low-mass stars} \label{sec:13-150comp}

\begin{figure*}
  \centering
  \includegraphics[width=0.99\textwidth]{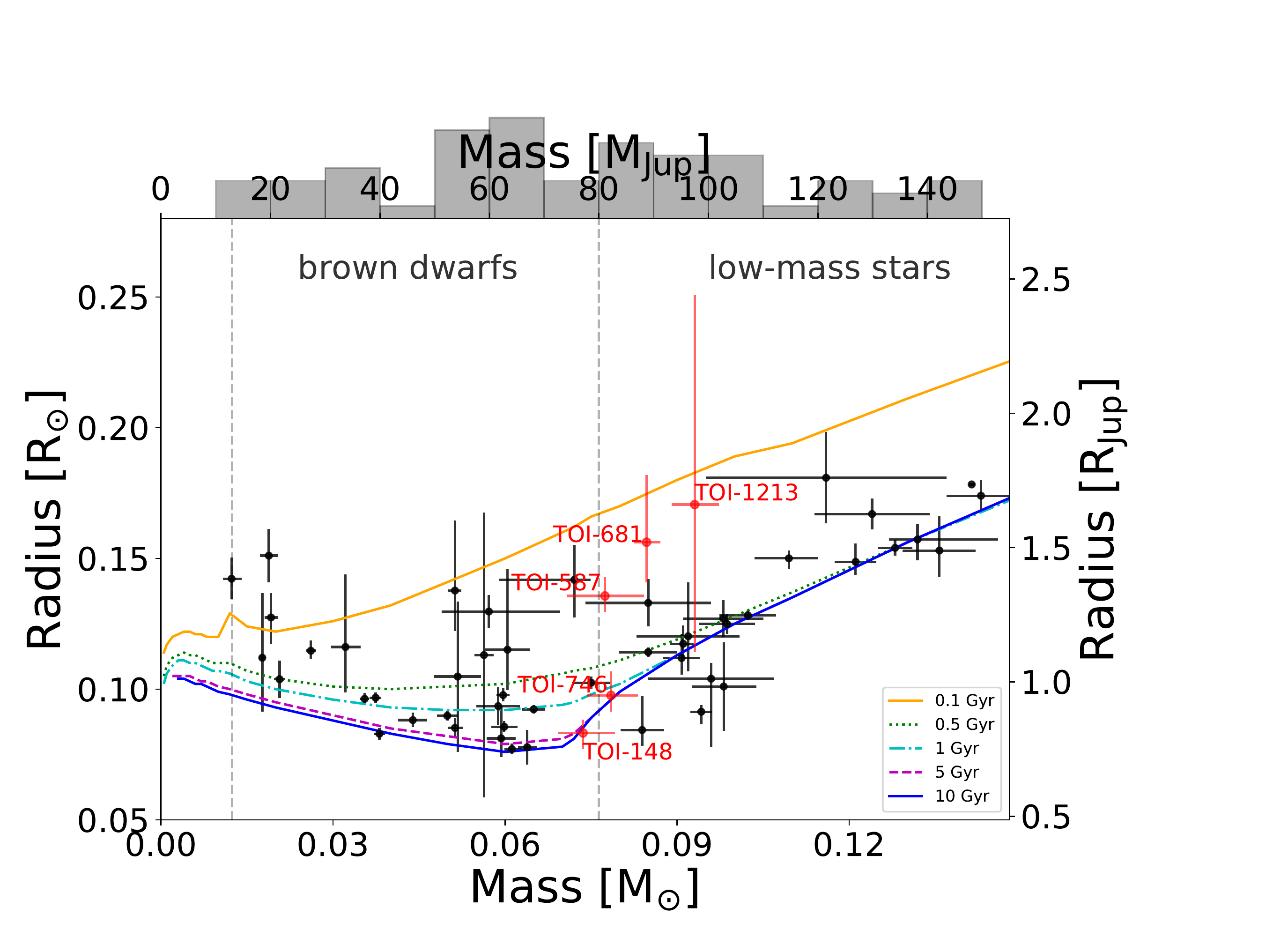}
  \caption{Radius-mass diagram for the 54 brown dwarfs and low-mass stars presented in Table \ref{tab:trancomp}. The five companions presented in this work are highlighted in red. The gray vertical dashed lines display the 13 and 80 \mjup approximate boundaries of the brown dwarf regime. The colored lines display isochrone models from \citet{Baraffe2003,Baraffe2015} for low mass stars and substellar objects at solar metallicity with ages of 0.1, 0.5, 1, 5, and 10 Gyr. The histogram at the top displays relative occurrence of these transiting objects. We note the brown dwarf RIK 72b is not shown because its radius is 3.1\,\rjup.}
  \label{fig:mr}
\end{figure*}

\begin{table*} 
\centering
List of Published 13-150 \mjup Transiting Companions as of June 2021
\tiny
\begin{tabular}{lccrcccccc}
\hline\hline
Name & M$_{\rm2}$ [\mjup] & R$_{\rm2}$ [\rjup] & $P$ [days] & $ecc$ & M$_{\rm1}$ [\msol] & R$_{\rm1}$ [\rsol] & \teff [K] & \feh & Reference \\

\hline

HATS-70b & 12.9$_{-1.6}^{+1.8}$ & 1.38$_{-0.07}^{+0.08}$ & 1.89 & <0.18 & 1.78\,$\pm$\,0.12 & 1.88$_{-0.07}^{+0.06}$ & 7930$_{-820}^{+630}$ & 0.04$_{-0.11}^{+0.10}$ & (1) \\ 
TOI-1278b & 18.5\,$\pm$\,0.5 & 1.09$_{-0.20}^{+0.24}$ & 14.48 & 0.013\,$\pm$\,0.004 & 0.54\,$\pm$\,0.02 & 0.57\,$\pm$\,0.01 & 3799\,$\pm$\,42 & -0.01\,$\pm$\,0.28 & (2) \\ 
GPX-1b & 19.7\,$\pm$\,1.6 & 1.47\,$\pm$\,0.10 & 1.74 & 0.000\,$\pm$\,0.000 & 1.68\,$\pm$\,0.10 & 1.56\,$\pm$\,0.10 & 7000\,$\pm$\,200 & 0.35\,$\pm$\,0.10 & (3) \\ 
Kepler-39b & 20.1$_{-1.2}^{+1.3}$ & 1.24$_{-0.10}^{+0.09}$ & 21.09 & 0.112\,$\pm$\,0.057 & 1.29$_{-0.07}^{+0.06}$ & 1.40\,$\pm$\,0.10 & 6350\,$\pm$\,100 & 0.10\,$\pm$\,0.14 & (4) \\ 
CoRoT-3b & 21.7\,$\pm$\,1.0 & 1.01\,$\pm$\,0.07 & 4.26 & 0 (fixed) & 1.37\,$\pm$\,0.09 & 1.56\,$\pm$\,0.09 & 6740\,$\pm$\,140 & -0.02\,$\pm$\,0.06 & (5) \\ 
KELT-1b & 27.4\,$\pm$\,0.9 & 1.12$_{-0.03}^{+0.04}$ & 1.22 & 0.010$_{-0.007}^{+0.010}$ & 1.33\,$\pm$\,0.06 & 1.47\,$\pm$\,0.04 & 6516\,$\pm$\,49 & 0.05\,$\pm$\,0.08 & (6) \\ 
NLTT 41135b & 33.7$_{-2.6}^{+2.8}$ & 1.13$_{-0.17}^{+0.27}$ & 2.89 & <0.02 & 0.19$_{-0.02}^{+0.03}$ & 0.21$_{-0.01}^{+0.02}$ & 3230\,$\pm$\,130 & -0.25\,$\pm$\,0.25 & (7) \\ 
WASP-128b & 37.2\,$\pm$\,0.8 & 0.94\,$\pm$\,0.02 & 2.21 & <0.007 & 1.16\,$\pm$\,0.04 & 1.15\,$\pm$\,0.02 & 5950\,$\pm$\,50 & 0.01\,$\pm$\,0.12 & (8) \\ 
CWW 89Ab & 39.2$_{-1.1}^{+0.9}$ & 0.94\,$\pm$\,0.02 & 5.29 & 0.189\,$\pm$\,0.002 & 1.10\,$\pm$\,0.04 & 1.03\,$\pm$\,0.02 & 5755\,$\pm$\,49 & 0.20\,$\pm$\,0.09 & (9) (10) \\ 
KOI-205b & 39.9\,$\pm$\,1.0 & 0.81\,$\pm$\,0.02 & 11.72 & <0.031 & 0.93\,$\pm$\,0.03 & 0.84\,$\pm$\,0.02 & 5237\,$\pm$\,60 & 0.14\,$\pm$\,0.12 & (11) \\ 
TOI-1406b & 46.0$_{-2.7}^{+2.6}$ & 0.86\,$\pm$\,0.03 & 10.57 & 0.026$_{-0.010}^{+0.013}$ & 1.18$_{-0.09}^{+0.08}$ & 1.35\,$\pm$\,0.03 & 6290\,$\pm$\,100 & -0.08\,$\pm$\,0.09 & (12) \\ 
EPIC 212036875b & 52.3\,$\pm$\,1.9 & 0.87\,$\pm$\,0.02 & 5.17 & 0.132\,$\pm$\,0.004 & 1.29$_{-0.06}^{+0.07}$ & 1.50\,$\pm$\,0.03 & 6238$_{-60}^{+59}$ & 0.01\,$\pm$\,0.10 & (10) (13) \\ 
TOI-503b & 53.7\,$\pm$\,1.2 & 1.34$_{-0.15}^{+0.26}$ & 3.68 & 0 (fixed) & 1.80\,$\pm$\,0.06 & 1.70$_{-0.04}^{+0.05}$ & 7650$_{-160}^{+140}$ & 0.30$_{-0.09}^{+0.08}$ & (14) \\ 
TOI-852b & 53.7$_{-1.3}^{+1.4}$ & 0.83\,$\pm$\,0.04 & 4.95 & 0.004$_{-0.003}^{+0.004}$ & 1.32$_{-0.04}^{+0.05}$ & 1.71\,$\pm$\,0.04 & 5768$_{-81}^{+84}$ & 0.33\,$\pm$\,0.09 & (15) \\ 
AD 3116b & 54.2\,$\pm$\,4.3 & 1.02\,$\pm$\,0.28 & 1.98 & 0.146\,$\pm$\,0.024 & 0.28\,$\pm$\,0.02 & 0.29\,$\pm$\,0.08 & 3184\,$\pm$\,29 & 0.16\,$\pm$\,0.10 & (16) \\ 
CoRoT-33b & 59.0$_{-1.7}^{+1.8}$ & 1.10\,$\pm$\,0.53 & 5.82 & 0.070\,$\pm$\,0.002 & 0.86\,$\pm$\,0.04 & 0.94$_{-0.08}^{+0.14}$ & 5225\,$\pm$\,80 & 0.44\,$\pm$\,0.10 & (17) \\ 
RIK 72b & 59.2$_{-6.7}^{+6.8}$ & 3.10\,$\pm$\,0.31 & 97.76 & 0.108$_{-0.006}^{+0.012}$ & 0.44\,$\pm$\,0.04 & 0.96\,$\pm$\,0.10 & 3349\,$\pm$\,142 & 0.00\,$\pm$\,0.10 & (18) \\ 
TOI-811b & 59.9$_{-8.6}^{+13.0}$ & 1.26\,$\pm$\,0.06 & 25.17 & 0.509\,$\pm$\,0.075 & 1.32$_{-0.07}^{+0.05}$ & 1.27$_{-0.09}^{+0.06}$ & 6107\,$\pm$\,77 & 0.40$_{-0.09}^{+0.07}$ & (15) \\ 
TOI-263b & 61.6\,$\pm$\,4.0 & 0.91\,$\pm$\,0.07 & 0.56 & 0.017$_{-0.010}^{+0.009}$ & 0.44\,$\pm$\,0.04 & 0.44\,$\pm$\,0.03 & 3471\,$\pm$\,33 & 0.00\,$\pm$\,0.10 & (19) \\ 
KOI-415b & 62.1\,$\pm$\,2.7 & 0.79$_{-0.07}^{+0.12}$ & 166.79 & 0.689\,$\pm$\,0.000 & 0.94\,$\pm$\,0.06 & 1.25$_{-0.10}^{+0.15}$ & 5810\,$\pm$\,80 & -0.24\,$\pm$\,0.11 & (20) \\ 
WASP-30b & 62.5\,$\pm$\,1.2 & 0.95$_{-0.02}^{+0.03}$ & 4.16 & 0 (fixed) & 1.25$_{-0.04}^{+0.03}$ & 1.39\,$\pm$\,0.03 & 6202$_{-51}^{+42}$ & 0.08$_{-0.05}^{+0.07}$ & (21) \\ 
LHS 6343c & 62.7\,$\pm$\,2.4 & 0.83\,$\pm$\,0.02 & 12.71 & 0.056\,$\pm$\,0.032 & 0.37\,$\pm$\,0.01 & 0.38\,$\pm$\,0.01 & 3130\,$\pm$\,20 & 0.04\,$\pm$\,0.08 & (22) \\ 
CoRoT-15b & 63.3\,$\pm$\,4.1 & 1.12$_{-0.15}^{+0.30}$ & 3.06 & 0 (fixed) & 1.32\,$\pm$\,0.12 & 1.46$_{-0.14}^{+0.31}$ & 6350\,$\pm$\,200 & 0.10\,$\pm$\,0.20 & (23) \\ 
TOI-569b & 64.1$_{-1.4}^{+1.9}$ & 0.75\,$\pm$\,0.02 & 6.56 & 0.002$_{-0.001}^{+0.002}$ & 1.21\,$\pm$\,0.05 & 1.48\,$\pm$\,0.03 & 5768$_{-92}^{+110}$ & 0.29$_{-0.08}^{+0.09}$ & (12) \\ 
EPIC 201702477b & 66.9\,$\pm$\,1.7 & 0.76\,$\pm$\,0.07 & 40.74 & 0.228\,$\pm$\,0.003 & 0.87\,$\pm$\,0.03 & 0.90\,$\pm$\,0.06 & 5517\,$\pm$\,70 & -0.16\,$\pm$\,0.05 & (24) \\ 
LP261-75b & 68.1\,$\pm$\,2.1 & 0.90\,$\pm$\,0.01 & 1.88 & <0.007 & 0.30\,$\pm$\,0.01 & 0.31\,$\pm$\,0.00 & 3100\,$\pm$\,50 & ... & (25) \\ 
NGTS-7Ab & 75.5$_{-13.7}^{+3.0}$ & 1.38$_{-0.14}^{+0.13}$ & 0.68 & 0 (fixed) & 0.24\,$\pm$\,0.03 & 0.61\,$\pm$\,0.06 & 3359$_{-89}^{+106}$ & 0.00\,$\pm$\,0.10 & (26) \\ 
KOI-189b & 78.6\,$\pm$\,3.5 & 1.00\,$\pm$\,0.02 & 30.36 & 0.275\,$\pm$\,0.004 & 0.76\,$\pm$\,0.05 & 0.73\,$\pm$\,0.02 & 4952\,$\pm$\,40 & -0.12\,$\pm$\,0.10 & (27) \\ 
TOI-148b & 77.1$_{-4.6}^{+5.8}$ & 0.81$_{-0.06}^{+0.05}$ & 4.87 & 0.005$_{-0.004}^{+0.006}$ & 0.97$_{-0.09}^{+0.12}$ & 1.20\,$\pm$\,0.07 & 5990\,$\pm$\,140 & -0.24\,$\pm$\,0.25 & this work \\ 
TOI-587b & 81.1$_{-7.0}^{+7.1}$ & 1.32$_{-0.06}^{+0.07}$ & 8.04 & 0.051$_{-0.036}^{+0.049}$ & 2.33\,$\pm$\,0.12 & 2.01\,$\pm$\,0.09 & 9800\,$\pm$\,200 & 0.08$_{-0.12}^{+0.11}$ & this work \\ 
TOI-746b & 82.2$_{-4.4}^{+4.9}$ & 0.95$_{-0.06}^{+0.09}$ & 10.98 & 0.199\,$\pm$\,0.003 & 0.94$_{-0.08}^{+0.09}$ & 0.97$_{-0.03}^{+0.04}$ & 5690\,$\pm$\,140 & -0.02\,$\pm$\,0.23 & this work \\
EBLM J0555-57Ab & 87.9\,$\pm$\,4.0 & 0.82$_{-0.06}^{+0.13}$ & 7.76 & 0.089\,$\pm$\,0.004 & 1.18\,$\pm$\,0.08 & 1.00$_{-0.07}^{+0.14}$ & 6386\,$\pm$\,124 & -0.04\,$\pm$\,0.14 & (28) \\ 
TOI-681b & 88.7$_{-2.3}^{+2.5}$ & 1.52$_{-0.15}^{+0.25}$ & 15.78 & 0.093$_{-0.019}^{+0.022}$ & 1.54$_{-0.05}^{+0.06}$ & 1.47\,$\pm$\,0.04 & 7440$_{-140}^{+150}$ & -0.08\,$\pm$\,0.05 & this work \\ 
OGLE-TR-123b & 89.0\,$\pm$\,11.5 & 1.29\,$\pm$\,0.09 & 1.80 & 0 (fixed) & 1.29\,$\pm$\,0.26 & 1.55\,$\pm$\,0.10 & 6700\,$\pm$\,300 & ... & (29) \\
TOI-694b & 89.0\,$\pm$\,5.3 & 1.11\,$\pm$\,0.02 & 48.05 & 0.521\,$\pm$\,0.002 & 0.97$_{-0.04}^{+0.05}$ & 1.00\,$\pm$\,0.01 & 5496$_{-81}^{+87}$ & 0.21\,$\pm$\,0.08 & (30) \\ 
KOI-607b & 95.1$_{-3.4}^{+3.3}$ & 1.09$_{-0.06}^{+0.09}$ & 5.89 & 0.395\,$\pm$\,0.009 & 0.99\,$\pm$\,0.05 & 0.92\,$\pm$\,0.03 & 5418$_{-85}^{+87}$ & 0.38$_{-0.09}^{+0.07}$ & (10) \\ 
J1219-39b & 95.4$_{-2.5}^{+1.9}$ & 1.14$_{-0.05}^{+0.07}$ & 6.76 & 0.055\,$\pm$\,0.000 & 0.83\,$\pm$\,0.03 & 0.81$_{-0.02}^{+0.04}$ & 5412$_{-65}^{+81}$ & -0.21\,$\pm$\,0.07 & (21) \\ 
OGLE-TR-122b & 96.3\,$\pm$\,9.4 & 1.17$_{-0.13}^{+0.20}$ & 7.27 & 0.205\,$\pm$\,0.008 & 0.98\,$\pm$\,0.14 & 1.05$_{-0.09}^{+0.20}$ & 5700\,$\pm$\,300 & 0.15\,$\pm$\,0.36 & (31) \\ 
TOI-1213b & 97.5$_{-4.2}^{+4.4}$ & 1.66$_{-0.55}^{+0.78}$ & 27.22 & 0.498$_{-0.002}^{+0.003}$ & 0.99$_{-0.06}^{+0.07}$ & 0.99\,$\pm$\,0.04 & 5590\,$\pm$\,150 & 0.25$_{-0.14}^{+0.13}$ & this work \\ 
K2-76b & 98.7\,$\pm$\,2.0 & 0.89$_{-0.05}^{+0.03}$ & 11.99 & 0.255$_{-0.006}^{+0.007}$ & 0.96\,$\pm$\,0.03 & 1.17$_{-0.06}^{+0.03}$ & 5747$_{-70}^{+64}$ & 0.01\,$\pm$\,0.04 & (32) \\ 
CoRoT 101186644 & 100.5\,$\pm$\,11.5 & 1.01$_{-0.25}^{+0.06}$ & 20.68 & 0.402\,$\pm$\,0.006 & 1.20\,$\pm$\,0.20 & 1.07\,$\pm$\,0.07 & 6090\,$\pm$\,200 & 0.20\,$\pm$\,0.20 & (33) \\ 
J2343+29Ab & 102.7\,$\pm$\,7.3 & 1.24\,$\pm$\,0.07 & 16.95 & 0.161$_{-0.003}^{+0.002}$ & 0.86\,$\pm$\,0.10 & 0.85$_{-0.06}^{+0.05}$ & 5150$_{-60}^{+90}$ & 0.10\,$\pm$\,0.14 & (34) \\ 
EBLM J0954-23Ab & 102.8$_{-5.9}^{+6.0}$ & 0.98\,$\pm$\,0.17 & 7.58 & 0.042\,$\pm$\,0.001 & 1.17\,$\pm$\,0.08 & 1.23\,$\pm$\,0.17 & 6406\,$\pm$\,124 & -0.01\,$\pm$\,0.14 & (28) \\ 
KOI-686b & 103.4\,$\pm$\,5.1 & 1.22\,$\pm$\,0.04 & 52.51 & 0.556\,$\pm$\,0.004 & 0.98\,$\pm$\,0.07 & 1.04\,$\pm$\,0.03 & 5834\,$\pm$\,100 & -0.06\,$\pm$\,0.13 & (27) \\ 
TIC 220568520b & 107.2\,$\pm$\,5.2 & 1.25\,$\pm$\,0.02 & 18.56 & 0.096\,$\pm$\,0.003 & 1.03\,$\pm$\,0.04 & 1.01\,$\pm$\,0.01 & 5589\,$\pm$\,81 & 0.26\,$\pm$\,0.07 & (30) \\ 
HATS551-016B & 114.7$_{-6.3}^{+5.2}$ & 1.46$_{-0.04}^{+0.03}$ & 2.05 & 0.080\,$\pm$\,0.020 & 0.97$_{-0.06}^{+0.05}$ & 1.22$_{-0.03}^{+0.02}$ & 6420\,$\pm$\,90 & -0.60\,$\pm$\,0.06 & (35) \\ 
OGLE-TR-106b & 121.5\,$\pm$\,22.0 & 1.76\,$\pm$\,0.17 & 2.54 & 0.000\,$\pm$\,0.020 & ... & 1.31\,$\pm$\,0.09 & ... & ... & (36) \\ 
EBLM J1431-11Ab & 126.9$_{-3.9}^{+3.8}$ & 1.45$_{-0.05}^{+0.07}$ & 4.45 & 0 (fixed) & 1.20\,$\pm$\,0.06 & 1.11$_{-0.03}^{+0.04}$ & 6161\,$\pm$\,124 & 0.15\,$\pm$\,0.14 & (28) \\ 
HAT-TR-205-013B & 129.9\,$\pm$\,10.5 & 1.62\,$\pm$\,0.06 & 2.23 & 0.012\,$\pm$\,0.021 & 1.04\,$\pm$\,0.13 & 1.28\,$\pm$\,0.04 & 6295$_{-335}^{+245}$ & ... & (37) \\ 
TIC 231005575b & 134.1\,$\pm$\,3.1 & 1.50\,$\pm$\,0.03 & 61.78 & 0.298$_{-0.001}^{+0.004}$ & 1.04\,$\pm$\,0.04 & 0.99\,$\pm$\,0.05 & 5500\,$\pm$\,85 & -0.44\,$\pm$\,0.06 & (38) \\ 
HATS551-021B & 138.2$_{-5.2}^{+14.7}$ & 1.53$_{-0.08}^{+0.06}$ & 3.64 & 0.060\,$\pm$\,0.020 & 1.10\,$\pm$\,0.10 & 1.20$_{-0.01}^{+0.08}$ & 6670\,$\pm$\,220 & -0.40\,$\pm$\,0.10 & (35) \\ 
EBLM J2017+02Ab & 142.2$_{-6.7}^{+6.6}$ & 1.49$_{-0.10}^{+0.13}$ & 0.82 & 0 (fixed) & 1.10\,$\pm$\,0.07 & 1.20$_{-0.05}^{+0.08}$ & 6161\,$\pm$\,124 & -0.07\,$\pm$\,0.14 & (28) \\ 
KIC 1571511B & 148.1\,$\pm$\,0.5 & 1.74$_{-0.01}^{+0.00}$ & 14.02 & 0.327\,$\pm$\,0.003 & 1.26$_{-0.03}^{+0.04}$ & 1.34\,$\pm$\,0.01 & 6195\,$\pm$\,50 & 0.37\,$\pm$\,0.08 & (39) \\ 
WTS 19g-4-02069B & 149.8\,$\pm$\,6.3 & 1.69\,$\pm$\,0.06 & 2.44 & 0 (fixed) & 0.53\,$\pm$\,0.02 & 0.51\,$\pm$\,0.01 & 3300\,$\pm$\,140 & ... & (40) \\ 
\hline
\end{tabular}
\caption{\tiny \textbf{References.}
(1) \citealt{Zhou2019};
(2) \citealt{Artigau2021}; 
(3) \citealt{Benni2020}; 
(4) \citealt{Bonomo2015}; 
(5) \citealt{Deleuil2008}; 
(6) \citealt{Siverd2012}; 
(7) \citealt{Irwin2010}; 
(8) \citealt{Hodzic2018}; 
(9) \citealt{Nowak2017}; 
(10) \citealt{Carmichael2019}; 
(11) \citealt{Diaz2013}; 
(12) \citealt{Carmichael2020}; 
(13) \citealt{Persson2019}; 
(14) \citealt{Subjak2020}; 
(15) \citealt{Carmichael2021}; 
(16) \citealt{Gillen2017}; 
(17) \citealt{Csizmadia2015}; 
(18) \citealt{David2019}; 
(19) \citealt{Palle2021}; 
(20) \citealt{Moutou2013}; 
(21) \citealt{Triaud2013}; 
(22) \citealt{Johnson2011}; 
(23) \citealt{Bouchy2011}; 
(24) \citealt{Bayliss2017}; 
(25) \citealt{Irwin2018}; 
(26) \citealt{Jackman2019}; 
(27) \citealt{Diaz2014}; 
(28) \citealt{vonBoetticher2019}; 
(29) \citealt{Pont2006}; 
(30) \citealt{Mireles2020}; 
(31) \citealt{Pont2005a}; 
(32) \citealt{Shporer2017}; 
(33) \citealt{Tal-Or2013}; 
(34) \citealt{Chaturvedi2016}; 
(35) \citealt{Zhou2014}; 
(36) \citealt{Pont2005b}; 
(37) \citealt{Beatty2007}; 
(38) \citealt{Gill2020};
(39) \citealt{Ofir2012}; 
(40) \citealt{Nefs2013}.
\textbf{Note.} The brown dwarf binary system from \citet{Stassun2006}, triple system from \citet{Triaud2020}, and white dwarf companions are not included.
}
\label{tab:trancomp}
\end{table*}

Here we place these five transiting companions into context by comparing them with other very low-mass stars and brown dwarfs using transiting brown dwarfs from the list compiled by \citet{Carmichael2021}. We also include 21 low-mass stars with masses between 80 and 150 \mjup from the list compiled by \citet{Mireles2020}. We exclude the brown dwarf binary system from \citet{Stassun2006} and triple system from \citet{Triaud2020} so as only to focus on transiting companions. We also exclude brown dwarfs and low-mass stars known to transit white dwarfs \citep[e.g.,][]{Parsons2012a,Parsons2012b,Parsons2012c}. We add the three recent transiting brown dwarf discoveries of GPX-1b \citep{Benni2020}, TOI-263b \citep{Palle2021}, and TOI-1278b \citep{Artigau2021}. This includes a total of 54 transiting companions in the mass range of 13\,-\,150\,\mjup presented in Table \ref{tab:trancomp}.




\subsubsection{13-150 \mjup transiting companion radius-mass relationship}


Figure \ref{fig:mr} places the five objects studied here in the radius-mass diagram with  transiting brown dwarfs and low-mass stars with masses up to 150 \mjup. We also plot the theoretical isochrones for solar metallicity at ages 0.1, 0.5, 1, 5, and 10 Gyr from \citet{Baraffe2003,Baraffe2015}. 
Theoretical isochrones for low-mass stars and brown dwarfs \citep[e.g.,][]{Baraffe2003,Baraffe2015} show the age of the object should affect its radius for a particular mass, with older objects having smaller radii. This is particularly true for brown dwarfs that do not have the long-term outward pressure of hydrogen fusion to balance out the crush of gravity. Using well-defined ages \citet{Carmichael2021} displayed a clear contrast between the radii of an old and a young brown dwarf; further systems with well-characterized ages are crucial to test this age-radius effect.

We derive young ages for TOI-681 (0.170\,$\pm$\,0.025\,Gyr from cluster membership) and TOI-587 (0.2\,$\pm$\,0.1\,Gyr from MIST isochrone stellar modeling of the hot A star) and see relatively enlarged radii for both of these objects. However, TOI-681b displays a slightly grazing transit and therefore the radius may be overestimated. TOI-681's age is notably close to the Pleiades cluster whose brown dwarf and very low-mass star candidates were spectroscopically classified as late M dwarfs \citep[e.g.,][]{Dahm2015}. 

As discussed in Section \ref{sec:grazing} TOI-1213b's radius is not well constrained from the grazing transit geometry and possibly overestimated given its age of  5.3$^{+4.2}_{-3.4}$\,Gyr. We find an older 7.7\,$\pm$\,3.7\,Gyr age for TOI-148 and with $R_b$=0.81$^{+0.05}_{-0.06}$\,\rjup TOI-148b has one of the smallest radii among 13-150\,\mjup transiting companions. We also find an older 6.5$^{+4.3}_{-3.9}$\,Gyr age for TOI-746 and a relatively small 0.95$^{+0.09}_{-0.06}$\,\rjup companion radius.

The histogram on top of Figure \ref{fig:mr} displays relative occurrence of these transiting objects. Currently more massive brown dwarfs seem to have a slightly higher occurrence than less massive brown dwarfs. However, we note this is an incomplete sample and more systems are needed to understand the occurrence of these objects. Additionally we note that there may be a bias of not publishing low-mass stars as many current transit and RV research teams emphasize lower mass companion discoveries.






\subsubsection{13-150 \mjup transiting companion eccentricity distribution.}

By interpreting the eccentricity distribution of known brown dwarf candidate companions around FGK-type stars, many of which only had lower-mass estimates ($M_{\rm{BD}}$\,sin\,$i$), \citet{MaGe2014} suggested that brown dwarfs could be split at $\sim$42.5\,\mjup. Brown dwarfs below 42.5\,\mjup may primarily form in the protoplanetary disk through core-accretion or disk gravitational instability and brown dwarfs above this mass may dominantly form like a stellar binary through molecular cloud fragmentation. This was suggested because lower-mass brown dwarfs showed a trend in their eccentricity distribution with lower maximum eccentricity with increasing mass, consistent with eccentricity scattering with other objects formed in the disk \citep[e.g., ‘planet–planet scattering’,][]{Rasio1996,Chatterjee2008,Ford2008}. Where higher-mass brown dwarfs showed more diverse eccentricities similar to that of stellar binaries \citep{MaGe2014}. This trend was again analyzed for known brown dwarf companions by \citet{Grieves2017} who found an overall similar distribution. Additionally, \citet{Kiefer2021} found the eccentricity distribution for brown dwarfs and low-mass M-dwarf companions around FGK stars at less than 60 pc of the Sun with a true mass measured to match that of \citet{MaGe2014}.

\begin{figure}
  \centering
  \includegraphics[width=0.49\textwidth]{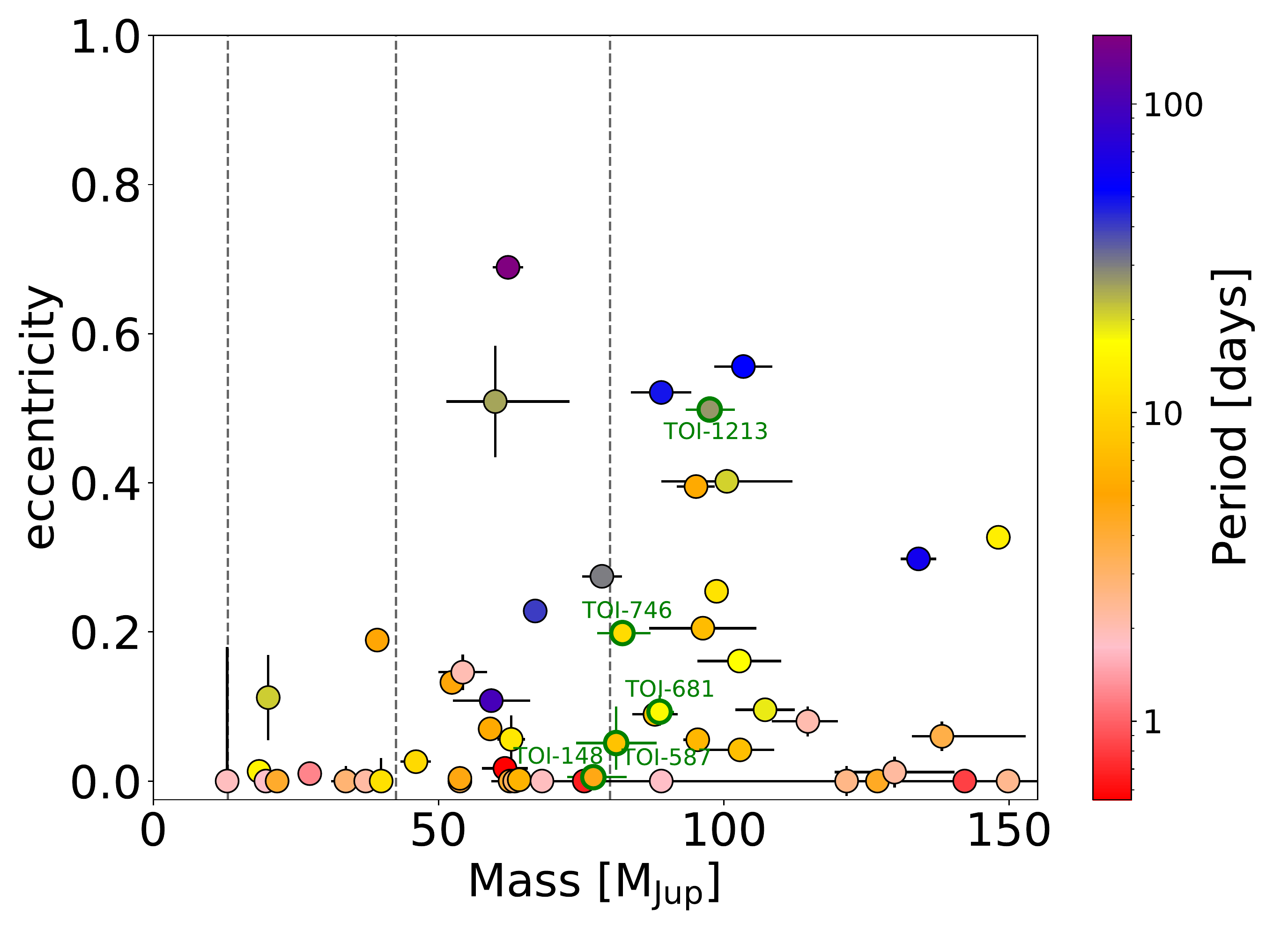}
  \caption{Eccentricity and mass for transiting brown dwarfs and low-mass stellar companions. The color of each circle denotes the period. Transiting companions presented in this work have green edges and error bars. The vertical dashed lines are on the masses of 13, 42.5, and 80 $\mjup$.}
  \label{fig:masseccper}
\end{figure}

We analyze the eccentricity distribution for transiting brown dwarfs and low-mass stars up to 150\,\mjup in Figure \ref{fig:masseccper}. Although the higher-mass brown dwarfs and low-mass stars display a larger range in eccentricities, the sample size is still too small to make a claim that these represent two separate populations. The symbols in Figure \ref{fig:masseccper} are colored by period, and as expected the companions with the highest eccentricities have relatively longer orbital periods.



\begin{figure}
  \centering
  \includegraphics[width=0.55\textwidth]{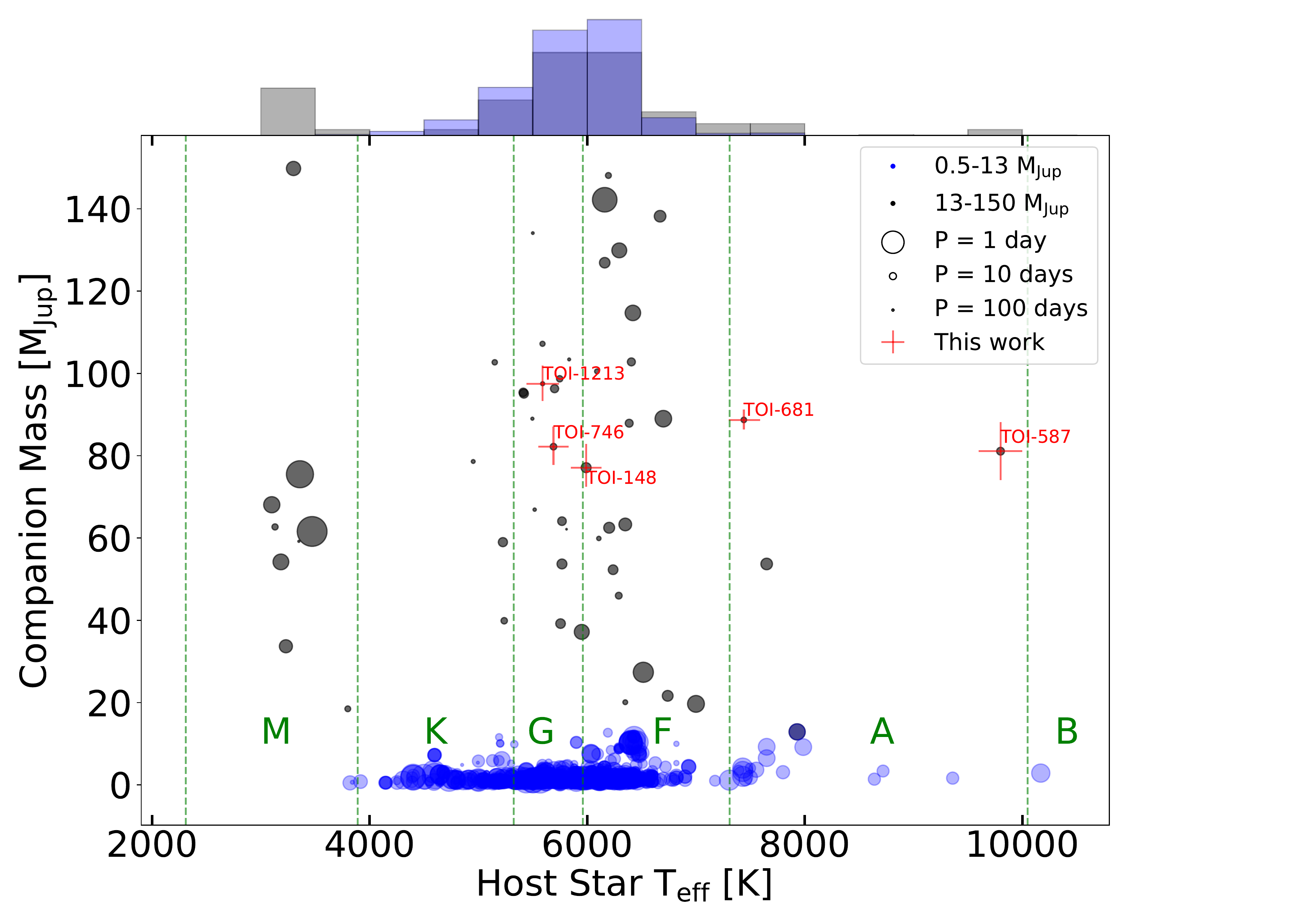}
  \caption{Companion mass and host star effective temperature $\teff$ for transiting brown dwarfs and low-mass stellar companions (gray circles) with the five new companions of this work shown with red error bars. We also display giant planets that have both RV and transit data from the NASA exoplanet archive\protect\footnote{https://exoplanetarchive.ipac.caltech.edu} (1089 giant planets) with blue circles. The circles are inversely sized by the companion period to highlight a higher proportion of brown dwarfs and low-mass stars in close orbit around F dwarfs. The histograms at the top display the number of relative companions in 500\,K $\teff$ bins. The vertical dashed green lines display approximate $\teff$ borders of stellar types.}
  \label{fig:massteff}
\end{figure}

%



\begin{figure}
  \centering
  \includegraphics[width=0.55\textwidth]{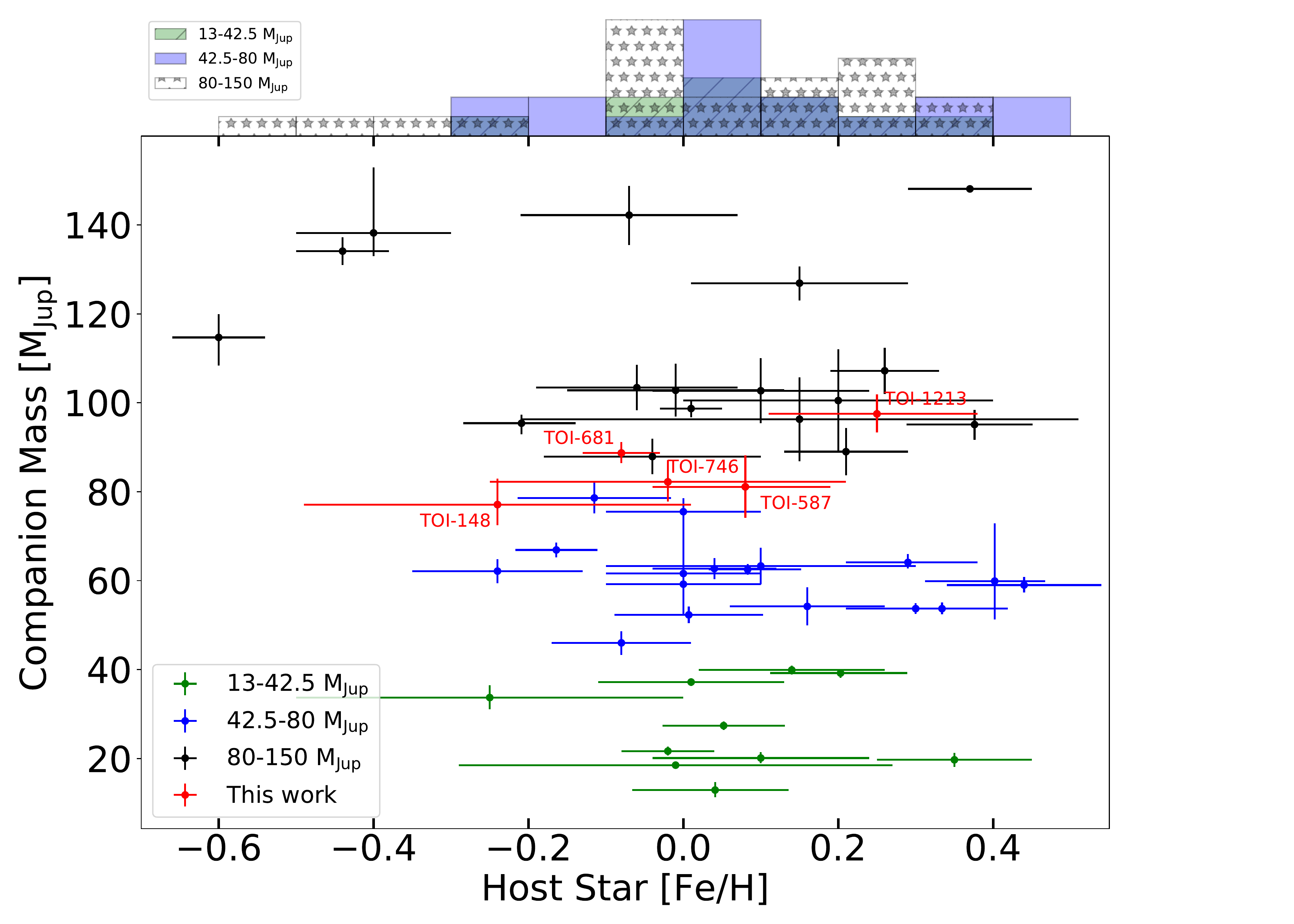}
  \caption{Companion mass and host star metallicity $\feh$ for transiting brown dwarfs and low-mass stellar companions. We split the companions between 13\,-\,42.5, 42.5\,-\,80, and 80\,-\,150 \mjup to visualize possible metallicity differences between low-mass and high-mass brown dwarfs and low-mass stars. The histograms at the top display the number of companions in 0.1 dex $\feh$ bins for these mass ranges.}
  \label{fig:massfeh}
\end{figure}


\subsubsection{13-150 \mjup transiting companion host stars}

We review the host star effective temperature and stellar type of transiting brown dwarfs and low-mass stars in Figure \ref{fig:massteff}. G and F-type host stars are the most common while there are only four hotter K-type stars hosting a transiting companion in the mass range 13 - 150\,\mjup. As noted by \citet{Carmichael2020} we see a relatively large percentage of transiting BDs with M dwarf host stars, which is in contrast to hot Jupiters that have a relatively smaller percentage of M dwarf host stars. TOI-587 is the hottest main-sequence host star of a transiting brown dwarf or low-mass star below 150 $\mjup$ with a $\teff$ = $9800^{+200}_{-200}$\,K. From the most recent table\footnote{\url{http://www.pas.rochester.edu/~emamajek/EEM_dwarf_UBVIJHK_colors_Teff.txt}} of mean dwarf star effective temperatures by \citet{PecautMamajek2013} we find stellar types of F9.5V for TOI-148, A0V for TOI-587, A9V for TOI-681, G4V for TOI-746, and G6V for TOI-1213.


In Figure~\ref{fig:massteff}, following \citet{Bouchy2011}, the symbol size is inversely proportional to the companion orbital period in order to highlight the higher proportion of brown dwarfs and low-mass stars found in close orbit around F dwarfs compared to those around G dwarfs. A plausible conjecture is that G dwarfs may have engulfed their close-in brown-dwarf companions while F dwarfs, which undergo less magnetic braking \citep[e.g.,][]{SadeghiArdestani+2017}, would have preserved them \citep{Guillot+2014}. The two companions orbiting G dwarfs from this study, TOI-746 and TOI-1213, are both relatively long-period (10.98 days and 27.22 days, respectively) and eccentric (see Figure~\ref{fig:masseccper}), indicating that tides have played a limited role in their dynamical evolution. On the other hand, TOI-148, the companion with the shortest orbital period in this study, 4.87 days, orbits around an F dwarf, and is circular, thus somewhat strengthening this tendency for close-in massive companions to be found preferentially around F dwarfs \citep{Guillot+2014}. A deeper, quantitative study is warranted. The relative lack of these companions around K dwarfs and their presence around M dwarfs as shown in Figure~\ref{fig:massteff} is unexplained. 

We display host star metallicity $\feh$ in Figure \ref{fig:massfeh} of transiting brown dwarfs and low-mass stars. Figure \ref{fig:massfeh} shows the most common metallicity for brown dwarf host stars is at $\feh$\,$\sim$\,0, which is consistent with previous findings that brown dwarfs are not preferentially found around metal rich stars like hot Jupiters and are more consistent with metallicity  distributions of stars without substellar companions or stars with low-mass planets \citep[e.g.,][]{MaGe2014,MataSanchez2014,Maldonado2019,Adibekyan2019}. \citet{Maldonado2017} found that stars with less massive brown dwarfs tend to have higher metallicities than stars with more massive brown dwarfs, which is consistent with the interpretations of \citet{MaGe2014} and \citet{Maldonado2019} that more massive brown dwarfs tend to form more like low-mass stars. \citet{Narang2018} also found that the metallicity of host stars with brown dwarf companions is lower compared to 1-4\,\mjup mass objects but similar to more massive ($>$4\,\mjup) planets, suggesting similar formation mechanisms for both. Figure \ref{fig:massfeh} has relatively few companions to make significant statistical claims that these populations have inherently different metallicites, but the distribution is still in agreement with a possible separate population for more massive ($\gtrsim$42.5\,\mjup) brown dwarfs having a metallicity distribution more similar to low-mass stars \citep{Maldonado2017,Kiefer2021}.

\section{Conclusion} \label{sec:conc}

We report the discovery and characterization of five transiting companions near the hydrogen-burning mass limit detected by TESS as objects of interest: TOI-148 (UCAC4 260-199322), TOI-587 (HD 74162), TOI-681 (TYC 8911-00495-1), TOI-746 (TYC 9177-00082-1), and TOI-1213 (TYC 8970-00020-1). We combine TESS photometry with ground-based photometry and spectra from the CORALIE, CHIRON, TRES, and FEROS spectrographs to find companion masses between 77 and 98\,$\mjup$ and companion radii between 0.81 and 1.66\,$\rjup$. We found young ages for TOI-587 (from isochrone stellar modeling) and TOI-681 (from cluster membership) and find their companion radii to be relatively larger compared to companions of similar masses, as expected from theoretical isochrone models. However, TOI-681b has a grazing transit making its companion radius not as well-constrained, and TOI-1213b also has a grazing transit that creates an even less well-constrained companion radius. TOI-148 and TOI-746 have relatively older ages and smaller companion radii. TOI-587 is the hottest main-sequence star ($\teff$ = 9800\,$\pm$\,200\,K) known to host a transiting brown dwarf or low-mass star below 150\,$\mjup$. We find evidence of spin-orbit synchronization for TOI-148 and TOI-746, tidal circularization for TOI-148, and possible pseudosynchronization at periastron for TOI-1213. The sample of transiting brown dwarfs and low-mass stars we analyzed is still too small to make significant statistical claims; however, their eccentricity and metallicity distributions are still consistent with previous suggestions of two separate populations for lower and higher mass brown dwarfs. These companions are all near the hydrogen-burning mass limit and add to the statistical sample needed to distinguish the population differences between brown dwarfs and low-mass stars. 



\begin{acknowledgements}
We thank the Swiss National Science Foundation (SNSF) and the Geneva University for their continuous support to our planet low-mass companion search programs. This work was carried out in the frame of the Swiss National Centre for Competence in Research (NCCR) $PlanetS$ supported by the Swiss National Science Foundation (SNSF). This publication makes use of The Data \& Analysis Center for Exoplanets (DACE), which is a facility based at the University of Geneva (CH) dedicated to extrasolar planet data visualization, exchange, and analysis. DACE is a platform of NCCR $PlanetS$ and is available at https://dace.unige.ch. This paper includes data collected by the TESS mission. Funding for the TESS mission is provided by the NASA Explorer Program. 

We are grateful to the CHIRON team: Todd Henry, Leonardo Paredes, Hodari James, Azmain Nisak, Rodrigo Hinojosa, Roberto Aviles, and Wei-Chun Jao, for carrying out the CHIRON observations and data reduction. 

This work has made use of data from the European Space Agency (ESA) mission $Gaia$, processed by the $Gaia$ Data Processing and Analysis Consortium (DPAC). Funding for the DPAC has been provided by national institutions, in particular the institutions participating in the $Gaia$ Multilateral Agreement. This work makes use of observations from the LCOGT network. LCOGT telescope time was granted by NOIRLab through the Mid-Scale Innovations Program (MSIP). MSIP is funded by NSF. 

We acknowledge support from the French and Italian Polar Agencies, IPEV and PNRA, and from Université Côte d’Azur under Idex UCAJEDI (ANR-15-IDEX-01). 

MEarth is funded by the David and Lucile Packard Fellowship for Science and Engineering, the National Science Foundation under grants AST-0807690, AST-1109468, AST-1004488 (Alan T. Waterman Award) and AST-1616624, and the John Templeton Foundation. This publication was made possible through the support of a grant from the John Templeton Foundation. The opinions expressed in this publication are those of the authors and do not necessarily reflect the views of the John Templeton Foundation.

This research received funding from the European Research Council (ERC) under the European Union's Horizon 2020 research and innovation programme (grant agreement n$^\circ$ 803193/BEBOP), and from the Science and Technology Facilities Council (STFC; grant n$^\circ$ ST/S00193X/1).

Resources supporting this work were provided by the NASA High-End Computing (HEC) Program through the NASA Advanced Supercomputing (NAS) Division at Ames Research Center for the production of the SPOC data products.

We acknowledge the use of public TESS Alert data from pipelines at the TESS Science Office and at the TESS Science Processing Operations Center.

SG has been supported by STFC through consolidated grants ST/L000733/1 and ST/P000495/1. 

KKM acknowledges support from the New York Community Trust Fund for Astrophysical Research.

AJ acknowledges support from FONDECYT project 1210718, and from ANID – Millennium Science Initiative – ICN12\_009.

LAdS is supported by funding from the European Research Council (ERC) under the European Union's Horizon 2020 research and innovation programme (project {\sc Four Aces} grant agreement No 724427).

\end{acknowledgements}

%
%

\bibliographystyle{aa}
\bibliography{bib}

\begin{appendix} 


\onecolumn

\section{Supplementary material}
\begin{table} 
    \centering
    \begin{tabular}{cccc}
        
        \hline\hline
        Time [BJD TDB] & RV [$\ms$] & RV error [$\ms$] & Instrument  \\
        \hline
        TOI-148 & & & \\
        \hline
        2458384.515064 & -30660.46 & 30.42 & CORALIE \\
        2458386.814005 & -37976.20 & 82.56 & CORALIE \\
        2458408.719335 & -20372.36 & 85.21 & CORALIE \\
        2458428.534714 & -20473.55 & 63.01 & CORALIE \\
        2458479.532552 & -38130.30 & 191.13 & CORALIE \\
        2458495.531823 & -25865.29 & 221.07 & CORALIE \\
        2458496.547259 & -20001.33 & 122.21 & CORALIE \\
        2458656.803753 & -20849.44 & 130.90 & CORALIE \\
        2458657.770975 & -23724.86 & 115.92 & CORALIE \\
        2458663.857660 & -35835.81 & 149.59 & CORALIE \\
        2458679.761215 & -33895.58 & 178.17 & CORALIE \\
        \hline
        TOI-587 & & & \\
        \hline
        2458589.719888 & 376.98 & 1201.42 & TRES \\
        2458594.622423 & -8024.60 & 637.52 & TRES \\
        2458597.701320 & 3.14 & 821.62 & TRES \\
        2458598.631813 & 1292.14 & 745.23 & TRES \\
        2458599.640043 & 0.00 & 821.62 & TRES \\
        2458601.636267 & -6391.02 & 604.17 & TRES \\
        2458603.630556 & -6396.90 & 1140.20 & TRES \\
        \hline
        TOI-681 & & & \\
        \hline
        2458622.513448 & 26180.44 & 88.22 & CORALIE \\
        2458654.485737 & 26094.93 & 78.95 & CORALIE \\
        2458818.766341 & 16662.76 & 114.76 & CORALIE \\
        2458819.812002 & 17277.56 & 102.12 & CORALIE \\
        2458842.699941 & 27731.61 & 128.99 & CORALIE \\
        2458850.641613 & 17162.30 & 84.73 & CORALIE \\
        2458852.779240 & 18827.26 & 102.06 & CORALIE \\
        2458857.727110 & 27269.20 & 85.20 & CORALIE \\
        2458923.549796 & 24577.10 & 130.60 & CORALIE \\
        2458641.493741 & 23125.7 & 349.9 & FEROS \\
        2458643.486061 & 20111.0 & 213.9 & FEROS \\
        2458645.490361 & 18394.8 & 331.0 & FEROS \\
        2458652.498401 & 29384.2 & 605.8 & FEROS \\
        \hline
        TOI-746 & & & \\
        \hline
        2458777.827622 & 26662.93 & 63.11 & CORALIE \\
        2458815.804367 & 39986.84 & 44.16 & CORALIE \\
        2458820.773420 & 25318.60 & 40.97 & CORALIE \\
        2458824.651474 & 35494.88 & 63.08 & CORALIE \\
        2458830.772597 & 27610.22 & 45.59 & CORALIE \\
        2458839.622606 & 38959.05 & 71.50 & CORALIE \\
        2458844.682083 & 29562.96 & 70.22 & CORALIE \\
        2458920.561788 & 26623.69 & 100.59 & CORALIE \\
        2458798.773421 & 25294.4 & 10.3 & FEROS \\
        2458804.702041 & 39780.9 & 7.9 & FEROS \\
        \hline
        TOI-1213 & & & \\
        \hline
        2458886.78385 & 19685.0 & 30.0 & CHIRON \\
        2458909.72687 & 21761.0 & 29.0 & CHIRON \\
        2458917.78559 & 19130.0 & 30.0 & CHIRON \\
        2458883.615306 & 22638.52 & 24.61 & CORALIE \\
        2458895.675990 & 34772.99 & 70.44 & CORALIE \\
        2458920.746399 & 27968.23 & 57.41 & CORALIE \\
        2458925.564278 & 31941.32 & 69.56 & CORALIE \\
        2458882.839021 & 23064.1 & 6.0 & FEROS \\
        2458908.787371 & 23728.5 & 7.6 & FEROS \\
        2458915.735351 & 20533.3 & 7.2 & FEROS \\
        \hline
        \hline
    \end{tabular}
    \caption{Radial Velocities}
    \label{tab:rv}
\end{table}

\begin{figure*}
  \centering
  \includegraphics[width=0.75\textwidth]{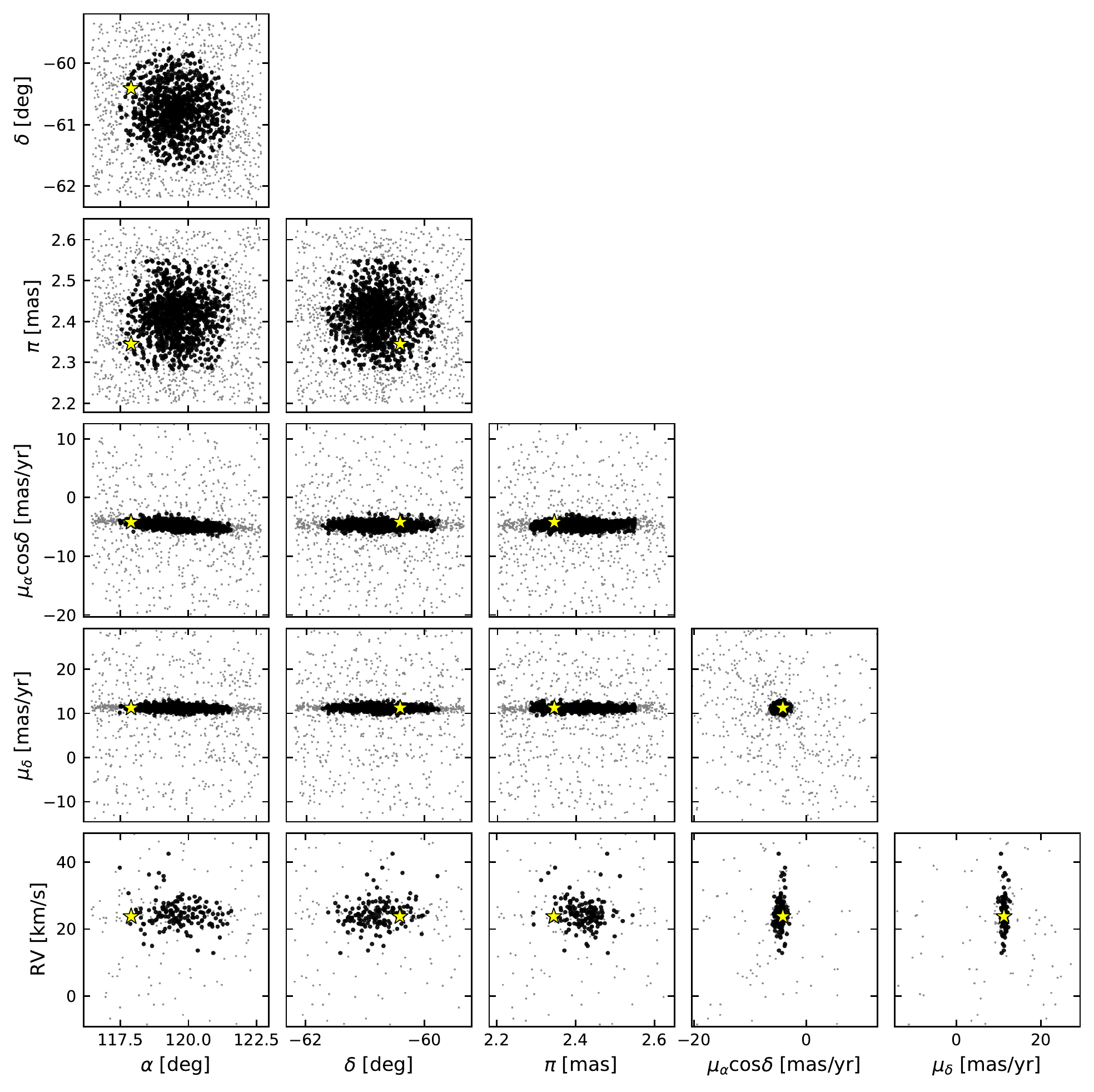}
  \caption{Kinematic analysis of TOI-681 exhibiting its membership to the open star cluster NGC 2516. Black points are NGC 2516 members reported by \citet{Cantat-Gaudin2018}. Gray points are randomly drawn from a cube
  in $\{ \alpha, \delta, \pi \}$ centered on the cluster, with
  side widths $\pm 4\sigma$, where $\sigma$ is the standard deviation of
  the \citet{Cantat-Gaudin2018} cluster member right ascension, declination, and parallax.
  }
  \label{fig:toi681kin}
\end{figure*}


\begin{table*}
\centering
Star Name \& Aliases: TOI 148, TIC 393940766, UCAC4 260-199322, 2MASS J22331696-3809349 \\
\begin{tabular}{llcc}
\hline\hline
Parameter & Unit & Source & Value \\
\hline
Stellar Parameters & & & \\
$\alpha$ & Right Ascension \text{(hh:mm:ss)} & TICv8 & 22:33:16.99 \\
$\delta$ & Declination \text{(deg:min:sec)} & TICv8 & -38:09:35.38 \\
$V_{mag}$ & V-band magnitude & TICv8 & 12.40 \\
$\varpi$ & Parallax (mas) & GAIA DR2 & $2.441\pm0.017$ \\
\vsini & Surface Rotational Velocity (\kms) & CORALIE & $10.1\pm0.8$ \\
\\
\hline
Parameter & Unit & Exofast & Value \\
 & & Priors & \\
\hline
\\
$M_*$ & Mass (\msol) & - & $0.97^{+0.12}_{-0.09}$ \\ 
$R_*$ & Radius (\rsol) & $\mathcal{G\text{[1.192,0.068]}}$ & $1.20^{+0.07}_{-0.07}$ \\
$L_*$ & Luminosity (\lsol) & - & $1.66^{+0.26}_{-0.23}$ \\
$\rho_*$ & Density (cgs) & - & $0.80^{+0.17}_{-0.12}$ \\
$\log{g}$ & Surface gravity (cgs) & - & $4.27^{+0.06}_{-0.06}$ \\
$T_{\rm eff}$ &Effective Temperature (K) & $\mathcal{G\text{[5975,150]}}$ & $5990^{+140}_{-140}$ \\
$[{\rm Fe/H}]$ &Metallicity (dex) & $\mathcal{G\text{[-0.28,0.28]}}$ & $-0.24^{+0.25}_{-0.25}$ \\
$Age$ & Age (Gyr) & - & $7.7^{+3.7}_{-3.7}$ \\
$EEP$ & Equal Evolutionary Point & - & $416^{+17}_{-41}$ \\
\\
Companion Parameters & & & \\
$M_b$ & Mass (\mj) & - & $77.1^{+5.8}_{-4.6}$ \\
$R_b$ & Radius (\rj) & - & $0.81^{+0.05}_{-0.06}$ \\
$P$ & Period (days) & - & $4.867103^{+0.000015}_{-0.000014}$ \\
$T_C$ & Time of conjunction (\bjdtdb) & - & $2458327.32980^{+0.00200}_{-0.00210}$ \\
$a$ & Semi-major axis (AU) & - & $0.0571^{+0.0021}_{-0.0017}$ \\
$i$ & Inclination (Degrees) & - & $86.85^{+0.65}_{-0.53}$ \\
$e$ & Eccentricity & - & $0.0052^{+0.0060}_{-0.0037}$ \\
$\omega_*$ & Argument of Periastron (Degrees) & - & $-63.0^{+87.0}_{-71.0}$ \\
$T_{eq}$ & Equilibrium temperature (K) & - & $1321^{+46}_{-49}$ \\
$K$ & RV semi-amplitude (m/s) & - & $8950.0^{+51.0}_{-52.0}$ \\
$\delta$ & Transit depth (fraction) & - & $0.00476^{+0.00017}_{-0.00017}$ \\
$\tau$ & Ingress/egress transit duration (days) & - & $0.01280^{+0.00170}_{-0.00170}$ \\
$T_{14}$ & Total transit duration (days) & - & $0.13790^{+0.00200}_{-0.00200}$ \\
$b$ & Transit Impact parameter & - & $0.565^{+0.061}_{-0.088}$ \\
$\rho_P$ & Density (cgs) & - & $183.0^{+45.0}_{-32.0}$ \\
log g$_P$ & Surface gravity & - & $5.471^{+0.066}_{-0.059}$ \\
$\fave$ & Incident Flux (\fluxcgs) & - & $0.69^{+0.10}_{-0.10}$ \\
\hline
\end{tabular}
\\
\begin{tabular}{llccc}
\\
Wavelength Parameters & & R & Sloani & TESS \\ 
$u_{1}$ & linear limb-darkening coeff & $0.314^{+0.056}_{-0.055}$ &$0.259^{+0.033}_{-0.031}$ &$0.262^{+0.041}_{-0.040}$ \\ 
$u_{2}$ & quadratic limb-darkening coeff & $0.304^{+0.051}_{-0.051}$ &$0.291^{+0.026}_{-0.026}$ &$0.294^{+0.035}_{-0.035}$ \\ 
\end{tabular}
\\
\begin{tabular}{llc}
 \\ 
RV Parameters & & CORALIE \\ $\gamma_{\rm rel}$ & Relative RV Offset (m/s) & $-29137^{+52}_{-53}$  \\ 
$\sigma_J$ & RV Jitter (m/s) & $69^{+23}_{-54} $ \\ 
$\sigma_J^2$ & RV Jitter Variance & $4800^{+3600}_{-4500} $ \\ 
\hline
\end{tabular}
\begin{tablenotes}
\item Priors are for the EXOFASTv2 global model only, where $\mathcal{G\text{[a,b]}}$ are Gaussian priors. The topmost parameters were not modeled with EXOFASTv2 and the source of their value is displayed.
\end{tablenotes}
\caption{\textbf{TOI-148} stellar and companion parameters.}
\label{tab:TOI-148}
\end{table*}
\begin{table*}
\centering
Star Name \& Aliases: TOI 587, TIC 294090620, HD 74162, HIP 42654, TYC 6024-00943-1, 2MASS J08413504-2211395 \\
\begin{tabular}{llcc}
\hline\hline
Parameter & Unit & Source & Value \\
\hline
Stellar Parameters & & & \\
$\alpha$ & Right Ascension \text{(hh:mm:ss)} & TICv8 & 08:41:35.02 \\
$\delta$ & Declination \text{(deg:min:sec)} & TICv8 & -22:11:39.47 \\
$V_{mag}$ & V-band magnitude & TICv8 & 	7.8 \\
$\varpi$ & Parallax (mas) & Gaia DR2 & 4.755$\pm$0.028 \\
\vsini & Surface Rotational Velocity (\kms) & TRES & 34.0$\pm$2.0 \\
\\
\hline
Parameter & Unit & Exofast & Value \\
 & & Priors & \\
\hline
\\
$M_*$ & Mass (\msol) & - & $2.33^{+0.12}_{-0.12}$ \\ 
$R_*$ & Radius (\rsol) & $\mathcal{G\text{[2.031,0.092]}}$ & $2.01^{+0.09}_{-0.09}$ \\
$L_*$ & Luminosity (\lsol) & - & $33.50^{+4.20}_{-3.90}$ \\
$\rho_*$ & Density (cgs) & - & $0.41^{+0.06}_{-0.05}$ \\
$\log{g}$ & Surface gravity (cgs) & - & $4.20^{+0.04}_{-0.04}$ \\
$T_{\rm eff}$ &Effective Temperature (K) & $\mathcal{G\text{[9800,200]}}$ & $9800^{+200}_{-200}$ \\
$[{\rm Fe/H}]$ &Metallicity (dex) & $\mathcal{G\text{[0.07,0.12]}}$ & $0.08^{+0.11}_{-0.12}$ \\
$Age$ & Age (Gyr) & - & $0.2^{+0.1}_{-0.1}$ \\
$EEP$ & Equal Evolutionary Point & - & $326^{+11}_{-16}$ \\
\\
Companion Parameters & & & \\
$M_b$ & Mass (\mj) & - & $81.1^{+7.1}_{-7.0}$ \\
$R_b$ & Radius (\rj) & - & $1.32^{+0.07}_{-0.06}$ \\
$P$ & Period (days) & - & $8.043450^{+0.000730}_{-0.000720}$ \\
$T_C$ & Time of conjunction (\bjdtdb) & - & $2458520.15829^{+0.00053}_{-0.00053}$ \\
$a$ & Semi-major axis (AU) & - & $0.1054^{+0.0018}_{-0.0018}$ \\
$i$ & Inclination (Degrees) & - & $87.93^{+1.00}_{-0.63}$ \\
$e$ & Eccentricity & - & $0.0510^{+0.0490}_{-0.0360}$ \\
$\omega_*$ & Argument of Periastron (Degrees) & - & $70.0^{+100.0}_{-100.0}$ \\
$T_{eq}$ & Equilibrium temperature (K) & - & $2062^{+55}_{-56}$ \\
$K$ & RV semi-amplitude (m/s) & - & $4580.0^{+360.0}_{-360.0}$ \\
$\delta$ & Transit depth (fraction) & - & $0.00459^{+0.00008}_{-0.00008}$ \\
$\tau$ & Ingress/egress transit duration (days) & - & $0.01660^{+0.00260}_{-0.00220}$ \\
$T_{14}$ & Total transit duration (days) & - & $0.22080^{+0.00240}_{-0.00200}$ \\
$b$ & Transit Impact parameter & - & $0.400^{+0.120}_{-0.200}$ \\
$\rho_P$ & Density (cgs) & - & $43.3^{+7.6}_{-6.5}$ \\
log g$_P$ & Surface gravity & - & $5.059^{+0.053}_{-0.054}$ \\
$\fave$ & Incident Flux (\fluxcgs) & - & $4.09^{+0.46}_{-0.42}$ \\
\hline
\end{tabular}
\\
\begin{tabular}{llc}
\\
Wavelength Parameters & & TESS \\ 
$u_{1}$ & linear limb-darkening coeff & $0.133^{+0.057}_{-0.058}$  \\ 
$u_{2}$ & quadratic limb-darkening coeff & $0.245^{+0.085}_{-0.086} $ \\ 
\end{tabular}
\\
\begin{tabular}{llc}
 \\ 
RV Parameters & & TRES \\ $\gamma_{\rm rel}$ & Relative RV Offset (m/s) & $-3280^{+330}_{-320}$  \\ 
$\sigma_J$ & RV Jitter (m/s) & $0^{+82}_{-0}$  \\ 
$\sigma_J^2$ & RV Jitter Variance & $-100^{+6900}_{-6700}$  \\ 
\hline
\end{tabular}
\begin{tablenotes}
\item Priors are for the EXOFASTv2 global model only, where $\mathcal{G\text{[a,b]}}$ are Gaussian priors. The topmost parameters were not modeled with EXOFASTv2 and the source of their value is displayed.
\end{tablenotes}
\caption{\textbf{TOI-587} stellar and companion parameters.}
\label{tab:TOI-587}
\end{table*}
\begin{table*}
\centering
Star Name \& Aliases: TOI 681, TIC 410450228, TYC 8911-00495-1, UCAC4 148-012283, 2MASS J07513479-6024448 \\
\begin{tabular}{llcc}
\hline\hline
Parameter & Unit & Source & Value \\
\hline
Stellar Parameters & & & \\
$\alpha$ & Right Ascension \text{(hh:mm:ss)} & TICv8 & 07:51:34.79 \\
$\delta$ & Declination \text{(deg:min:sec)} & TICv8 & -60:24:44.6 \\
$V_{mag}$ & V-band magnitude & TICv8 & 10.885 \\
$\varpi$ & Parallax (mas) & GAIA DR2 & 2.442$\pm$0.013 \\
\vsini & Surface Rotational Velocity (\kms) & GALAH & $30.80\pm0.78$ \\
\\
\hline
Parameter & Unit & Exofast & Value \\
 & & Priors & \\
\hline
\\
$M_*$ & Mass (\msol) & - & $1.54^{+0.06}_{-0.05}$ \\ 
$R_*$ & Radius (\rsol) & $\mathcal{G\text{[1.586,0.067]}}$ & $1.47^{+0.04}_{-0.04}$ \\
$L_*$ & Luminosity (\lsol) & - & $5.98^{+0.65}_{-0.58}$ \\
$\rho_*$ & Density (cgs) & - & $0.68^{+0.05}_{-0.05}$ \\
$\log{g}$ & Surface gravity (cgs) & - & $4.29^{+0.02}_{-0.02}$ \\
$T_{\rm eff}$ &Effective Temperature (K) & $\mathcal{G\text{[7390,150]}}$ & $7440^{+150}_{-140}$ \\
$[{\rm Fe/H}]$ &Metallicity (dex) & $\mathcal{G\text{[-0.12,0.05]}}$ & $-0.08^{+0.05}_{-0.05}$ \\
$Age$ & Age (Gyr) & $\mathcal{G\text{[0.170,0.025]}}$ & $0.170^{+0.025}_{-0.025}\dagger$ \\ 
$EEP$ & Equal Evolutionary Point & - & $264^{+6}_{-7}$ \\
\\
Companion Parameters & & & \\
$M_b$ & Mass (\mj) & - & $88.7^{+2.5}_{-2.3}$ \\
$R_b$ & Radius (\rj) & [0,5] & $1.52^{+0.25}_{-0.15}$ \\
$P$ & Period (days) & - & $15.778482^{+0.000026}_{-0.000026}$ \\
$T_C$ & Time of conjunction - 2450000 (\bjdtdb) & - & $2458546.47759^{+0.00068}_{-0.00069}$ \\
$a$ & Semi-major axis (AU) & - & $0.1449^{+0.0018}_{-0.0017}$ \\
$i$ & Inclination (Degrees) & - & $87.62^{+0.09}_{-0.11}$ \\
$e$ & Eccentricity & - & $0.0930^{+0.0220}_{-0.0190}$ \\
$\omega_*$ & Argument of Periastron (Degrees) & - & $-86.5^{+7.2}_{-5.6}$ \\
$T_{eq}$ & Equilibrium temperature (K) & - & $1143^{+26}_{-25}$ \\
$K$ & RV semi-amplitude (m/s) & - & $5209.0^{+72.0}_{-67.0}$ \\
$\delta$ & Transit depth (fraction) & - & $0.01120^{+0.00340}_{-0.00170}$ \\
$\tau$ & Ingress/egress transit duration (days) & - & $0.07207^{+0.00085}_{-0.00084}$ \\
$T_{14}$ & Total transit duration (days) & - & $0.14410^{+0.00170}_{-0.00170}$ \\
$b$ & Transit Impact parameter & - & $0.958^{+0.025}_{-0.017}$ \\
$\rho_P$ & Density (cgs) & - & $31.0^{+11.0}_{-11.0}$ \\
log g$_P$ & Surface gravity & - & $4.979^{+0.086}_{-0.130}$ \\
$\fave$ & Incident Flux (\fluxcgs) & - & $0.38^{+0.04}_{-0.03}$ \\
\hline
\end{tabular}
\\
\begin{tabular}{llccccc}
\\
Wavelength Parameters & & B & I & R & Sloang & TESS \\ 
$u_{1}$ & linear limb-darkening coeff & $0.362^{+0.048}_{-0.048}$ & $0.180^{+0.034}_{-0.034}$ & $0.132^{+0.046}_{-0.046}$ & $0.376^{+0.045}_{-0.045}$ & $0.153^{+0.033}_{-0.033}$ \\ 
$u_{2}$ & quadratic limb-darkening coeff & $0.336^{+0.049}_{-0.049}$ & $0.350^{+0.034}_{-0.035}$ & $0.278^{+0.047}_{-0.048}$ & $0.370^{+0.046}_{-0.047}$ & $0.320^{+0.033}_{-0.034}$ \\ 
\end{tabular}
\\
\begin{tabular}{llcc}
 \\ 
RV Parameters & & CORALIE & FEROS \\
$\gamma_{\rm rel}$ & Relative RV Offset (m/s) & $22120^{+96}_{-87}$ & $23750^{+140}_{-140}$ \\ 
$\sigma_J$ & RV Jitter (m/s) & $95^{+3}_{-7}$ & $0^{+0}_{-0}$ \\ 
$\sigma_J^2$ & RV Jitter Variance & $9040^{+710}_{-1400}$ & $-23000^{+21000}_{-16000}$ \\ 
\hline
\end{tabular}
\begin{tablenotes}
\item Priors are for the EXOFASTv2 global model only, where $\mathcal{G\text{[a,b]}}$ are Gaussian priors and priors in brackets represent hard limit priors. The topmost parameters were not modeled with EXOFASTv2 and the source of their value is displayed. \\ $\dagger$Age of cluster NGC 2516.
\end{tablenotes}
\caption{\textbf{TOI-681} stellar and companion parameters.}
\label{tab:TOI-681}
\end{table*}
\begin{table*}
\centering
Star Name \& Aliases: TOI 746, TIC 167418903, TYC 9177-00082-1, UCAC4 112-012172, 2MASS J06382899-6738563 \\
\begin{tabular}{llcc}
\hline\hline
Parameter & Unit & Source & Value \\
\hline
Stellar Parameters & & & \\
$\alpha$ & Right Ascension \text{(hh:mm:ss)} & TICv8 & 06:38:28.99 \\
$\delta$ & Declination \text{(deg:min:sec)} & TICv8 & -67:38:56.22 \\
$V_{mag}$ & V-band magnitude & TICv8 & 11.807 \\
$\varpi$ & Parallax (mas) & GAIA DR2 & 4.167$\pm$0.014 \\
\vsini & Surface Rotational Velocity (\kms) & CORALIE & 6.1$\pm$1.2 \\
\\
\hline
\\
\hline
Parameter & Unit & Exofast & Value \\
 & & Priors & \\
\hline
\\
$M_*$ & Mass (\msol) & - & $0.94^{+0.09}_{-0.08}$ \\ 
$R_*$ & Radius (\rsol) & $\mathcal{G\text{[0.957,0.051]}}$ & $0.97^{+0.04}_{-0.03}$ \\
$L_*$ & Luminosity (\lsol) & - & $0.89^{+0.12}_{-0.11}$ \\
$\rho_*$ & Density (cgs) & - & $1.45^{+0.12}_{-0.13}$ \\
$\log{g}$ & Surface gravity (cgs) & - & $4.44^{+0.03}_{-0.03}$ \\
$T_{\rm eff}$ &Effective Temperature (K) & $\mathcal{G\text{[5700,150]}}$ & $5690^{+140}_{-140}$ \\
$[{\rm Fe/H}]$ &Metallicity (dex) & $\mathcal{G\text{[0.01,0.29]}}$ & $-0.02^{+0.23}_{-0.23}$ \\
$Age$ & Age (Gyr) & - & $6.5^{+4.3}_{-3.9}$ \\
$EEP$ & Equal Evolutionary Point & - & $368^{+30}_{-32}$ \\
\\
Companion Parameters & & & \\
$M_b$ & Mass (\mj) & - & $82.2^{+4.9}_{-4.4}$ \\
$R_b$ & Radius (\rj) & - & $0.95^{+0.09}_{-0.06}$ \\
$P$ & Period (days) & - & $10.980303^{+0.000011}_{-0.000011}$ \\
$T_C$ & Time of conjunction - 2450000 (\bjdtdb) & - & $2458335.77067^{+0.00054}_{-0.00054}$ \\
$a$ & Semi-major axis (AU) & - & $0.0973^{+0.0028}_{-0.0026}$ \\
$i$ & Inclination (Degrees) & - & $87.03^{+0.11}_{-0.14}$ \\
$e$ & Eccentricity & - & $0.1985^{+0.0029}_{-0.0031}$ \\
$\omega_*$ & Argument of Periastron (Degrees) & - & $116.5^{+1.0}_{-1.0}$ \\
$T_{eq}$ & Equilibrium temperature (K) & - & $867^{+26}_{-25}$ \\
$K$ & RV semi-amplitude (m/s) & - & $7565.0^{+21.0}_{-22.0}$ \\
$\delta$ & Transit depth (fraction) & - & $0.01014^{+0.00140}_{-0.00091}$ \\
$\tau$ & Ingress/egress transit duration (days) & - & $0.04163^{+0.00071}_{-0.00380}$ \\
$T_{14}$ & Total transit duration (days) & - & $0.08340^{+0.00130}_{-0.00130}$ \\
$b$ & Transit Impact parameter & - & $0.911^{+0.014}_{-0.010}$ \\
$\rho_P$ & Density (cgs) & - & $119.0^{+24.0}_{-27.0}$ \\
log g$_P$ & Surface gravity & - & $5.355^{+0.052}_{-0.075}$ \\
$\fave$ & Incident Flux (\fluxcgs) & - & $0.12^{+0.01}_{-0.01}$ \\
\hline
\end{tabular}
\\
\begin{tabular}{llcccc}
\\
Wavelength Parameters & & Sloan $u$ & Sloan $z$ & TESS & V \\ 
$u_{1}$ & linear limb-darkening coeff & $0.830^{+0.088}_{-0.095}$ &$0.254^{+0.042}_{-0.042}$ &$0.312^{+0.045}_{-0.045}$ &$0.469^{+0.067}_{-0.065}$ \\ 
$u_{2}$ & quadratic limb-darkening coeff & $0.023^{+0.091}_{-0.088}$ &$0.267^{+0.036}_{-0.036}$ &$0.271^{+0.037}_{-0.038}$ &$0.248^{+0.056}_{-0.057}$ \\ 
\end{tabular}
\\
\begin{tabular}{llcc}
 \\ 
RV Parameters & & CORALIE & FEROS \\
$\gamma_{\rm rel}$ & Relative RV Offset (m/s) & $33547^{+20}_{-22}$ & $33509^{+44}_{-43}$  \\ 
$\sigma_J$ & RV Jitter (m/s) & $0^{+54}_{-0}$ & $62^{+26}_{-32}$  \\ 
$\sigma_J^2$ & RV Jitter Variance & $-400^{+3300}_{-1000}$ & $3900^{+3900}_{-2900}$  \\ 
\hline
\end{tabular}
\begin{tablenotes}
\item Priors are for the EXOFASTv2 global model only, where $\mathcal{G\text{[a,b]}}$ are Gaussian priors. The topmost parameters were not modeled with EXOFASTv2 and the source of their value is displayed.
\end{tablenotes}
\caption{\textbf{TOI-746} stellar and companion parameters.}
\label{tab:TOI-746}
\end{table*}
\begin{table*}
\centering
Star Name \& Aliases: TOI 1213, TIC 399144800, TYC 8970-00020-1, UCAC4 114-034370, 2MASS J10524799-6723161 
\begin{tabular}{llcc}
\hline\hline
Parameter & Unit & Source & Value \\
\hline
Stellar Parameters & & & \\
$\alpha$ & Right Ascension \text{(hh:mm:ss)} & TICv8 & 10:52:47.72 \\
$\delta$ & Declination \text{(deg:min:sec)} & TICv8 & -67:23:14.9 \\
$V_{mag}$ & V-band magnitude & TICv8 & 11.54 \\
$\varpi$ & Parallax (mas) & GAIA DR2 & 6.197$\pm$0.014 \\
\vsini & Surface Rotational Velocity (\kms) & CORALIE & 4.0$\pm$1.2 \\
\\
\hline
Parameter & Unit & Exofast & Value \\
 & & Priors & \\
\hline
\\
$M_*$ & Mass (\msol) & - & $0.99^{+0.07}_{-0.06}$ \\ 
$R_*$ & Radius (\rsol) & $\mathcal{G\text{[0.951,0.059]}}$ & $0.99^{+0.04}_{-0.04}$ \\
$L_*$ & Luminosity (\lsol) & - & $0.86^{+0.14}_{-0.12}$ \\
$\rho_*$ & Density (cgs) & - & $1.43^{+0.20}_{-0.13}$ \\
$\log{g}$ & Surface gravity (cgs) & - & $4.44^{+0.04}_{-0.03}$ \\
$T_{\rm eff}$ &Effective Temperature (K) & $\mathcal{G\text{[5675,175]}}$ & $5590^{+150}_{-150}$ \\
$[{\rm Fe/H}]$ &Metallicity (dex) & $\mathcal{G\text{[0.28,0.16]}}$ & $0.25^{+0.13}_{-0.14}$ \\
$Age$ & Age (Gyr) & - & $5.3^{+4.2}_{-3.4}$ \\
$EEP$ & Equal Evolutionary Point & - & $355^{+38}_{-30}$ \\
\\
Companion Parameters & & & \\
$M_B$ & Mass (\mj) & - & $97.5^{+4.4}_{-4.2}$ \\
$R_B$ & Radius (\rj) & [0,3] & $1.66^{+0.78}_{-0.55}$ \\
$P$ & Period (days) & - & $27.215250^{+0.000150}_{-0.000140}$ \\
$T_C$ & Time of conjunction (\bjdtdb) & - & $2458576.18670^{+0.00160}_{-0.00160}$ \\
$a$ & Semi-major axis (AU) & - & $0.1819^{+0.0041}_{-0.0039}$ \\
$i$ & Inclination (Degrees) & - & $88.85^{+0.12}_{-0.12}$ \\
$e$ & Eccentricity & - & $0.4983^{+0.0025}_{-0.0022}$ \\
$\omega_*$ & Argument of Periastron (Degrees) & - & $-63.2^{+0.2}_{-0.2}$ \\
$T_{eq}$ & Equilibrium temperature (K) & - & $628^{+21}_{-21}$ \\
$K$ & RV semi-amplitude (m/s) & - & $7199.0^{+23.0}_{-22.0}$ \\
$\delta$ & Transit depth (fraction) & - & $0.02900^{+0.03200}_{-0.01500}$ \\
$\tau$ & Ingress/egress transit duration (days) & - & $0.08320^{+0.00170}_{-0.00150}$ \\
$T_{14}$ & Total transit duration (days) & - & $0.16640^{+0.00330}_{-0.00300}$ \\
$b$ & Transit Impact parameter & - & $1.067^{+0.088}_{-0.066}$ \\
$\rho_P$ & Density (cgs) & - & $26.0^{+62.0}_{-18.0}$ \\
log g$_P$ & Surface gravity & - & $4.940^{+0.350}_{-0.340}$ \\
$\fave$ & Incident Flux (\fluxcgs) & - & $0.03^{+0.00}_{-0.00}$ \\
\hline
\end{tabular}
\\
\begin{tabular}{llccc}
\\
Wavelength Parameters & & R & Sloani & TESS \\ 
$u_{1}$ & linear limb-darkening coeff & $0.476^{+0.053}_{-0.053}$ & $0.316^{+0.054}_{-0.052}$ & $0.337^{+0.052}_{-0.054}$ \\ 
$u_{2}$ & quadratic limb-darkening coeff & $0.303^{+0.048}_{-0.049}$ & $0.231^{+0.046}_{-0.049}$ & $0.249^{+0.050}_{-0.051}$ \\ 
\end{tabular}
\\
\begin{tabular}{llccc}
 \\ 
RV Parameters & & CHIRON & CORALIE & FEROS \\
$\gamma_{\rm rel}$ & Relative RV Offset (m/s) & $24533^{+16}_{-17}$ & $25958^{+36}_{-39}$ & $25976^{+33}_{-33}$  \\ 
$\sigma_J$ & RV Jitter (m/s) & $34^{+28}_{-22}$ & $51^{+31}_{-39}$ & $52^{+30}_{-26}$  \\ 
$\sigma_J^2$ & RV Jitter Variance & $1200^{+2700}_{-1000}$ & $2600^{+4200}_{-2500}$ & $2700^{+4000}_{-2100}$  \\ 
\hline
\end{tabular}
\begin{tablenotes}
\item Priors are for the EXOFASTv2 global model only, where $\mathcal{G\text{[a,b]}}$ are Gaussian priors. The topmost parameters were not modeled with EXOFASTv2 and the source of their value is displayed.
\end{tablenotes}
\caption{\textbf{TOI-1213} stellar and companion parameters.}
\label{tab:TOI-1213}
\end{table*}

\end{appendix}

\end{document}